\documentclass[12pt,dvips]{article}
\usepackage{rotating}
\usepackage{axodraw}
\usepackage{epsfig}
\usepackage{color}
\usepackage{latexsym}
\newcommand{\nc}{\newcommand}
\nc{\postscript}[2] 
{\setlength{\epsfxsize}{#2\hsize}\centerline{\epsfbox{#1}}}
\nc{\bg}{B. Grzadkowski}
\nc{\non}{\nonumber}
\nc{\hc}{\hbox {h.c.}} \nc{\re}{\hbox {Re}} 
\nc{\mev}{\hbox {MeV}} \nc{\gev}{\;\hbox {GeV}} \nc{\tev}{\;\hbox {TeV}}
\def\lsim{\mathrel{\raise.3ex\hbox{$<$\kern-.75em\lower1ex\hbox{$\sim$}}}}
\def\gsim{\mathrel{\raise.3ex\hbox{$>$\kern-.75em\lower1ex\hbox{$\sim$}}}}

\nc{\prd}[3]{{\it Phys.\ Rev.}\ {{\bf D{#1}} (#2), #3}}
\nc{\prl}[3]{{\it Phys.\ Rev.\ Lett.}\ {{\bf {#1}} (#2), #3}}
\nc{\plb}[3]{{\it Phys.\ Lett.}\ {{\bf B{#1}} (#2), #3}}
\nc{\npb}[3]{{\it Nucl.\ Phys.}\ {{\bf B{#1}} (#2), #3}}
\nc{\ptp}[3]{{\it Prog.\ Theor.\ Phys.}\ {{\bf {#1}} (#2), #3}}
\nc{\zfp}[3]{{\it Z.\ Phys.}\ {{\bf C{#1}} (#2), #3}}
\nc{\epj}[3]{{\it Eur.\ Phys.\ J.}\ {{\bf C{#1}} (#2), #3}}
\nc{\mpla}[3]{{\it Mod.\ Phys.\ Lett.}\ {{\bf A{#1}} (#2), #3}}
\nc{\rmp}[3]{{\it Rev.\ Mod.\ Phys.}\ {{\bf {#1}} (#2), #3}}
\nc{\ijmpa}[3]{{\it Int.\ J.\ of\ Mod.\ Phys.}\
               {{\bf A{#1}} (#2), #3}}
\nc{\Lsp}{\;\;\;\;\;\;\;\;\;\;}  \nc{\LLLsp}{\lspace \lspace}
\nc{\lsp}{\;\;\;\;\;\;}
\nc{\spac}{\;\;\;}
\nc{\noi}{\noindent}
\nc{\beq}{\begin{equation}}   \nc{\eeq}{\end{equation}}
\nc{\bea}{\begin{eqnarray}}   \nc{\eea}{\end{eqnarray}}
\nc{\baa}{\begin{array}}      \nc{\eaa}{\end{array}}
\nc{\bit}{\begin{itemize}}    \nc{\eit}{\end{itemize}}
\nc{\ben}{\begin{enumerate}}  \nc{\een}{\end{enumerate}}
\nc{\bce}{\begin{center}}     \nc{\ece}{\end{center}}

\def\epem{e^+e^-}
\def\mplv{M_{Pl\,5}}
\def\hsm{h_{SM}}
\def\mhsm{m_{\hsm}}
\def\vhat{\widehat v}
\def\Hhat{\widehat H}
\def\vo{v_0}
\def\ho{h_0}
\def\mho{m_{\ho}}
\def\phio{\phi_0}
\def\mphio{m_{\phio}}
\def\fbi{~\mbox{fb$^{-1}$}}
\def\ie{{\it i.e.}}
\def\anti{\overline}
\def\mz{m_Z}
\def\mw{m_W}
\def\wp{W^+}
\def\wm{W^-}
\def\gam{\gamma}
\def\h{h}
\def\mh{m_{h}}
\def\mphi{m_\phi}
\def\what{\widehat}
\def\lwh{\widehat\Lambda_W}

\def\lphi{\Lambda_\phi}

\def\mphi{m_\phi}

\def\hbar{\overline h}
\def\lam{\lambda}
\def\mpl{M_{Pl}}

\def\ifmath#1{\relax\ifmmode #1\else $#1$\fi}
\def\half{\ifmath{{\textstyle{1 \over 2}}}}

\def\quarter{\ifmath{{\textstyle{1 \over 4}}}}

\def\call{{\cal L}}

\def\sig{\sigma}
\def\eps{\epsilon}
\begin{document}
\pagestyle{plain}
\pagestyle{empty} \setlength{\footskip}{2.0cm}
\setlength{\oddsidemargin}{0.5cm} \setlength{\evensidemargin}{0.5cm}
\renewcommand{\thepage}{-- \arabic{page} --}
\def\mib#1{\mbox{\boldmath $#1$}}
\def\bra#1{\langle #1 |}      \def\ket#1{|#1\rangle}
\def\vev#1{\langle #1\rangle} \def\dps{\displaystyle}
\nc{\tb}{\stackrel{{\scriptscriptstyle (-)}}{t}}
\nc{\bb}{\stackrel{{\scriptscriptstyle (-)}}{b}}
\nc{\fb}{\stackrel{{\scriptscriptstyle (-)}}{f}}
\nc{\pp}{\gamma \gamma}
\nc{\pptt}{\pp \to \ttbar}
\nc{\barh}{\overline{h}}
   \def\thebibliography#1{\centerline{REFERENCES}
     \list{[\arabic{enumi}]}{\settowidth\labelwidth{[#1]}\leftmargin
     \labelwidth\advance\leftmargin\labelsep\usecounter{enumi}}
     \def\newblock{\hskip .11em plus .33em minus -.07em}\sloppy
     \clubpenalty4000\widowpenalty4000\sfcode`\.=1000\relax}\let
     \endthebibliography=\endlist
   \def\sec#1{\addtocounter{section}{1}\section*{\hspace*{-0.72cm}
     \normalsize\bf\arabic{section}.$\;$#1}\vspace*{-0.3cm}}
\vspace*{-2cm}
\begin{flushright}
$\vcenter{
\hbox{IFT-02-10}
\hbox{UCD-02-08} 
\hbox{DFF-385-03-02}
\hbox{hep-ph/0206192}
}$
\end{flushright}
\vskip .2cm
\begin{center}
{\large\bf The Scalar Sector of the Randall-Sundrum Model}
\end{center}

\vspace*{.2cm}
\begin{center}
\renewcommand{\thefootnote} {\alph{footnote})}
{\sc Daniele DOMINICI$^{\:1),\:}$}\footnote{E-mail address:
\tt dominici@fi.infn.it}
{\sc Bohdan GRZADKOWSKI$^{\:2),\:}$}\footnote{E-mail address:
\tt bohdan.grzadkowski@fuw.edu.pl}\\
{\sc John F. GUNION$^{\:3),\:}$}\footnote{E-mail address:
\tt jfgucd@higgs.ucdavis.edu} {\sc Manuel TOHARIA$^{\:3),\:}$}\footnote{E-mail address:
\tt toharia@physics.ucdavis.edu}
\end{center}

\vspace*{.2cm}
{\sl
\bce
 $^1$ Dipartimento di Fisica, Florence University and INFN, \\
 Via Sansone 1, 50019 Sesto. F. (FI), ITALY\\
\vskip 0.1cm
$^2$ Institute of Theoretical Physics, Warsaw 
University,\\
 Ho\.za 69, PL-00-681 Warsaw, POLAND\\
\vskip 0.1cm
$^3$ Davis Institute for High Energy Physics, 
University of California Davis, \\ Davis, CA 95616-8677, USA \\
\ece
}
\vskip 0.2cm

\centerline{ABSTRACT} \vspace*{0.2cm} \baselineskip=20pt plus 0.1pt
minus 0.1pt We consider the scalar sector of the Randall-Sundrum
model.  We derive the effective potential for the Standard Model
Higgs-boson sector interacting with Kaluza-Klein excitations of the
graviton ($h_\mu^{\nu\, n}$) and the radion ($\phi$) and show that
{\it only} the Standard Model vacuum solution of $\partial V(h)/\partial
h =0$ ($\h$ is the Higgs field) is allowed.  
We then turn to our main focus: the consequences of the
curvature-scalar mixing $\xi\, R\, \Hhat^\dagger \Hhat$ 
(where $\Hhat$ is a Higgs doublet field on the visible brane), which 
causes the  physical mass eigenstates $\h$ and $\phi$
to be mixtures of the original Higgs and radion fields.
First, we discuss the theoretical constraints on the allowed parameter space.
Next, we give precise procedures for computing the $\h$
and $\phi$ couplings given the {\it physical} eigenstate masses, $\mh$
and $\mphi$, $\xi$ and the new physics scales of the model.
Relations among these new-physics scales are discussed and a set of
values not far above the smallest values required by 
precision electroweak constraints and
RunI data is chosen.
A simple result for the sum of the $ZZh$ and $ZZ\phi$ squared couplings 
relative to the $ZZ\hsm$ squared coupling is derived.
We demonstrate that this sum rule 
in combination with LEP/LEP2 data implies that 
not both the $h$ and $\phi$ can be light. We present explicit
results for the still allowed region in the $(\mh,\mphi)$ plane 
that remains after
imposing the appropriate LEP/LEP2 upper limits coming from the Higgs-strahlung
channel. In the remaining allowed region of parameter
space, we examine numerically
the couplings and branching ratios of the $h$ and $\phi$
for several cases with $\mh=120\gev$ and $\mphi\leq 300\gev$.
The resulting prospects for detection of the $\h$
and $\phi$ at the LHC, a future LC and a $\gam\gam$ collider
are reviewed. For moderate $|\xi|$, 
both the anomalous $h\to gg$ coupling
and (when $\mh>2\mphi$) the non-standard decay channel $h \to \phi\phi$
can substantially impact $h$ discovery.  
Presence of the latter is a direct signature for non-zero $\xi$. 
We find that $BR(h \to \phi\phi)$ as large as $30 \div 40\, \%$ is possible 
when $|\xi|$ is large. Conversely, if $\mphi>2\mh$ then
$BR(\phi\to hh)$ is generally large.
Detection of a light $\phi$ might require the LC.
Detection of a heavy $\phi$ might need to take into account the $\phi\to hh$
channel. The feasibility of
experimentally measuring the anomalous $gg$ and $\gam\gam$ couplings
of the $\h$ and $\phi$ is examined.
\vfill



\newpage
\renewcommand{\thefootnote}{\arabic{footnote}}
\pagestyle{plain} \setcounter{footnote}{0}
\baselineskip=21.0pt plus 0.2pt minus 0.1pt
\section{Introduction}

The Standard Model (SM) of electroweak interactions describes
successfully almost all existing experimental data. However the model
suffers from many theoretical drawbacks. One of these is the hierarchy
problem: namely, the SM can not consistently accommodate the weak
energy scale ${\cal O}(1\tev)$ and a much higher scale such as the
Planck mass scale ${\cal O}(10^{19}\gev)$.  Therefore, it is commonly
believed that the SM is only an effective theory emerging as the
low-energy limit of some more fundamental high-scale theory that
presumably could contain gravitational interactions.  In the last few
years there have been many models proposed that involve extra
dimensions. These models have received tremendous attention since they
could provide a solution to the hierarchy problem. One of the most
attractive attempts has been formulated by Randall and
Sundrum~\cite{rs}, who postulated a 5D universe with two 4D surfaces
(``3-branes''). In the simplest version,
all the SM particles and forces with the exception of
gravity are assumed to be confined to one of the 3-branes called the
visible brane.  Gravity lives on the visible brane, on the second
brane (the ``hidden brane'') and in the bulk.  All mass scales in the
5D theory are of the order of the Planck mass.  By placing the SM
fields on the visible brane, all the order Planck mass terms are
rescaled by an exponential suppression factor (the ``warp factor'')
$\Omega_0\equiv e^{-m_0 b_0/2}$, which reduces them down to the weak
scale ${\cal O}(1 \tev)$ on the visible brane without any severe fine
tuning. A ratio of $1\tev/\mpl$ (where $\mpl$
is the reduced Planck mass, $\mpl\sim 2.4\times 10^{18}\gev$)
corresponds to $m_0 b_0 /2\sim 35$. 
This is a great improvement compared to the original problem
of accommodating both the weak and the Planck scale within a single theory.

In order to obtain a consistent solution to the Einstein
equations corresponding to a low-energy effective theory 
on the visible  brane with a flat metric,
the branes must have
equal but opposite cosmological constants and these must
be precisely related to the bulk cosmological constant.
The model is defined by the 5D action:
\vspace*{-.1in}
\bea
S&=&-\int d^4x\, dy \sqrt{-\what g}\left({R\over16\pi G_5}+\Lambda\right)
\non\\
&&+\int d^4x\,\sqrt{-g_{hid}}({\cal L}_{hid}-V_{hid})
+\int d^4x\,\sqrt{-g_{vis}}({\cal L}_{vis}-V_{vis}),
\label{actionbasic}
\eea
where  $\what g^{\what\mu\what\nu}$ ($\what\mu,\what\nu=0,1,2,3,4$,
where $4$ refers to the $y$ coordinate)
is the bulk metric and 
$g_{hid}^{\mu\nu}(x)\equiv \what g^{\mu \nu}(x,y=0)$ and
$g_{vis}^{\mu\nu}(x)\equiv \what g^{\mu \nu}(x,y=1/2)$ ($\mu,\nu=0,1,2,3$)
are the induced metrics on the branes. We will use the notation
$\eps^2=16\pi G_5=1/\mplv^3$.~\footnote{Our $\mplv$ is the same
as the $M$ of \cite{Csaki:1999mp}.}
One finds that if the bulk and brane
cosmological constants are related by
$\Lambda/m_0=-V_{hid}=V_{vis}=-12m_0/\eps^2$ and if 
periodic boundary conditions identifying $(x,y)$ with $(x,-y)$ are imposed, 
then the 5D Einstein equations lead to the following metric:
\beq
ds^2=e^{-2\sigma(y)}\eta_{\mu\nu}dx^\mu dx^\nu-b_0^2dy^2, 
\label{metricz}
\eeq
where
$\sigma(y)=m_0b_0\left[y(2\theta(y)-1)-2(y-1/2)\theta(y-1/2)\right]$;
$b_0$ is a constant parameter that is not determined by the 
action, Eq.~(\ref{actionbasic}).
Gravitational fluctuations around the above background metric
will be defined through the replacements:
\beq
\eta_{\mu\nu} \to \eta_{\mu\nu}+\epsilon h_{\mu\nu}(x,y)\lsp b_0\to b_0+b(x)\,.
\label{metric}
\eeq
Below we will be expanding in powers of $\eps h_{\mu\nu} $
 and eventually $b(x)/b_0$ as well.

The paper is organized as follows. First, 
in Sec.~\ref{secpotential} we describe the
basic framework for our analysis and derive 
the effective potential
for the SM Higgs-boson sector interacting with  
Kaluza-Klein excitations of the graviton  ($h_\mu^{\nu\, n}$)
and the radion ($\phi$). We discuss the need to retain
a full form in order to show that the only consistent
minimum of this effective potential is the standard one.
In Sec.~\ref{secmixing}, we introduce the 
the curvature-scalar mixing $\xi\, R\, \Hhat^\dagger \Hhat$  and discuss its
consequences for couplings and interactions. Here, $\Hhat$ is
the Higgs field on the visible brane before any rescalings required
for canonical normalization. 
In Sec.~\ref{secpheno5}, we detail the phenomenology
of the scalar sector, including the particularly important
possibility of $h\to\phi\phi$ decays, assuming that
the new physics scale is large, $\lphi=5\tev$ (where $\lphi$
specifies the strength of the radion interactions with matter). 
We consider detection of the $h$ and $\phi$ at both
the Large Hadron Collider (LHC) and a future 
linear collider (LC), as well as in $\gam\gam$ collisions at the latter.
In Sec.~\ref{secpheno1}, we discuss the even more dramatic features
that would arise if $\lphi=1\tev$, a choice that might be
excluded with additional analysis of RunI Tevatron data
and/or precision electroweak constraints.
We summarize our results in Sec.~\ref{secfinal}.
The Appendix presents a complete tabulation of the Feynman rules we employ. 

There is  already an extensive
literature on the scalar sector phenomenology
of the Randall-Sundrum model. Studies in the absence of mixing ($\xi=0$)
include
Refs.~\cite{Bae:2000pk,Davoudiasl:1999jd,Cheung:2000rw,Davoudiasl:2000wi,Park:2000xp}.
Some aspects of $\xi\neq 0$
phenomenology appear in 
Refs.~\cite{wells_mix,csaki_mix,Han:2001xs,Chaichian:2001rq}.
In this paper, we focus 
especially on the impacts of the tri-linear couplings that
emerge only when $\xi\neq 0$ mixing is present.

\section{The effective potential}
\label{secpotential}

Our first goal is to determine the effective potential that is defined
as a collection of all non-derivative contributions to the 4D effective 
Lagrangian density. We wish to demonstrate that the the standard
vacuum defined by the stationary point of the Higgs potential is
the unique potential minimum.  It turns out that this requires
using a very complete form for the full effective potential.

In order to show that standard 4D gravity is 
reproduced by the model, and to identify scalar
degrees of freedom related to fluctuations of $b_0$,
let us assume temporarily that $h_{\mu\nu}$ is only a function 
of $x$.\footnote{In other words 
we consider here contributions from the massless zero Kaluza-Klein mode, 
see Eq.~(\ref{kk}).}
Integrating the bulk Lagrangian over the 5-th dimension one finds 
a contribution to the effective action 
(see, for example, \cite{Csaki:1999mp}\footnote{We note that
our $\eps^2$ is related to the $\kappa^2$ of \cite{Csaki:1999mp}
by $\eps^2=2\kappa^2$.}):
\beq
S_{eff}=\int d^4x\,\sqrt{- g}\left(- 
{(1-\Omega_b^2) \over \eps^2 m_0}\, 
R^{(4)}(g) +{6 \over \eps^2 m_0}
 \,(\partial_\mu \Omega_b) \, (\partial^\mu \Omega_b) 
+ \cdots \right)
\label{rad_lag}
\eeq 
where $g_{\mu\nu}(x)$ denotes the $\eta_{\mu\nu}+\epsilon
h_{\mu\nu}$ part of the metric, $R^{(4)}(g)$ is the 4D Ricci scalar
and $\Omega_b(x)\equiv e^{-m_0[b_0+b(x)]/2}$.  
Standard 4D gravity is reproduced by requiring 
\beq 
{M_{Pl}^2\over 2}={1-\Omega_0^2  \over \eps^2 m_0}.
\label{mplkappam0}  
\eeq 
where $\Omega_0\equiv e^{-m_0 b_0/2}$ is known as the warp factor
and $\mpl\sim 2.4\times 10^{18}\gev$ is the reduced Planck mass
defined as $1/\sqrt{8\pi G_4}$.
The canonically normalized
massless radion field $\phio(x)$ is defined by: 
\beq \phio(x) \equiv
\left({12 \over \eps^2 m_0}\right)^{1/2}\Omega_b(x) \simeq
\sqrt{6}\mpl\Omega_b(x)\,.  
\label{radiondef}
\eeq 
For the Lagrangian of 
Eq.~(\ref{rad_lag}), the radion is massless and there is
no potential leading to a definite vacuum expectation
value for the radion field.
This result is already apparent at the level of the RS
solution of the Einstein equations, Eq.~(\ref{metricz}), where $b_0$
appeared as a free parameter.  Therefore, some potential, $V(\phio)$, for
the radion field is necessary~\cite{GW} in order to determine its vacuum
expectation value and in consequence stabilize the distance between
the branes: 
$\langle \phio \rangle \equiv \lphi=\sqrt{6}\mpl\Omega_0$.

The SM action for the Higgs doublet $\Hhat$ on the visible brane is
\beq
S_{vis} \equiv \int d^4x\,\sqrt{-g_{vis}}({\cal L}_{vis}-V_{vis})=
\int d^4x\sqrt{-g}\left[\Omega_b^2 
D_\mu\Hhat^\dagger D^\mu \Hhat-\Omega_b^4V(\Hhat) \right],\non
\label{visaction}
\eeq
where we will show
that $V(\Hhat)$ must vanish at the potential minimum,
implying $V(\Hhat)=\lam \left(\Hhat^\dagger \Hhat-\half \hat v^2\right)^2$,
and in the first term the $\Omega_b^4$ from $\sqrt{-g_{vis}}$
is partially canceled by the $\Omega_b^{-2}$ from $g_{vis}$
in $g_{vis}^{\mu\nu} \partial_\mu\Hhat^\dagger\partial_\nu\Hhat$.
(In the final form of Eq.~(\ref{visaction}) and in subsequent
equations the flat metric $\eta_{\mu\nu}$ will be assumed 
whenever repeated indices are summed.) 

Incorporating $\Omega_0$ into the definition of the Higgs doublet
by the rescaling 
$H_0\equiv \Omega_0 \Hhat$, 
and employing the radion field $\phio$ of Eq.~(\ref{radiondef}),
we can rewrite $S_{vis}$ as
\beq
S_{vis}=\int d^4x \sqrt{-g}\left[\left({\phio\over\lphi}\right)^2 D_\mu
 H_0^\dagger D^\mu H_0-\left({\phio\over\lphi}\right)^4 V(H_0) \right],
\label{svisform}
\eeq
where the above form of $V(\Hhat)$ implies that
$V(H_0)=\lam \left(H_0^\dagger H_0- \half \vo^2\right)^2 $ with 
$\vo\equiv\Omega_0 \vhat$. Expanding around the 
(presumed) vacuum expectation values for the  
radion, $\phio \to \lphi + \phio$, and for the Higgs-boson, 
$H_0 \to {1\over \sqrt 2}(\vo+h_0+ia_0)$,
and dropping terms involving the Goldstone boson $a_0$,
one gets the following contribution to the
 effective action involving $\phio$ and $\ho$:\footnote{We will, 
of course, be dropping the derivative terms in 
Eq.~(\ref{leffphioho}) when discussing the effective potential.}
\beq
\int d^4x \sqrt{-g}\left\{
\call^{SM}(\ho)-{\phio\over\lphi}T_\mu^\mu(\ho)+\left({\phio\over\lphi}\right)^2
\left[\half\partial_\mu \ho\partial^\mu \ho-6V(\ho)\right]+
{\cal O}\left[\left({\phio\over \lphi}\right)^3\right]\right\},
\label{leffphioho}
\eeq
where $\call^{SM}(\ho)$ denotes the SM piece,
$T_\mu^\mu(\ho)=4V(\ho)-\partial_\mu \ho\partial^\mu \ho$ and
the form of $V(\ho)$ corresponding to the above $V(H_0)$ is
\beq
V(\ho)=\lam\left(\vo\ho+{\ho^2\over 2}\right)^2\,.
\label{vho}
\eeq
That $V(\ho)$ must indeed vanish as $\ho\to 0$ will be shown shortly.

From the form of the rescaled potential 
$V(\ho)$ in Eq.~(\ref{vho}), it is clear that even if the typical scale of
the 5D theory is of the order of the reduced Planck scale, $\vhat \sim M_{Pl}$, 
then $\vo=  e^{-m_0 b_0/2} \vhat \sim 1\tev$ 
for moderate values of model parameters:
$m_0 b_0 /2 \sim 35$. Therefore, the existence of the warp factor
provides a solution to the hierarchy problem, 
as it explains the large ratio of $\mpl/1\tev$. 

Keeping in mind that $h_{\mu\nu}(x,y)$ depends both on $x$ and $y$, we  
use the KK expansion in the extra dimension 
\beq
h_{\mu\nu}(x,y)=\sum_n h_{\mu\nu}^n(x){\chi^n(y)\over \sqrt b_0}
\label{kk}
\eeq
on the visible brane ($y=1/2$) to obtain
\bea
\sqrt{-g}&=& 1+{\eps\over 2}h_\rho^\rho -{\eps^2\over 4}
\left(h_\sigma^\rho h_\rho^\sigma-{1\over 2}h_\rho^\rho h_\sigma^\sigma\right)
+\mathcal{O}(\eps^3)
\non\\
&=&1+{1\over\lwh}\sum_n h_\rho^{\rho\,n}-
\left({1\over \widehat \Lambda_W}\right)^2\sum_{n,m}
\left(h_\sigma^{\rho\, n}h_\rho^{\sigma\, m}-\half 
h_\rho^{\rho\, n}h_\sigma^{\sigma\, m}\right)+ 
{\cal O}\left[\left({h^n\over \lwh}\right)^3\right]\,,\non\\
\label{rootg}
\eea
where 
\beq
\lwh \equiv {2 \sqrt{b_0}\over \eps \chi^n(1/2)}\sim
\sqrt 2 \mpl \Omega_0 \,,
\eeq
where we used $\chi^n(1/2)\sim\sqrt{m_0b_0}/\Omega_0$.
The full effective potential for radion plus Higgs is constructed
by using the expansion (\ref{rootg}) to obtain the effective potential parts of
of Eq.~(\ref{visaction}) and by including a stabilizing 
potential for the radion parameterized by a radion mass, $\mphio$:
\bea
V_{eff}^{brane}&=&\left[1+{1\over\lwh}\sum_n h_\rho^{\rho\,n}-
\left({1\over \widehat \Lambda_W}\right)^2\sum_{n,m}
\left(h_\sigma^{\rho\, n}h_\rho^{\sigma\, m}-\half 
h_\rho^{\rho\, n}h_\sigma^{\sigma\, m}\right)+\cdots\right]\times\non\\
&&\left[
\left(1+{\phio\over\lphi}
\right)^4 V(\ho)+\half \mphio^2\phio^2\right]
\label{vbrane0}
\eea
where the
shift to the perturbative $\phio$ fluctuation field $\phio\to \vev{\phio}+\phio=\lphi+\phio$
has been performed. In particular, for the radion stabilization, 
it is enough to assume some non-zero 
vacuum expectation value $b_0$ and to introduce
the mass term $\half \mphio^2\phio^2$  for the fluctuation field.

Since we will later 
investigate the vacuum structure of the theory, it will be useful
to restrict ourself to the trace part of 
$h_{\mu\nu}^n \sim \quarter 
\eta_{\mu\nu}\hbar^n$:\footnote{4D Lorentz invariance 
requires that the vacuum expectation value of
$h_{\mu\nu}^n$ be of the form $\vev{h_{\mu\nu}^n}\propto \eta_{\mu\nu}$.}
\beq
V_{eff}^{brane}=\left(1+{1\over\lwh}
\sum_n\hbar^n+{1\over 4\lwh^2}\sum_n\sum_m\hbar^n\hbar^m+\cdots\right)\left[
\left(1+{\phio\over\lphi}
\right)^4 V(\ho)+\half \mphio^2\phio^2\right].
\label{vbrane}
\eeq

In order to derive the remaining contributions to the effective potential 
coming from
the gravity fluctuation $h_{\mu\nu}$,
we temporarily drop derivatives of $b(x)$ in Eq.~(\ref{actionbasic}) 
and expand 
in powers of $\eps h_{\mu\nu}$. The leading contribution 
($\propto\epsilon^{-1}$) 
to the 5D Lagrangian density reads:
\beq
\left[\call_{branes}+\call_{bulk}\right] \longrightarrow
-{1\over \eps[b_0+b(x)] }\partial_y\left[e^{-4\sigma(y)}\left(
\partial_yh_\rho^\rho-h_\rho^\rho\sigma'\right)\right]\,
\eeq
where $\call_{bulk}$ and $\call_{branes}$ denote the bulk and brane 
Lagrangian densities, respectively.
Thus, by proper matching of bulk and brane contributions, we obtain
a total derivative, which vanishes upon integration over $y$.
The values of $V_{hid}$ and $V_{vis}$ required to get this total
derivative form are, of course,\footnote{Linear terms in an expansion around 
a solution of the equations of motion 
should vanish since the solution corresponds to 
an extremum of the action.} the same as required by the general
relativity equations. 

After some algebra, the $\mathcal{O}(1)$ term is found to be:
\beq
\left[\call_{branes}+\call_{bulk}\right]\longrightarrow
-{1\over4 [b_0+b(x)]}
e^{-4\sigma}\left\{(\partial_yh_{\mu\nu})(\partial_yh^{\mu\nu})
-(\partial_yh_\rho^\rho)^2\right\}\,.
\label{full}
\eeq
In order to find the corresponding contribution to the 4D effective potential,
 one has to expand $h_{\mu\nu}$
in KK modes and then integrate over $y$. The KK modes satisfy the following 
orthogonality conditions:
\beq
\int_{-1/2}^{1/2}e^{-2\sigma(y)|_{b(x)=0}}\chi^n(y)\chi^m(y)dy=\delta_{mn}\,,
\eeq
To apply the orthogonality relations, we will expand Eq.~(\ref{full})
in powers of $b(x)/b_0$. It is useful to keep in mind
the expression for  $b(x)$ in terms of the radion fluctuation $\phio(x)$:
\beq
{b(x)\over b_0}=-{2\over m_0b_0}\ln
\left[1+{\phio(x)\over \lphi}\right]\,.
\label{radionfluctuation}
\eeq 
Since the solution of the hierarchy problem requires $2/(m_0b_0)\simeq 1/35$,
we will make the approximation of dropping 
powers of $b(x)/b_0$ relative to $1$. 
Moreover, we will later expand in powers of 
$\phio/ \lphi\ll 1$, which 
provides an extra justification for neglecting  $b(x)/b_0$.
With this approximation,
after utilizing the orthogonality relations we find
the final result for the KK-graviton mass 
term\footnote{Without the expansion in 
  powers of $b(x)/b_0$, interpretation of
Eq.~(\ref{full})  in terms of a simple mass
term for the gravitons would be much more difficult.
The $h_{\mu\nu}^n$'s could not be interpreted as
the physical gravitons obeying the standard equations of
 motion for spin 2 particles. In order to recover the canonical
  graviton degrees of freedom one would have to redefine $h_{\mu\nu}^n$
by a $b(x)$ field-dependent factor.
  However, since in our case it is legitimate to expand in powers of
  $b(x)/b_0$, and keep only the very first constant term, we will not
  discuss this issue further here.}  in the effective Lagrangian
density\footnote{If we had used the parameterization of the metric
  proposed in \cite{csaki_mix, Charmousis:2000rg}:
\bea
ds^2 = e^{-2 \sigma(y)- 2\epsilon\ e^{2\sigma(y)} b(x)  }
\ (\eta_{\mu \nu} + \epsilon\ h_{\mu\nu})\ dx^\mu dx^\nu -
b_0^2(1+ 2 \epsilon\ e^{2\sigma(y)} b(x))^2 dy^2, \non
\eea
we would not need to expand in powers of $b(x)/b_0$ in order
to derive the interaction quadratic in the
graviton field. However, as we have
checked, the form of the graviton mass terms is the same in both approaches,
therefore we adopt here the straightforward definition given by 
Eqs.~(\ref{metricz}, \ref{metric}).}:
\beq
V_{eff}^{KK}=
{1\over4}\sum_nm_n^2\left\{h_{\mu\nu}^n h^{\mu\nu\, n}
-h_{\mu}^{\mu\,n}h_\nu^{\nu\,n}\right\}\,,
\label{vkk}
\eeq
where the KK-graviton masses are given by $m_n=m_0 x_n\Omega_0$, 
with $x_n$ denoting the 
zeroes of the Bessel function $J_1(x)$ and $\Omega_0\equiv e^{-m_0b_0/2}$.
Keeping in mind that the vacuum expectation value should satisfy
$h_{\mu\nu}^n \sim \quarter \eta_{\mu\nu} \hbar^n $ we get
\beq
V_{eff}^{KK}=
-{3\over 16}\sum_n m_n^2 (\hbar^n)^2\,.
\label{vkkmin}
\eeq
That completes the determination of the total 4D effective potential:
\bea
V_{eff}&=&V_{eff}^{brane}+V_{eff}^{KK}=\non\\
&&\left(1+{1\over\lwh}
\sum_n\hbar^n+{1\over 4\lwh^2}\sum_n\sum_m\hbar^n\hbar^m+\cdots\right)\left[
\left(1+{\phio\over\lphi}
\right)^4 V(\ho)+\half \mphio^2\phio^2\right]\non\\
&&-{3\over 16}\sum_n m_n^2 (\hbar^n)^2+\cdots\,,
\label{vefftot}
\eea
where the dots refer to terms of the order of ${\cal O}[(\eps h_{\mu\nu})^3]$.
Restricting ourself to the perturbative regime we will look for the 
minimum of $V_{eff}$ that satisfies
$\sum_n\hbar^n/\lwh\ll 1$ and $b(x)/b_0\ll 1$,
the latter being equivalent to $\phio(x)/\lphi \ll 1$.
Keeping all the terms\footnote{Since the KK-graviton mass term 
originates from contributions
of the order of $1/\lwh^2$, 
for consistency we keep the same approximation while 
expanding $\sqrt{g}$ in Eq.~(\ref{vefftot}).}
shown explicitly 
in Eq.~(\ref{vefftot}) the extremum conditions are as follows:
\bea
{\partial V_{eff}\over \partial \hbar^n}=0 &\; \Rightarrow \;&
\left({1\over\lwh}+\frac{1} {2 \lwh^2}\sum_n\hbar^n\right)
\left[ \left(1+{\phio\over\lphi} 
\right)^4V(\ho) +\frac12 \mphio^2\phio^2\right] -{3\over 8} m_n^2\hbar^n=0
\non\\ \label{min1}\\
{\partial V_{eff}\over \partial \phio}=0 &\;\Rightarrow \;&
\left(1+{1\over \lwh}\sum_n\hbar^n+
{1\over 4\lwh^2}\sum_n\sum_m\hbar^n\hbar^m\right)
4\left(1+{\phio\over\lphi}\right)^3 {V(\ho)\over \lphi} +\mphio^2\phio=0
\non\\
\label{min2}
\\
{\partial V_{eff}\over \partial h}=0 &\; \Rightarrow \;&
\left(1+{1\over \lwh}\sum_n\hbar^n+
{1\over 4\lwh^2}\sum_n\sum_m\hbar^n\hbar^m
\right) \left(1+{\phio\over\lphi}\right)^4
{\partial V(\ho)\over \partial \ho}=0\,.
\non\\
\label{min3}
\eea
There is only one solution of Eq.~(\ref{min3})
consistent with $\phio/\lphi\ll 1$ and $\hbar^n/\lwh\ll 1$: 
namely, ${\partial V(\ho)\over\partial \ho}\vert_{\vev{\ho}=0}=0$. 
For consistency of the RS model we must also require that $V(\vev{\ho})=0$.
If $V(\vev{\ho})\neq 0$, then the visible brane tension would
be shifted away from the very finely tuned RS solution to the Einstein
equations. With these two ingredients, Eq.~(\ref{min2})
requires that $\vev{\phio}=0$ at the minimum, implying that we
have chosen the correct expansion point for $\phio$, and
Eq.~(\ref{min1}) then leads to $\vev{\hbar^n}=0$, \ie\
we have expanded about the correct point in the $\hbar^n$ fields.
However, it is only if $\mphio^2>0$ that $\vev{\phio}=0$ is required
by the minimization conditions.
If $\mphio=0$, then Eq.~(\ref{min1}) still requires $\vev{\hbar^n}=0$
but all equations are satisfied for any $\vev{\phio}$.

We note that if one were to use the form
$V_{eff}=
\left(1+{1\over\lwh}\sum_n\hbar^n+4{\phio\over\lphi}\right)
V(\ho)+\left(1+{1\over\lwh}\sum_n\hbar^n\right)\half \mphio^2\phio^2
-{3\over 16}\sum_n m_n^2 (\hbar^n)^2$, then the linear term in $\hbar^n$
could be used to compensate the linear term in $\phio$ to obtain
an extremum that is apparently deeper than the standard minimum,
which minimum turns out to be tachyonic and, therefore, unphysical.
The full form with the positive definite $(1+\phio/\lphi)^4$
factor makes such a deeper extremum impossible.

Finally, we note that since $\partial V(\ho)/ \partial \ho =0$ at the minimum
(even after including interactions with the radion and KK gravitons)
there are no terms in the potential that are linear in
the Higgs field $\ho$ as shown in Eq.~(\ref{vho}).  
We will return to this observation in the next
section of the paper.

\section{The curvature-scalar mixing}
\label{secmixing}

Having determined the vacuum structure of the model,
 we are in a position to discuss the
possibility of mixing between gravity and the electroweak sector.
The simplest example of
the mixing is described by the following action~\cite{johum}: 
\beq
S_\xi=\xi \int d^4 x \sqrt{g_{\rm vis}}R(g_{\rm vis})\Hhat^\dagger \Hhat\,,
\eeq
where $R(g_{\rm vis})$ is the Ricci scalar for the metric induced 
on the visible brane,
$g^{\mu\nu}_{\rm vis}=\Omega_b^2(x)(\eta^{\mu\nu}+\eps h^{\mu\nu})$,
and we recall that 
$\Hhat$ is the Higgs field in the 5-D context before rescaling
to canonical normalization on the brane.
Using $H_0=\Omega_0 \Hhat$ and $\Omega_b(x)=\Omega_0\Omega(x)$
as before,  one obtains~\cite{csaki_mix}
\beq
\xi\sqrt{g_{\rm vis}}R(g_{\rm vis})\Hhat^\dagger \Hhat=6\xi\Omega(x)\left(-\Box\Omega(x)+
\eps h_{\mu\nu}\partial^\mu
\partial^\nu \Omega(x)+\cdots \right)H_0^\dagger H_0\,.
\label{ksiphi}
\eeq
To isolate the kinetic energy terms we again use the expansions 
\beq
H_0={1\over \sqrt 2}(\vo+\ho)\,,\quad \Omega(x)=1+{\phi_0\over \lphi}\,.
\label{expansion}
\eeq
The $h_{\mu\nu}$ term of Eq.~(\ref{ksiphi}) 
does not contribute to the kinetic energy
since a partial integration would lead to 
$h_{\mu\nu}\partial^\mu\partial^\nu \Omega=
 -\partial^\mu h_{\mu\nu}\partial^\nu\Omega=0$ by virtue
 of the gauge choice, $\partial^\mu h_{\mu\nu}=0$. 
We thus find the following kinetic energy terms:
\beq
\call=-\half\left\{1+6\gamma^2 \xi \right\}\phi_0\Box\phi_0
-\half\phi_0 \mphio^2\phi_0-\half h_0 (\Box+\mho^2)h_0-6\gamma \xi \phi_0\Box h_0\,,
\label{keform}
\eeq
where 
\beq \gamma\equiv \vo/\lphi\,.
\label{gamdef}
\eeq
In the above,
\beq
\mho^2=2\lam \vo^2\,,
\label{mhodef} 
\eeq
and $\mphio^2$ are the Higgs and radion masses before mixing.
Eq.~(\ref{keform}) differs from 
Ref.~\cite{wells_mix} by the extra $\phi_0\Box\phi_0$ 
piece proportional to $\xi$.

We define the mixing angle $\theta$ by
\beq
\tan 2\theta\equiv 12 \gam \xi Z {\mho^2\over \mphio^2-\mho^2(Z^2-36\xi^2\gam^2)}\,,
\label{theta}
\eeq
where
\beq
Z^2\equiv 1+6\xi\gam^2(1-6\xi)\equiv \beta-36\xi^2\gam^2\,.
\label{z2}
\eeq
In terms of these quantities, the states that diagonalize the kinetic energy
and have canonical normalization are $h$ and $\phi$ with: 
\bea
h_0&=&\left (\cos\theta -{6\xi\gam\over Z}\sin\theta\right)h
+\left(\sin\theta+{6\xi\gam\over Z}\cos\theta\right)\phi\equiv d h+c\phi
\label{hform}\\
\phi_0&=&-\cos\theta {\phi\over Z}+\sin\theta {h\over Z}\equiv a\phi+bh\,. \label{phiform}
\eea
(Our sign convention for $\phi_0$ is opposite that chosen for $r$
in Ref.~{\cite{csaki_mix}.)
To maintain positive definite kinetic energy terms for the $h$ and $\phi$,
we must have $Z^2>0$.  (Note that this implies that 
$\beta>0$, see Eq.~(\ref{z2}), is implicitly required.)
The corresponding mass-squared eigenvalues are~\footnote{Note that
the quantity inside the square root is positive definite so long
as $\mho^2\mphio^2>0$.}
\beq
m_\pm^2={1\over 2 Z^2}\left(\mphio^2+\beta \mho^2\pm\left\{
[\mphio^2+\beta \mho^2]^2-4Z^2\mphio^2\mho^2\right\}^{1/2}\right)\,.
\label{emasses}
\eeq
We will identify the larger of $[\mh,\mphi]$ with $m_+$.
This equation can be inverted to obtain
\beq
[\beta \mho^2,\mphio^2]={Z^2\over 2}\left[
m_+^2+m_-^2\pm\left\{(m_+^2+m_-^2)^2-{4\beta m_+^2 m_-^2\over Z^2}\right\}^{1/2}\right]\,.
\label{inversion}
\eeq
Using the symmetry of the inversion under $m_+^2\leftrightarrow m_-^2$,
we could equally well write Eq.~(\ref{inversion}) using $\mh^2$
and $\mphi^2$.
Note that 
for the quantity inside the square root appearing in Eq.~(\ref{inversion})
to be positive, we require that:~\footnote{Since $m_+>m_-$ by definition,
the second solution for the positivity condition is irrelevant.}
\beq
{m_+^2\over m_-^2}>1+{2\beta\over Z^2}\left(1-{Z^2\over\beta}\right)+{2\beta\over Z^2}\left[1-{Z^2\over \beta}\right]^{1/2}\,,
\label{rootconstraint}
\eeq
where $1-Z^2/\beta=36\xi^2\gam^2/\beta>0$.
In other words, since we will identify $m_+$ with either $\mh$
or $\mphi$, the physical states $h$ and $\phi$ 
cannot be too close to being degenerate
in mass, depending on the precise values of $\xi$ and $\gam$;
extreme degeneracy is allowed only for small $\xi$ and/or $\gam$.
We also note that
\beq
\beta \mho^2+\mphio^2=Z^2(m_+^2+m_-^2)\,,\quad
\beta \mho^2\mphio^2=Z^2\beta m_+^2m_-^2\,.
\label{inversion2}
\eeq
This leaves a two-fold ambiguity in solving for $\beta \mho^2$
and $\mphio^2$, corresponding to which we take to be the larger.
We resolve this ambiguity by requiring that $\mho^2\to \mh^2$
in the $\xi\to 0$ limit.  This means that for $\beta \mho^2$
we take the $+$ ($-$)
sign in Eq.~(\ref{inversion}) for $\mh>\mphi$ ($\mh<\mphi$),
\ie\ for $\mh=m_+$ ($\mh=m_-$), respectively.

Given this choice,
we complete the inversion by writing out the kinetic energy of
Eq.~(\ref{keform}) using the substitutions of Eqs.~(\ref{hform})
and (\ref{phiform}) and demanding that the coefficients of $-\half h^2$
and $-\half \phi^2$ 
agree with the given input values for $\mh^2$ and $\mphi^2$.
By using Eqs.~(\ref{inversion2}), it is easy to show that these
requirements are equivalent and imply
\bea
\sin 2\theta&=&{12\gam\xi  \mho^2
\over Z\left(\mphi^2-\mh^2\right)}\,.
\label{s2thform}
\eea
Note that the sign of $\sin2\theta$ depends upon whether $\mh^2>\mphi^2$
or vice versa.
It is convenient to rewrite the result for $\tan 2\theta$
of Eq.~(\ref{theta}) using Eq.~(\ref{inversion2})
in the form
\beq
\tan 2\theta={12\gam\xi \mho^2\over Z\left(\mphi^2+\mh^2-2\mho^2\right)}\,.
\label{tan2thform}
\eeq
In combination, Eqs.~(\ref{s2thform}) and (\ref{tan2thform})
are used to determine $\cos 2\theta$.
Together, $\sin 2\theta$ and $\cos 2\theta$ 
give a unique solution for $\theta$.
As a useful point of reference, we note that $\mphi^2=0$
corresponds to $\mphio^2=0$, $\beta\mho^2=Z^2\mh^2$, 
$\sin2\theta=-12\gam\xi Z/\beta$,
$\cos2\theta=-(\beta-2Z^2)/\beta$, 
$\sin\theta=-6\xi\gam/\sqrt{\beta}$, and $\cos\theta=Z/\sqrt{\beta}$.

Using this inversion, for given $\xi$, $\gam$, $\mh$ and $\mphi$ 
we compute $Z^2$ from Eq.~(\ref{z2}), $\mho^2$
and $\mphio^2$ from Eq.~(\ref{inversion}), and then
$\theta$ from Eq.~(\ref{theta}). With this
input, we can then obtain $a,b,c,d$ as defined in Eqs.~(\ref{hform}) and 
(\ref{phiform}). 

Altogether, when $\xi\neq0$ there are four 
independent~\footnote{Aside from the constraints that derive from
requiring that $Z^2>0$ and the constraint of Eq.~(\ref{rootconstraint}).}
 parameters that 
must be specified to completely fix the state mixing
parameters $a,b,c,d$ of Eqs.~(\ref{hform}) and (\ref{phiform})
defining the mass eigenstates. These are:
\beq
\xi\,,\quad \lphi\,,\quad \mh\,,\quad \mphi\,,
\eeq
where we recall that $\gam\equiv \vo/\lphi$ with $\vo=246\gev$.
Two additional parameters are required to completely fix
the phenomenology of the scalar sector, including all possible decays.
These are
\beq
\lwh\,,\quad m_1\,,
\eeq
where $\lwh$ will determine KK-graviton couplings to the $h$ and $\phi$
and $m_1$ is the mass of the first KK graviton excitation.
The parameter $\lwh$ is fixed in terms of $\lphi$ while $m_1$
depends upon $\lphi$ and the curvature parameter, $m_0/\mpl$.
We summarize the relations among all these parameters as given by
our earlier formulae:
\bea
\lwh &\equiv& {2 \sqrt{b_0}\over \eps \chi^n(1/2)}\simeq
\sqrt 2 \mpl \Omega_0 \,,\non\\
m_n&=&m_0 x_n\Omega_0\,,\non\\ \lphi&=&\sqrt6\mpl\Omega_0=\sqrt 3\lwh\,,
\label{params}
\eea
where $\Omega_0\mpl=e^{-m_0b_0/2}\mpl$ should be of order a TeV
to solve the hierarchy problem.
In Eq.~(\ref{params}), the $x_n$ are the 
zeroes of the Bessel function $J_1$ ($x_1\sim 3.8$, $x_2\sim 7.0$).
A useful relation following from the above equations is:
\beq
m_1=x_1 {m_0\over\mpl} {\lphi\over\sqrt 6}\,.
\label{m1form}
\eeq

To set the scale of $m_0$ independently of $b_0$ requires additional argument.
One line of reasoning is that of Ref.~\cite{Davoudiasl:1999jd}.
There it is argued that the 3-brane tension, 
$|V_{vis}|={12m_0\over\eps^2}$
with $\eps^2 m_0\sim 2 \mpl^{-2}$, see Eq.~(\ref{mplkappam0}),
should be roughly the same as the tension, $\tau_3$,
of a $D$ 3-brane in the heterotic string theory: 
$\tau_3={M_s^4\over g(2\pi)^3}$, where $g\sim 1$ is 
the string coupling constant and the string
scale is $M_s\sim g_{YM}\mpl$. Setting $|V_{vis}|=\tau_3$ gives
\beq
{m_0\over\mpl}\sim {g_{YM}^2 \over \sqrt 6 (2\pi)^{3/2}}\sim 0.013\,,
\label{m0mpl}
\eeq 
using $g_{YM}\sim 0.7$. Although this precise value should
probably not be taken too seriously, a reasonable range to
consider is $0.01\lsim {m_0\over\mpl}\lsim 0.1$. This guarantees
that the ratio of the bulk curvature $m_0$
to $\mplv$, ${m_0\over \mplv}\sim \left[{m_0\over \mpl}\right]^{2/3}$, 
is small, as required for reliability of the Randall-Sundrum approach.

In choosing parameters for a more detailed phenomenological study
of the scalar sector, we must be careful to avoid current bounds
deriving from RunI Tevatron data and from precision electroweak
constraints.  These have been examined in Ref.~\cite{Davoudiasl:2000wi} ---
see their Fig.~22. The smallest possible value for $m_1$ for which
it is clear that KK excitation corrections
to precision electroweak observables are 
not in conflict with existing bounds 
while at the same time all RunI 
bounds on KK excitations are satisfied is $m_1=450\gev$, for which
${m_0\over\mpl}\sim 0.05$ is required for 
simultaneous consistency. 
Inserting these values into Eq.~(\ref{m1form}) gives
$\lphi\sim 5.8\tev$. At higher $m_0/\mpl$, the naive RunI Tevatron
restriction becomes much stronger than the precision electroweak
constraint. Thus, for example, at $m_0/\mpl\sim 0.1$ we employ
the RunI Tevatron constraint of $m_1\gsim 600\gev$ from Fig.~22 of
\cite{Davoudiasl:2000wi}
to obtain $\lphi\gsim 4\tev$.  In our detailed study, we will
employ $m_0/\mpl=0.1$ and $\lphi=5\tev$, corresponding 
[see Eq.~(\ref{m1form})] to $m_1=750\gev$.  We note that this large
mass for the first KK excitation means that light (mass $\lsim 300\gev$) Higgs
bosons and radions cannot decay into KK excitations.
The full phenomenology of this scenario is explored in Sec.~\ref{secpheno5}.

Let us consider further the implications of our choice
of $\lphi=5\tev$.
From Eq.~(\ref{params}), this choice gives
$\mpl\Omega_0\sim 2.04\tev$ and thence 
$\Omega_0\sim{2000\over 2.4\times 10^{18}}\sim 0.85\times 10^{-15}$.  
This value is equivalent to $m_0b_0\sim 69$.
Again using Eq.~(\ref{params}), $\lphi=5\tev$ implies $\lwh\sim 3\tev$.
For $\mh$ and $\mphi$ we will consider a range of possibilities, but
with some prejudice towards $\mphi<\mh$.  Indeed, in
Ref.~\cite{csaki_mix} (see also \cite{Kribs:2001ic}) it is argued that 
$\mphio\sim ({\rm backreaction}){\Omega_0\mpl\over 35}$, with 
$({\rm backreaction})<1$ needed for consistency of 
their expansion. Inserting $\Omega_0\mpl\sim 2\tev$, as estimated
above for $\lphi=5\tev$, this
would correspond to $\mphio < 57\gev$.
In Ref.~\cite{GW}, it is argued that $\mphio\sim \eps \tev$
where $\eps \ll 1$ makes the radion stabilization model most natural.
This would again suggest the possibility of quite a light radion.
In fact, we shall find that the case of a light radion $\phi$ eigenstate
(which, even after mixing, still roughly corresponds to small $\mphio$)
presents a particularly rich phenomenology.  

Although large $\lphi>4\div 5\tev$ is guaranteed to avoid
conflict with all existing constraints from LEP/LEP2 and RunI
Tevatron data, it is by no means certain that such a large value
is required. For example, if $\lphi=1\tev$, 
\beq
m_1={m_0\over\mpl}1.55\tev
\eeq
ranges from $\sim 75\gev $ to $\sim 1.55\tev$
as $m_0/\mpl$ ranges from .05 to 1.  For this case, if we 
take $m_0/\mpl$ to be of order 1,
then $m_1\sim 1.55\tev$ and there are no precision
electroweak or RunI constraints. In fact, even RunII would
not probe this scenario (see Fig.~13 of \cite{Davoudiasl:2000wi}).
Of course, $m_0/\mpl>0.1$ implies large 5-dimensional curvature,
implying that corrections to the naive RS solution might be large.
Nonetheless,  in Sec.~\ref{secpheno1} 
we shall present results for $\lphi=1\tev$ first assuming that
$m_1$ is large. However, for $\lphi=1\tev$
it is also very interesting to consider small $m_0/\mpl$
and, hence, small $m_1$. As suggested in Ref.~\cite{Davoudiasl:2000wi},
the $h^1$ and subsequent resonances are very narrow 
for small $m_1$ and might have been
missed at the Tevatron.  Further, it is not clear that precision
electroweak data rules out this kind of scenario. 
In principle, one should perform an analysis
of precision electroweak constraints simultaneously taking into
account the KK excitation effects and the
radion and Higgs contributions.  Compensation between these two
classes of effects might be possible. Such an analysis
is beyond the scope of this paper.  However, we find it useful
to entertain several such scenarios at $\lphi=1\tev$
in order to explore the possible importance of 
Higgs decays to KK excitations.  
Thus, at the very end of Sec.~\ref{secpheno1} we will consider the values
$m_0/\mpl=0.065$ and $0.195$ corresponding to $m_1=100\gev$ and $300\gev$,
respectively. Referring to Fig.~22 of \cite{Davoudiasl:2000wi},
we see that both choices are  well within the RunI Tevatron nominally
excluded area, but would correspond to such narrow KK spikes
that they might have been missed.  The first choice also
leads to $S$ and $T$ electroweak observable corrections that are
too large on their own and would have to be compensated by
other contributions. The second choice leads to KK excitation
corrections to $S$ and $T$ that are small enough to be acceptable.

We now turn to the important interactions of the $h$, $\phi$ and 
$h_{\mu\nu}^n$. We begin with the $ZZ$ couplings of the $h$ and $\phi$.
The $h_0$ has standard $ZZ$ coupling while the $\phi_0$ has $ZZ$
coupling deriving from the interaction $-{\phi_0\over \lphi}T_\mu^\mu$
using the covariant derivative  portions of $T_\mu^\mu(h_0)$.  
After rewriting these interactions in terms of the mass eigenstates,
the $\eta_{\mu\nu}$ portion of the $ZZ$ couplings is given by:
\beq
\overline g_{ZZh}=\frac{g\,\mz}{c_W}\left(d+\gamma b\right)
\equiv \frac{g\,\mz}{c_W}g_{ZZ\h}\,, \quad
\overline g_{ZZ\phi}=\frac{g\,\mz}{c_W}\left(c+\gamma a\right)
\equiv \frac{g\,\mz}{c_W}g_{ZZ\phi}\,,
\label{vvcoups}
\eeq
where $g$ and  $c_W$ denote the $SU(2)$ gauge coupling 
and cosine of the Weinberg angle, respectively,
and we have adopted a notation in
which the $g$'s without the bar denote the `reduced' coupling strength
{\it relative to SM strength}. 
The $WW$ couplings are obtained by replacing $g\mz/c_W$ by $g\mw$.
As noted in \cite{csaki_mix}, there are additional contributions
to the $ZZh$ and $ZZ\phi$ couplings coming from $-{\phi_0\over\lphi}T_\mu^\mu$
for the gauge fixing portions of $T_{\mu\nu}$.  These terms vanish
when contracted with on-shell $W$ or $Z$ polarizations, which
is the physical situation we are interested in. In addition, these
extra couplings vanish in the unitary gauge. Thus, we do not
write these additional terms explicitly.
Notice also an absence of $Zh\phi$ tree level couplings.

Next, we consider the fermionic couplings of the $h$ and $\phi$.
The $h_0$ has standard fermionic couplings and the fermionic
couplings of the $\phi_0$ derive from $-{\phi_0\over \lphi}T_\mu^\mu$
using the Yukawa interaction contributions to $T_\mu^\mu$.
One obtains results in close analogy to the $VV$ couplings just
considered:
\beq 
\overline g_{f\bar{f}h}=-\frac{g\,m_f}{2\,\mw}g_{ZZ\h}\,, \quad
\overline g_{f\bar{f}\phi}=-\frac{g\,m_f}{2\,\mw}g_{ZZ\phi}\,;
\label{yuk}
\eeq
\ie\ the $f\anti f$ couplings are related to the SM couplings
by the same factors as are the $VV$ couplings.
These results for the $VV$ and $f\anti f$ couplings are summarized in
Fig.~\ref{vvfffig} of the Appendix.

For small values of $\gam$, the reduced couplings $g_{ZZh}$ and $g_{ZZ\phi}$ 
have the expansions: 
\beq
g_{ZZh}=1+{\cal O}(\gamma^2)\,, \;\;\;
g_{ZZ\phi}=
-\gamma\left(1+\frac{6\xi m_\phi^2}{\mh^2-m_\phi^2}\right) +
{\cal O}(\gamma^3)\,.
\eeq
We note that if $1-6\xi>0$ (\ie\ for $\xi$ smaller than the conformal limit
of $\xi=1/6$), then it is always possible to choose parameters so that
the $\phi$ decouples from $f\anti f$ and $VV$:
$c+\gam a=0$.  This is achieved by taking 
\beq
m_+^2={\rm max}\left\{{1\over Z^2},{1\over 1-6\xi}\right\}\mho^2\,,\quad
m_-^2={\rm min}\left\{{1\over Z^2},{1\over 1-6\xi}\right\}\mho^2\,,
\eeq
which corresponds to $\mphio^2=\mho^2/(1-6\xi)$.

The following simple and exact sum rules (independently noted
in \cite{Han:2001xs}) follow from the definitions
of $a,b,c,d$:
\beq
{\overline g_{ZZh}^2+\overline g_{ZZ\phi}^2\over \left(\frac{g\,\mz}{c_W}\right)^2}=
{\overline g_{f\anti f h}^2+\overline g_{f\anti f \phi}^2\over 
\left(\frac{gm_f}{2\mw}\right)^2}=g_{ZZ\h}^2+g_{ZZ\phi}^2=
\left[1+{\gamma^2(1-6\xi)^2\over Z^2}\right]\equiv R^2\,.
\label{summ_Z}
\eeq 
Note that $R^2>1$ is a result of the
non-orthogonality of the relations Eq.~(\ref{hform}) and Eq.~(\ref{phiform}). 
Of course, $R^2=1$ in the conformal limit, $\xi=1/6$.
It is important to note that $Z\to 0$ 
would lead to divergent $ZZ$ and $f\anti f$ couplings for the $\phi$.
As noted earlier, this was to be anticipated since $Z\to 0$ corresponds to 
vanishing of the radion kinetic term before going to canonical
normalization. 
After the rescaling that guarantees the canonical normalization, 
if $Z\to 0$ the radion coupling constants blow up: 
$g_{ZZ\phi} \sim (c+\gamma a) \simeq 1/(6 \xi \gamma Z) + {\cal O}(Z)$. 
To have $Z^2>0$, $\xi$ must lie in the region:
\beq
\frac{1}{12}\left(1-\sqrt{1+\frac{4}{\gamma^2}}\right)
\leq \xi \leq
\frac{1}{12}\left(1+\sqrt{1+\frac{4}{\gamma^2}}\right)\,.
\label{xilim}
\eeq
As an example, for $\lphi=5\tev$, $Z^2>0$ corresponds to the range
$-3.31\leq\xi\leq 3.47$.  Of course, if we choose $\xi$ sufficiently
close to the limits, $Z^2\to 0$ implies that the couplings, as
characterized by $R^2$ will become very large.  Thus, we 
should impose bounds on $\xi$ that keep $R^2$ moderate in size.
For example, for $\lphi=5\tev$,  
$R^2$ in Eq.~(\ref{summ_Z}) takes the values 2.48 and 1.96  at $\xi=-2.5$
and $\xi=2.5$, respectively.  We will impose an overall
restriction of $R\leq 5$. In practice, this bound seldom plays a role, being
almost always superseded by the bound of Eq.~(\ref{rootconstraint}) or by
constraints from precision electroweak corrections related to the $\h$
and/or $\phi$, which we roughly incorporate as described later.

Also of considerable phenomenological importance are the $h$
and $\phi$ couplings to $gg$ and $\gam\gam$.
As shown in \cite{csaki_mix}, these have anomalous contributions
in addition to the usual one-loop contributions.
(The latter must be computed after rescaling the $f\anti f$
and $VV$ couplings by $(d+\gam b)$ for the $h$
and $(c+\gam a)$ for the $\phi$.)  These anomalous
contributions can very significantly enhance the $gg$ coupling
in particular.  The Feynman rules for these vertices appear
in the Appendix and some of their phenomenological 
implications will be discussed in the next section.

The final crucial ingredient for the phenomenology that we shall consider
is the tri-linear interactions among the $h$ and $\phi$ and $h_{\mu\nu}^n$
fields.  In particular, these are crucial for 
the decays of these three types of particles. The tri-linear interactions
derive from four basic sources.
\begin{enumerate}
\item
First, we have the cubic interactions coming from 
\beq
\call\ni -V(H_0)= - \lambda (H_0^\dagger H_0 - \frac 1 2 \vo^2)^2=
-\lambda  (\vo^2 h_0^2 + \vo h_0^3 + \frac 1 4 h_0^4)\,,
\label{vhexpansion}
\eeq
after substituting $H_0=\frac {1} {\sqrt{2}} (\vo + \ho)$.
Here, the first term above implies that 
$\lambda$ is related to the bare Higgs mass
as in Eq.~(\ref{mhodef}). 
The $h_0^3$ interaction can then be expressed as
\beq
\call \ni -{\mho^2\over 2 \vo}h_0^3\,.
\eeq
\item
Second, there is the interaction of the radion 
$\phi_0$ with the stress-energy momentum tensor trace:
\beq
\call\ni -{\phi_0\over\lphi}T_\mu^\mu(h_0)=-{\phi_0\over\lphi}\left(-\partial^\rho h_0\partial_\rho h_0 + 4\lam \vo^2h_0^2\right)\,.
\eeq
\item
Thirdly, we have the interaction of the KK-gravitons with the 
contribution to the stress-energy
momentum tensor coming from the $h_0$ field:
\beq
\call\ni -{\eps\over 2}
 h_{\mu\nu}T^{\mu\nu}\ni -{1\over\lwh}\sum_n h_{\mu\nu}^n 
\partial^\mu h_0\partial^\nu h_0
\,,
\label{kkhiggs}
\eeq
where we have kept only the derivative contributions and
we have dropped (using the gauge $h_\mu^{\mu\,n}=0$) 
the $\eta^{\mu\nu}$ parts of $T^{\mu\nu}$.
\item
Finally, we have
the $\xi$-dependent tri-linear components of Eq.~(\ref{ksiphi}):
\bea
&&6\xi\Omega(x)\left(-\Box\Omega(x)+
\eps h_{\mu\nu}\partial^\mu
\partial^\nu \Omega(x)\right)H_0^\dagger H_0\ni\Bigl[- 3 \frac  \xi \lphi h_0^2 \Box \phi_0 - 6 \xi \frac {\vo} {\lphi^2}
h_0 \phi_0 \Box \phi_0
\non\\
&&\quad  - 12 \xi \frac {\vo}{\lwh \lphi}
\sum_n h_{\mu\nu}^n\partial^\mu\phi_0\partial^\nu h_0
-6 \xi \frac {\vo^2}{\lwh \lphi^2}
\sum_n h_{\mu\nu}^n\partial^\mu\phi_0\partial^\nu \phi_0\Bigr]
\label{ksitri}
\eea
where we have  employed $\partial^\mu h_{\mu\nu}^n=0$, used
the traceless gauge condition $h_\mu^{\mu\, n}=0$, and also
used the symmetry of $h_{\mu\nu}$. 
\end{enumerate}
We discuss briefly why several kinds of tri-linear interactions are absent.
First, there are no $h^n h_0h_0$ tri-linear vertices
other than that appearing in Eq.~(\ref{kkhiggs}). 
Other possible sources are zero in the gauge we employ. In particular,
consider the $h^n \ho\ho$ 
interactions that arise in Eqs.~(\ref{svisform}) [after expanding
$\sqrt{-g}$ as in Eq.~(\ref{vbrane0})] from the kinetic energy
derivative terms and from expanding $V(H_0)$
about the minimum as in Eq.~(\ref{vhexpansion}).  The Lorentz
structure of these (and other such tri-linear terms)
can only be of the form  $h_\rho^{\rho\, n} h_0^2$ or $h_\rho^{\rho\,
 n} \partial_\mu h_0 \partial^\mu h_0$, both of which
 are absent in the $h_\rho^{\rho\, n}=0$ gauge.
Next, there is the possibility of $\phi_0^3$ interactions.  In
our derivations we have considered only interactions
generated after including the stabilizing $\half\mphio^2\phio^2$
radion mass term. We have examined the expectation for the $\phio^3$
interaction in the context of the Goldberger-Wise stabilization mechanism
\cite{GW}. Carrying their procedure to the $\phio^3$ level gives
an interaction of strength $\sim {\mphio^2\over \mpl}\phio^3$.
Thus, there seems to be at least one approach in which there is excellent
justification for neglecting $\phio^3$ interactions in our treatment.

There is another generic class of 
tri-linear interaction term that can arise, involving two
$h^n$'s and one $\phi$. 
For example, such an interaction arises if we  
retain the $b(x)/b_0$ term in the expansion of Eq.~(\ref{full}) [
$(b_0+b(x))^{-1}\sim b_0^{-1}(1-b(x)/b_0+\ldots)$]
and use [see Eq.~(\ref{radionfluctuation})],
${b(x)\over b_0}\sim {-2\over m_0b_0}
{\phi_0(x)\over \lphi}$.  The resulting contribution to the Lagrangian
takes the form
\beq
-{1\over 4}{1\over \lphi }{2\over m_0b_0}
\sum_nm_n^2\left\{h_{\mu\nu}^n h^{\mu\nu\, n}
-h_{\mu}^{\mu\,n}h_\nu^{\nu\,n}\right\}\phi_0(x)\,.
\eeq
Using our earlier numerical estimates,  the effective coupling
for this interaction is of order:
\beq
{m_1^2\over 2\lphi  m_0 b_0}
\sim {(0.75\tev)^2\over  10\tev\times 69}\sim 0.8\gev\,.
\eeq
Keeping this interaction small is a natural result
of having a small value of $m_0/\mpl$. 
There are actually many other sources of $h^nh^n\phi$ interactions
that could be retained by a more exact treatment
of the various Lagrangian contributions.  As another example,
in the reduction to Eq.~(\ref{rad_lag}), one approximates
$g_{\mu\nu}$ by $\eta_{\mu\nu}$ in obtaining the second term.
If one instead inserts the full expansion of $\sqrt{g_5}$
in terms of the $h_{\mu\nu}(x,y)$ fields out to
order $\eps^2$, and uses the eigenexpansion
of Eq.~(\ref{kk}), $h^nh^n\phi_0$ interactions are generated
with a coefficient magnitude similar in size to that estimated above.
Note that there has been some discussion of the possible nature of the
$h^nh^n\phi$ coupling in Ref.~\cite{Delbourgo:2000nq}, where it is stated
that it can only appear at one loop. The coupling generated
in the ways mentioned above does not conform to their assumptions. 
In any case, 
we saw earlier that the $h^n$ KK excitations must be
very massive for choices of $m_0/\mpl$ and $\lphi\gsim 4\div 5\tev$
that clearly satisfy the combined constraints from RunI Tevatron data
and precision electroweak constraints. Even for the $\lphi=1\tev$
choice discussed earlier, which requires relaxing the naive RunI and precision
electroweak constraints, an $m_1$ value below $100\gev$ would
be highly improbable. As a result, $h\to \h^n\h^n$
or $\phi\to h^n h^n$ decays (that would be induced by the
above interactions after `rotating' to the mass eigenstates, 
$\phi$ and $h$) are not relevant for the modest
$\mh$ and $\mphi$ values explored in the bulk of this paper. Thus, 
we have not worked out a full expression for this vertex.

To proceed with the tri-linear interactions enumerated earlier,
we substitute for $h_0$ and $\phi_0$ 
in terms of the physical $h$ and $\phi$
states using Eqs.~(\ref{hform}) and (\ref{phiform}), respectively.
The results for the tri-linear vertices 
generated, after this substitution 
into the enumerated interactions, appear in Fig.~\ref{trilinearfig}
in the Appendix.
Note that the Feynman rules generated are specified in part 
by terms containing the parameter $\mho^2$; $\mho^2$ must be computed
from $\mh^2$ and $\mphi^2$ using the inversion procedure given earlier.
Since the effective potential shown in Eq.~(\ref{vefftot})
does not contain any interactions linear in the Higgs field, vertices
like  $\phi^2 h$ and $h^n\phi h$ 
are a clear indication for the curvature-Higgs mixing.
As we shall see, they could also be of considerable phenomenological
importance. It is also useful to note that 
since the $\xi$-mixing angle $\theta$ for $\gamma \ll 1$ 
is proportional to $\gamma$, Eq.~(\ref{theta}), 
the interaction terms, Eqs.~(\ref{kkhiggs}) and 
(\ref{ksitri}), are suppressed by at least 
one power of $1/\lphi$
or $1/\lwh$ and as a result the related 
couplings will be of the order of $1/\lphi^2$ 
or $1/(\lphi\lwh)$. A useful reference is the small $\gam$ limits
of the couplings. For instance, the two couplings that vanish
linearly as $\xi\to 0$ have the limits: 
\bea 
\anti g_{n\phi h}\lwh&=&12\gamma\xi\left(-3+{2\over x}\right)+{\cal
O}(\gamma^3)\,,
\label{gnphlim} \\ 
\anti g_{\phi \phi\h}\lphi&=& 12\gamma\xi \mh^2\left[
-x(1-6\xi)+(4-21\xi)-{3(1-8\xi)\over x}-{9\xi \over x^2}\right] 
+{\cal O}(\gamma^2)\,,
\label{gpphlim}
\eea 
where we have employed the results for $\anti g_{n\phi h}$ 
and $\anti g_{\phi\phi\h}$
given in Fig.~\ref{trilinearfig} of the 
Appendix and defined $x\equiv 1-{\mphi^2/\mh^2}$.

\section{Phenomenology for {\boldmath $\lphi=5\tev$}}
\label{secpheno5}

We begin by discussing the
restrictions on the $h,\phi$ sector imposed by LEP Higgs-boson searches.
 LEP/LEP2 provides 
an upper limit for the coupling of a $ZZ$ pair to a scalar~($s$) 
as a function of the scalar mass. Because the decays of the $h$
and $\phi$ can be strongly influenced by the $\xi$ mixing,
it is necessary to consider  
limits that are obtained both with and without making use of $b$ tagging.
The most recent paper 
on the `flavor-blind' limits obtained without $b$ tagging 
is Ref.~\cite{Abbiendi:1998rd}.\footnote{There is
a much earlier paper \cite{Buskulic:1993gi} which claims much stronger
limits at low scalar masses $\lsim 20\gev$ in the 
case where the scalar decays to any of a certain class of modes.
In particular, \cite{Buskulic:1993gi} gives increasingly
strong limits on $g_{ZZ\phi}^2$ as $\mphi$ decreases below $8\gev$,
the 95\% CL limits being $\sim 0.005$ at $\mphi\sim 0$ and
$\sim 0.02$ at $\mphi\sim 10\gev$ (using the curve in which
the scalar is assumed to decay to the final states to which
a SM Higgs boson would decay, but not necessarily with
the same branching ratios). The caveat is that  $\phi\to gg$
decays are dominant in this region and it is unclear whether
or not the limits of Ref.~\cite{Buskulic:1993gi} apply.
In particular, the $gg$ final state might have a higher
multiplicity of pions at modest $\mphi$ than allowed for
in the analysis. For this reason,
we do not employ the results of \cite{Buskulic:1993gi}.  Even if
employed, they do not result 
in any additional excluded parameter regions in the case of $\lphi=5\tev$.}  
Next, there is a preliminary OPAL note \cite{pn495}
in which decay-mode-independent limits on the $ZZs$ coupling
are obtained that  are considerably stronger 
than those of \cite{Abbiendi:1998rd}, but not as strong as those of 
\cite{Buskulic:1993gi}.
For scalar masses above $60\gev$, the flavor-blind 
limits of the above references are superseded by the results
found on the LEPHIGGS working group homepage \cite{teixeiradias},
which extend up to $m_s\lsim 113\gev$.
We have chosen to employ \cite{Abbiendi:1998rd} for $m_s<60\gev$
and \cite{teixeiradias} for $60\gev\leq m_s\leq 113\gev$.
Including the stronger limits of \cite{Buskulic:1993gi} and/or \cite{pn495}
would have no impact on the plots presented.
Next, we have the limits on $g_{ZZs}^2$ obtained using $b$ tagging
and assuming that $BR(s\to b\anti b)=BR(\hsm\to b\anti b)$. 
The best limits that we have found are those contained 
in \cite{Abbiendi:1998rd} for $m_s<60\gev$ and in
\cite{:2001xw} for $60\gev \leq m_s\leq 115\gev$. In implementing
these limits, we correct the the difference between $BR(\h\to b\anti b)$
or $BR(\phi\to b\anti b)$ compared to $BR(\hsm\to b\anti b)$
computed assuming $\mhsm=\mh$ or $\mhsm=\mphi$, respectively,
and using $\lphi=5\tev$.

\begin{figure}[p!]
\vspace*{-.1in}
\begin{center}
\includegraphics[width=4in]{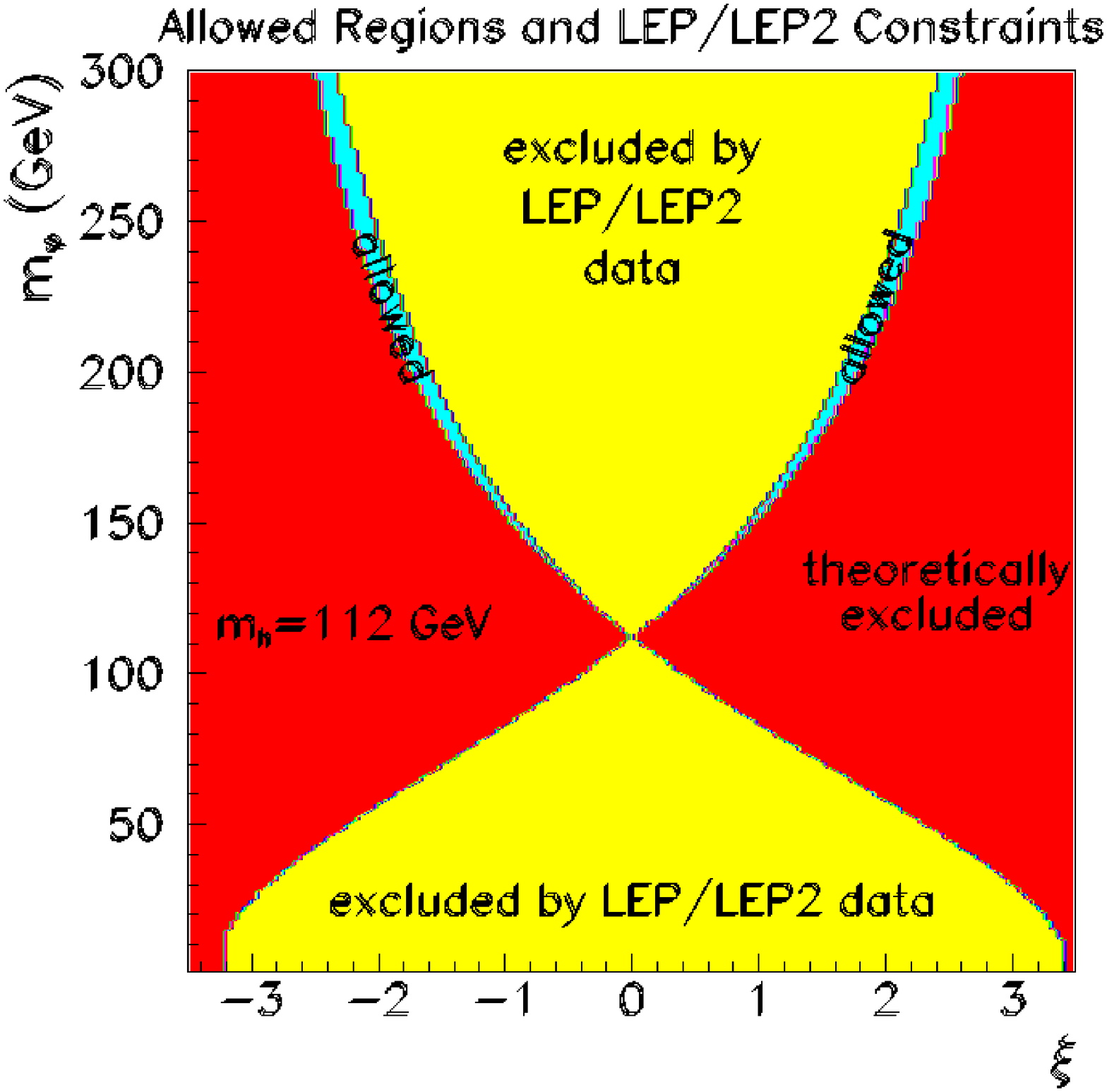}
\end{center}
\vspace*{-.2in}
\caption{Allowed regions (see text)
in $(\xi,\mphi)$ parameter space for $\lphi=5\tev$
and $\mh=112\gev$. The dark red portion of parameter
space is theoretically disallowed. The light yellow  portion
is eliminated by LEP/LEP2 constraints
on the $ZZs$ coupling-squared $g_{ZZs}^2$ or on
$g_{ZZs}^2BR(s\to b\anti b)$, with $s=\h$ or $s=\phi$. 
}
\label{allowed_mh112}
%
\begin{center}
\includegraphics[width=4in]{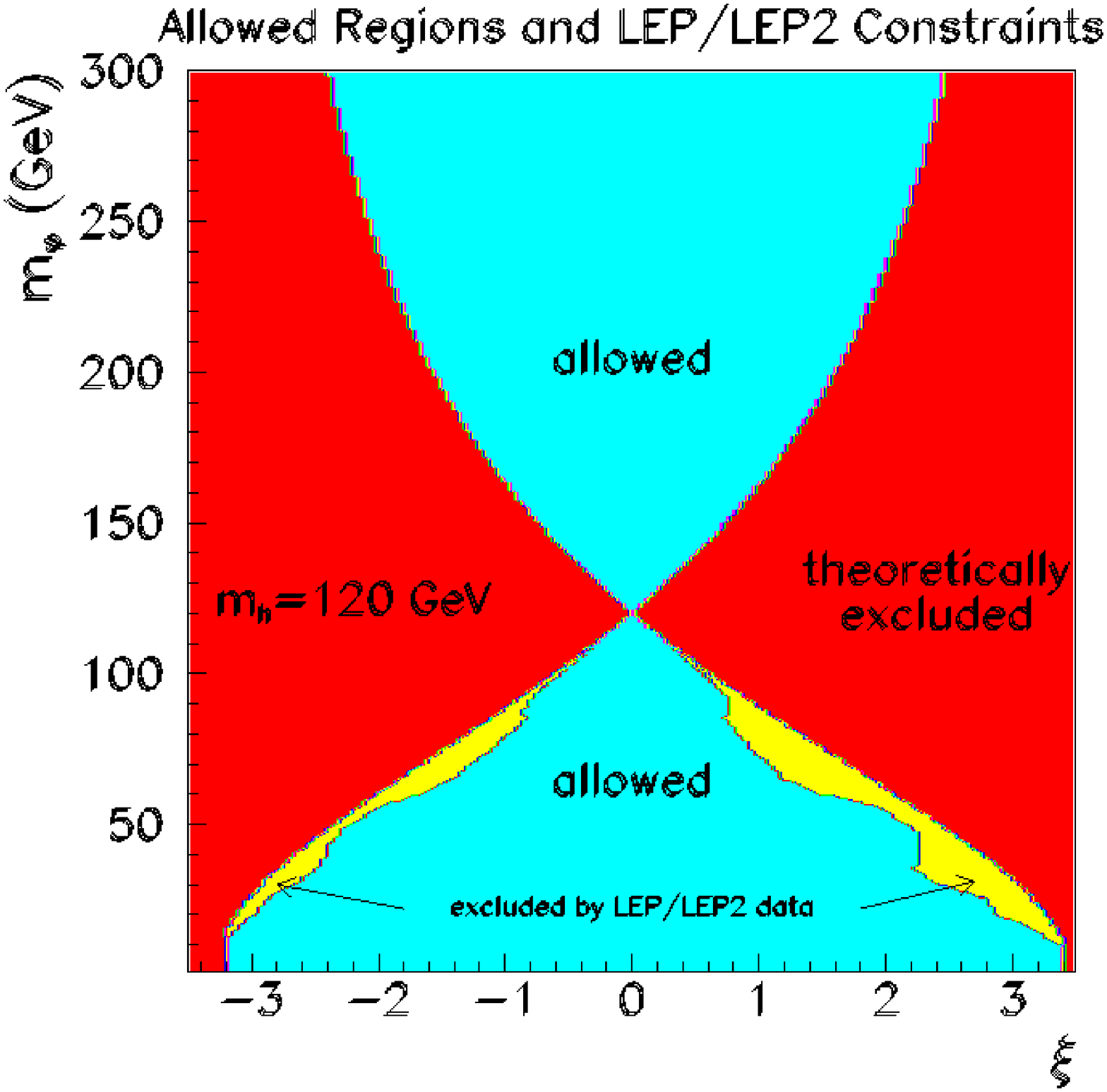}
\end{center}
\vspace*{-.3in}
\caption{As in Fig.~\ref{allowed_mh112} but for
$\mh=120\gev$. 
}
\label{allowed_mh120}
\end{figure}

The first question that arises is whether {\it both} the $\phi$ and the $h$
could be light without either having been detected at LEP and LEP2.
The sum rule of Eq.~(\ref{summ_Z}) implies that this is impossible
since the couplings of
the $h$ and $\phi$ to $ZZ$ cannot both be suppressed.  
For any given value of $\mh$ and $\mphi$, the range of $\xi$ is limited
by: (a) the constraint of Eq.~(\ref{rootconstraint})
limiting $\xi$ according to the degree of $\mh$--$\mphi$ degeneracy; 
(b) the constraint that $Z^2>0$, Eq.~(\ref{z2}); and (c)
the requirements that $g_{ZZh}^2$ and $g_{ZZ\phi}^2$ both lie
below any relevant LEP/LEP2 limit.  The regions in the $(\xi,\mphi)$ plane
consistent with the first two constraints as well as $R<5$ are shown in
Figs.~\ref{allowed_mh112} and \ref{allowed_mh120} for $\mh=112\gev$
and $\mh=120\gev$, respectively, 
assuming a value of $\lphi=5\tev$.  
For the most part, it is the degeneracy constraint (a)
that defines the theoretically acceptable regions shown.
The regions within the theoretically acceptable
regions that are excluded by the LEP/LEP2
limits are shown by the yellow  shaded regions,
while the allowed regions are in blue.
For $\mh=112\gev$, the LEP/LEP2 limits
exclude a large portion of the theoretically consistent parameter space.
For $\mh=110\gev$ (not plotted), the sum rule of Eq.~(\ref{summ_Z})
results in all of the
theoretically allowed parameter space being excluded by LEP/LEP2 constraints.
For $\mh=120\gev$, the LEP/LEP2 limits do not apply to the $\h$
and it is only for $\mphi\lsim 115\gev$ and significant $g_{ZZ\phi}^2$
(requiring large $|\xi|$) that some points are ruled out
by the LEP/LEP2 constraints. As a result, the allowed region is dramatically
larger than for $\mh=112\gev$.
The precise regions shown are somewhat sensitive to the $\lphi$ choice,
but the overall picture is always similar to that presented here
for $\lphi=5\tev$. This is illustrated in Sec.~\ref{secpheno1}, 
where the allowed regions for $\mh=120\gev$ are shown
in the case of $\lphi=1\tev$.

\begin{figure}[p!]
\begin{center}
\includegraphics[width=6in]{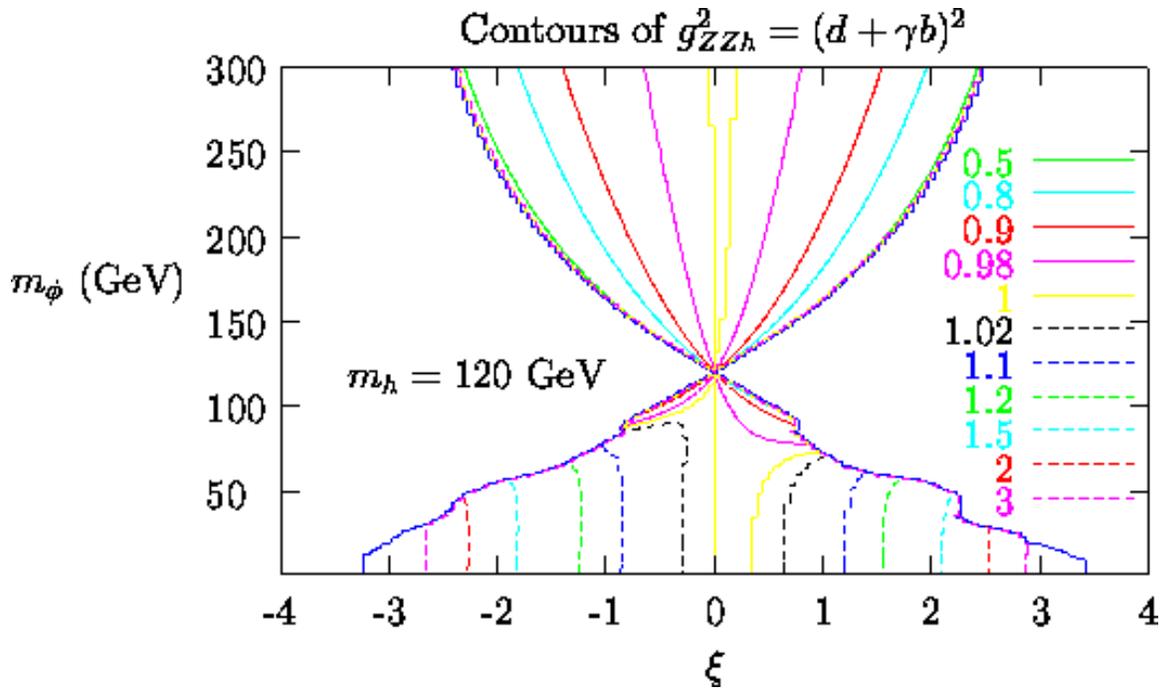}
\vspace*{.2in}

\includegraphics[width=4in,angle=90]{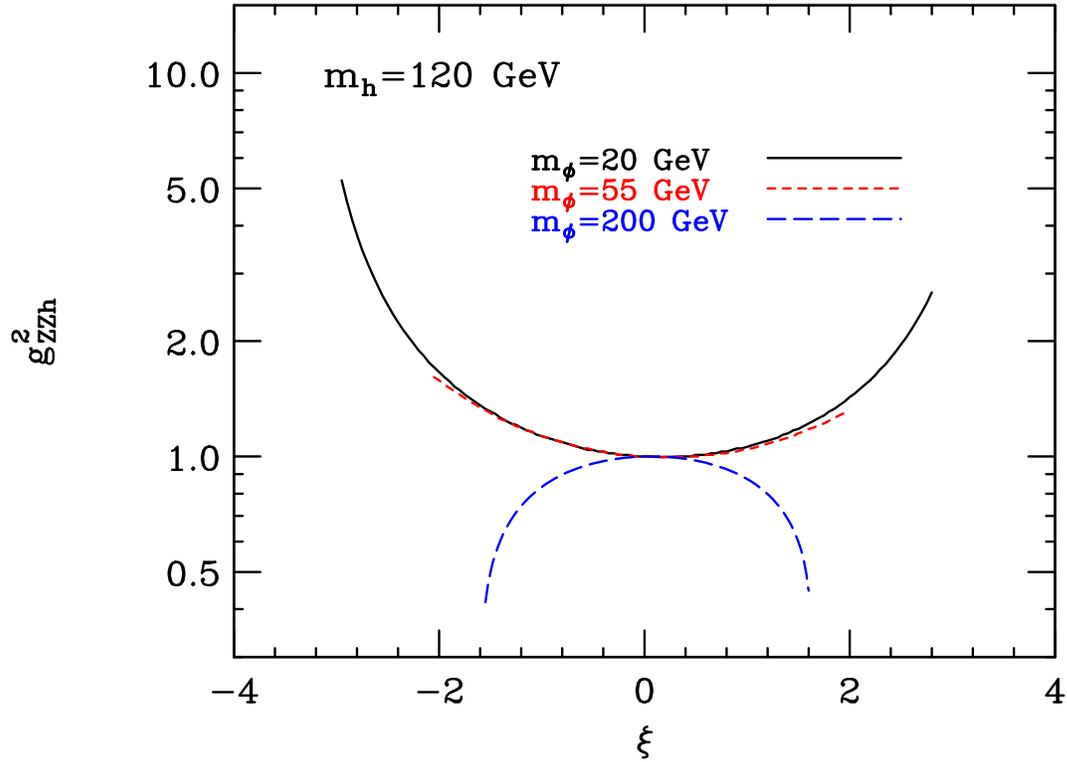}
\end{center}
\caption{For $\mh=120\gev$ and $\lphi=5\tev$, we plot 
the quantity $g_{ZZ\h}^2=(d+\gam b)^2$ 
which specifies the ratio of the $h$'s 
$f\anti f$ and $VV$ couplings squared to the corresponding values
for the SM Higgs boson, taking $\mhsm=\mh$. 
In the upper figure we show contours; line colors/textures drawn actually
on the boundary should be ignored. The lower figure presents
results for $\mphi=20$, $55$ and $200\gev$.}
\label{couplingsh}
\end{figure} 

\begin{figure}[p!]
\begin{center}
\includegraphics[width=6in]{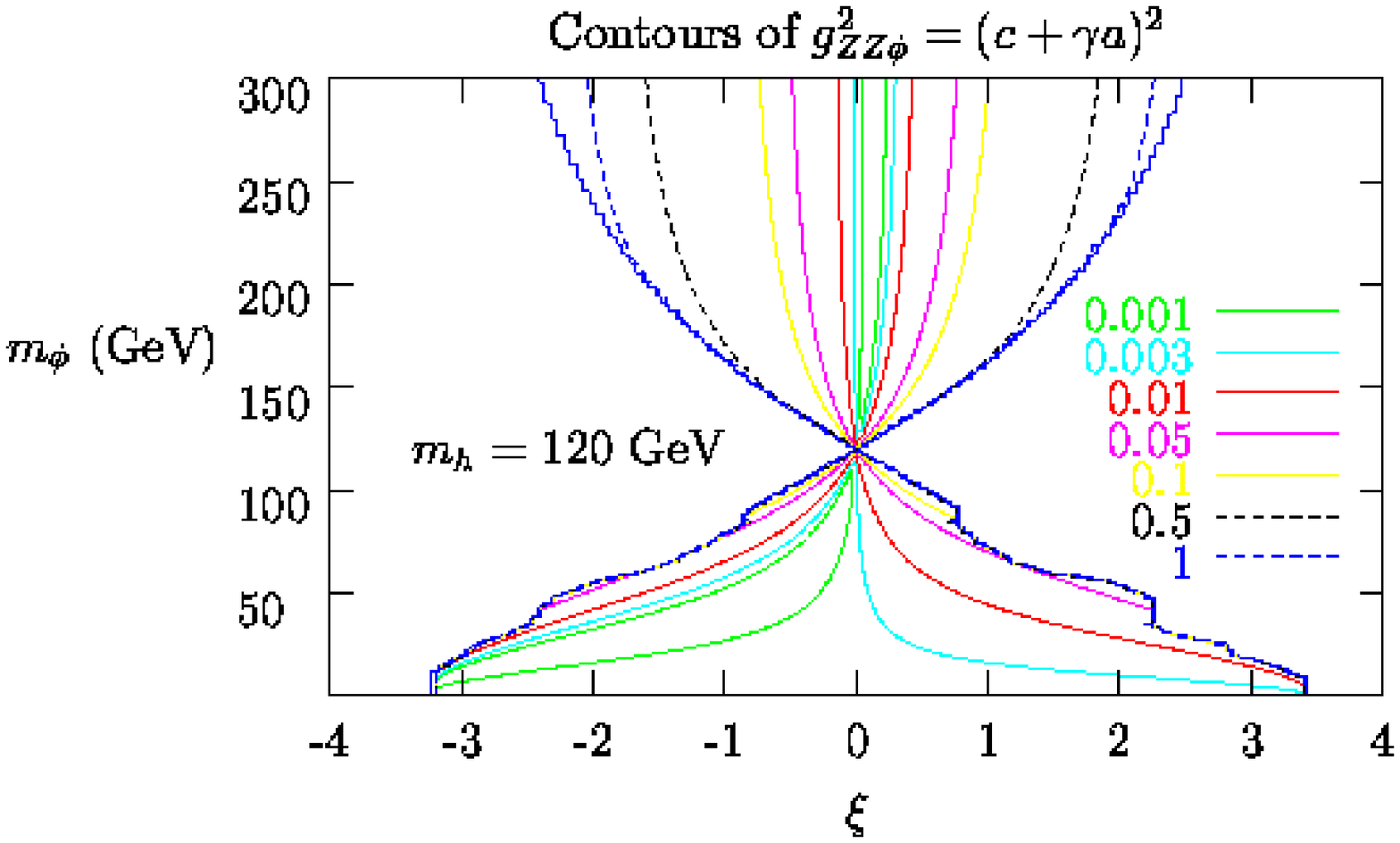}
\vspace*{.2in}

\includegraphics[width=4in,angle=90]{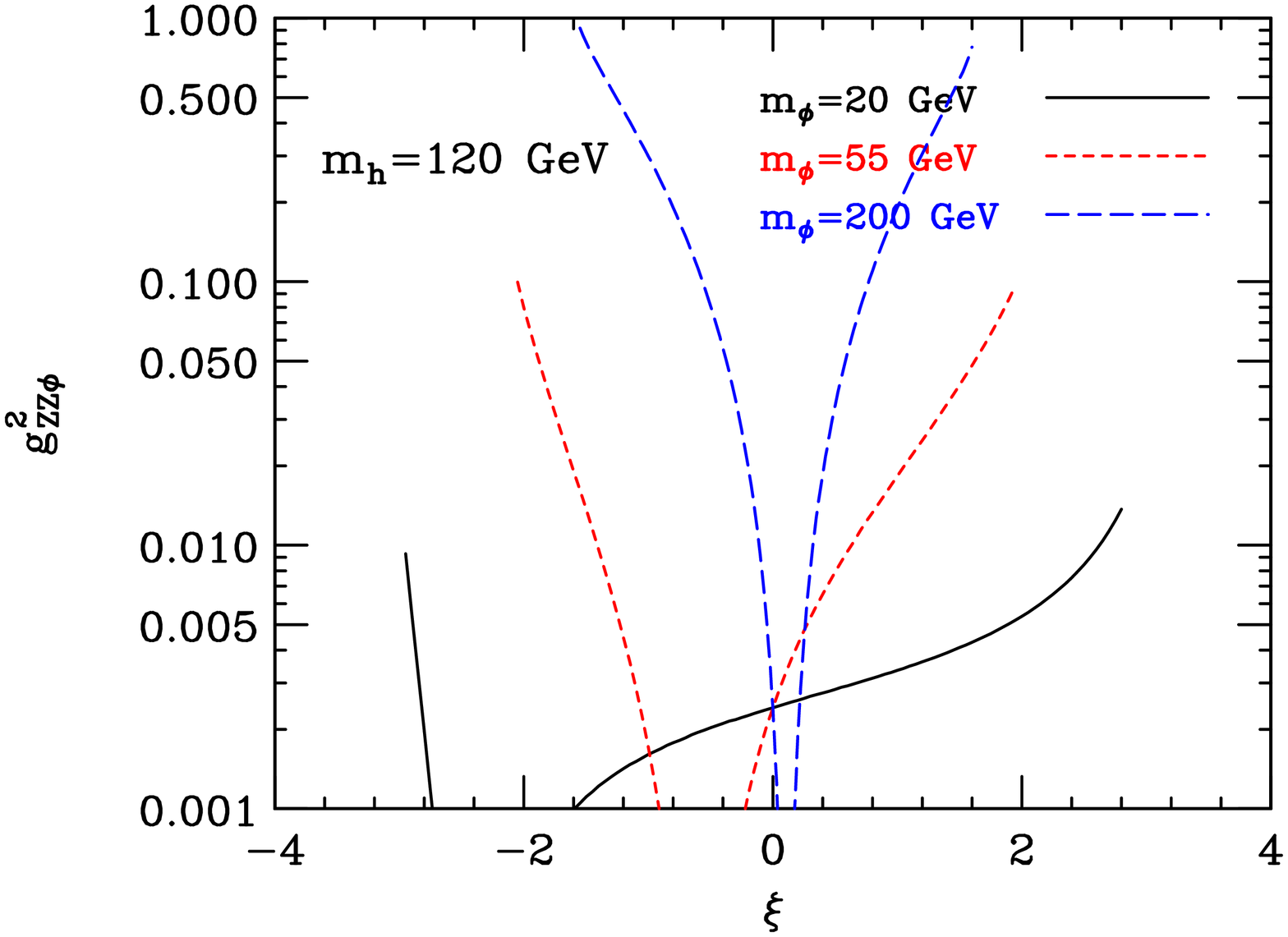}
\end{center}
\caption{As in Fig.~\ref{couplingsh}, except for the $\phi$.
Note the zeroes in the middle of the allowed $\xi$ ranges.}
\label{couplingsphi}
\end{figure}

Next, we discuss the couplings of the $\h$ and $\phi$.
We begin with their $f\anti f$ and $VV$ couplings-squared.
These are illustrated in Figs.~\ref{couplingsh} and \ref{couplingsphi}.  
There we consider $\mh=120\gev$
and $\lphi=5\tev$ (for which the allowed region was plotted
in Fig.~\ref{allowed_mh120}) and plot 
(in the upper figures) contours of $g_{ZZh}^2\equiv (d+\gam b)^2$
and $g_{ZZ\phi}^2\equiv (c+\gam a)^2$.  
As in Eqs.~(\ref{vvcoups}) and (\ref{yuk}), these quantities 
specify the `reduced' couplings
squared of the $h$ and $\phi$, respectively, to $f\anti f$ and $VV$
with respect to the squared coupling strength of the SM Higgs boson.
The lower figures show the variation of these couplings with
$\xi$ at fixed $\mphi=20$, $55$ and $200\gev$.
Large enhancements of $(d+\gam b)^2$ are possible for small $\mphi<\mh$
as are large suppressions when $\mphi>\mh$. For the $\phi$,
$(c+\gam a)^2$ is smaller than 1 except for the largest
$|\xi|$ values at high $\mphi$.   Indeed, $(c+\gam a)^2\ll 1$
is the norm in the $\mphi<\mh$ portion of parameter space
and for small $|\xi|$ when $\mphi>\mh$. In particular, there
is a line along which $(c+\gam a)^2=0$ between the paired contour
lines corresponding to $(c+\gam a)^2=0.001$; these zeroes are also apparent
from the lower plot of Fig.~\ref{couplingsphi}.

\begin{figure}[p!]
\includegraphics[width=5.7in]{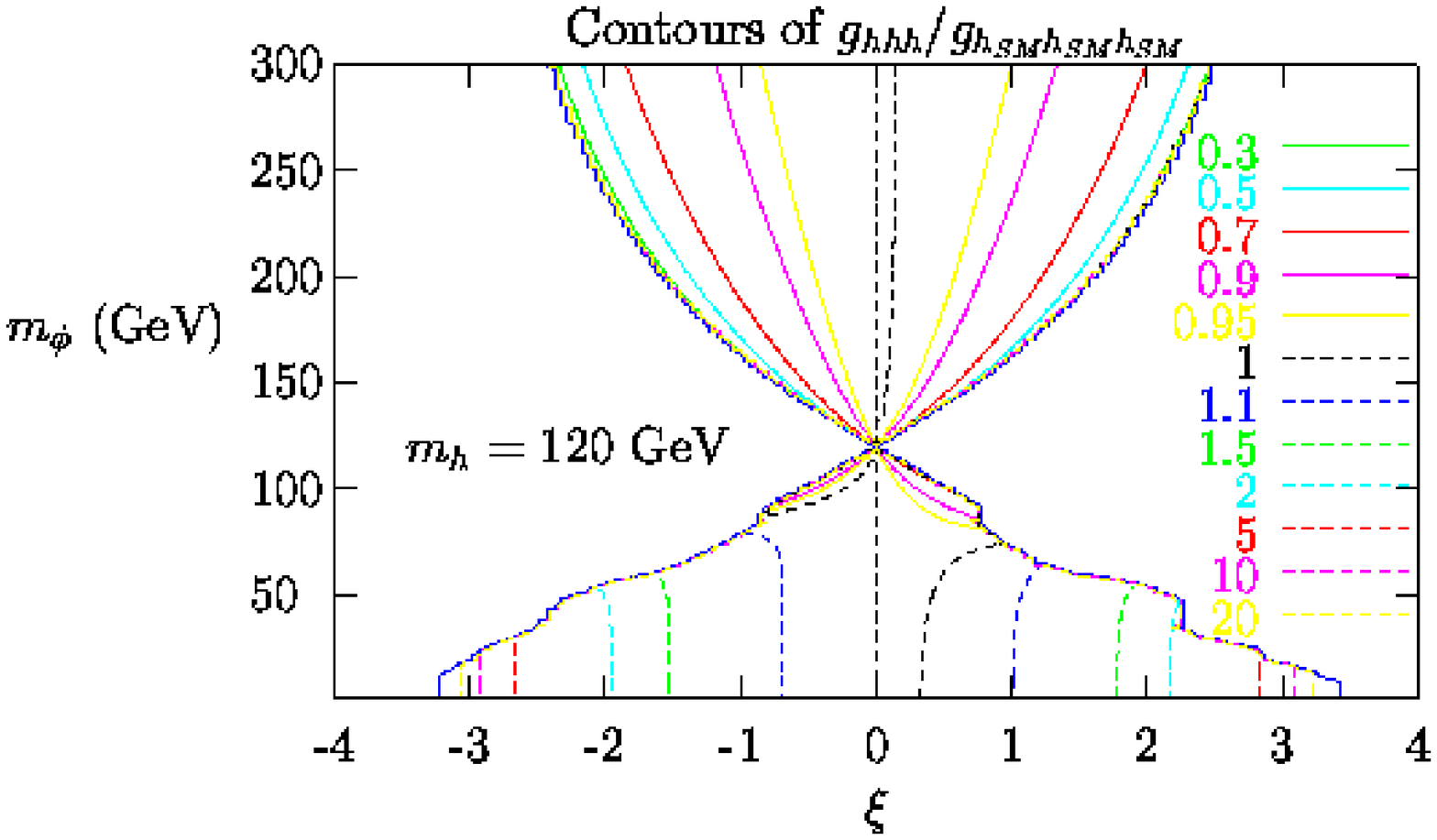}
\vspace*{.2in}

\includegraphics[width=4in,angle=90]{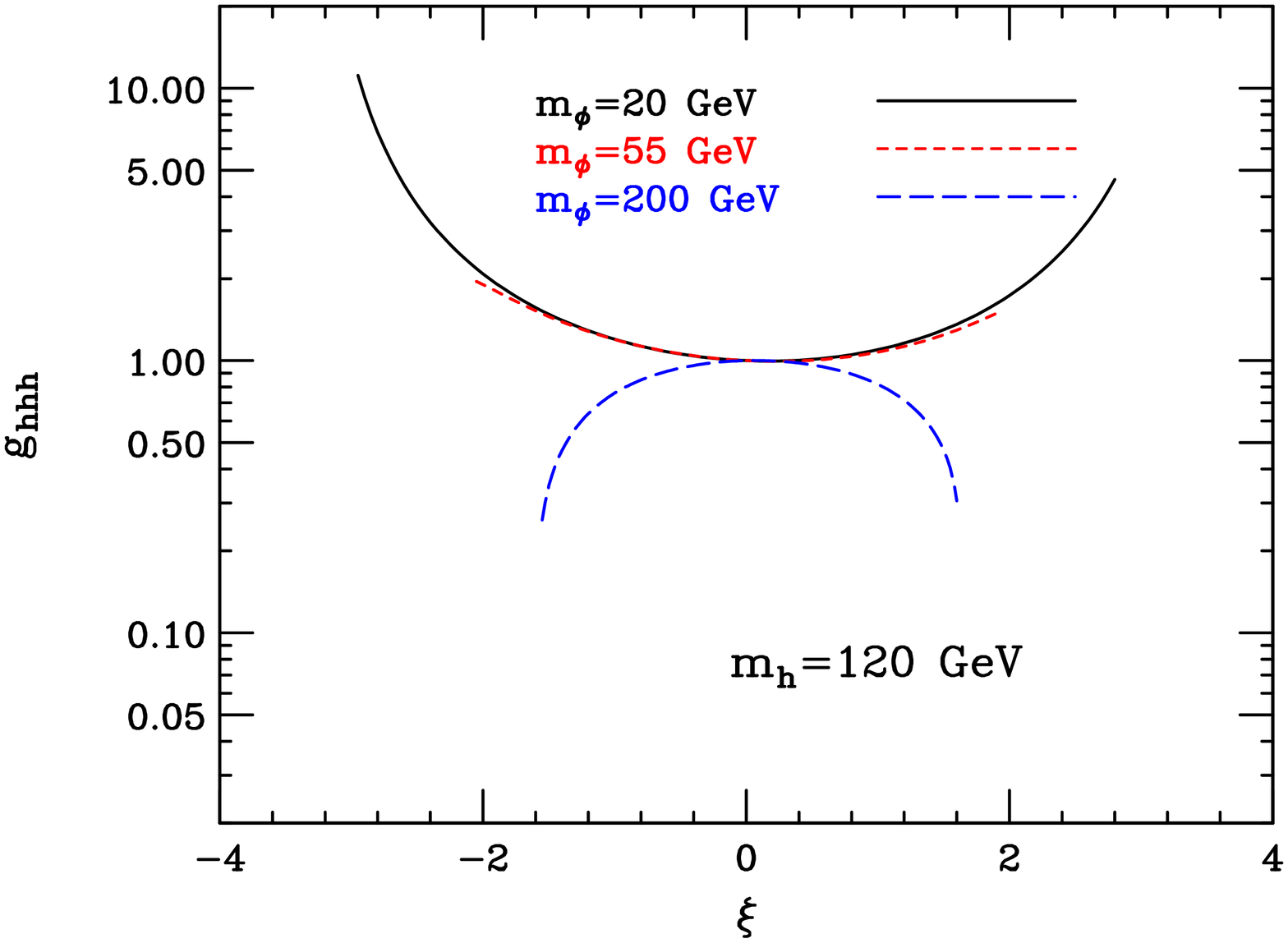}
\vspace*{-.1in}
\caption{For $\mh=120\gev$ and $\lphi=5\tev$, we plot 
the ratio, $g_{hhh}=\overline g_{hhh}/\overline g_{\hsm\hsm\hsm}$, 
of the $h^3$ self coupling to the SM prediction for the $\hsm^3$
coupling for $\mhsm=\mh$. The first plot gives contours,
while the 2nd plot shows results 
at the fixed values of $\mphi=20$, $55$ and $200\gev$.   
}
\label{ghhh_mh120}
\end{figure}
\begin{figure}[p!]
\includegraphics[width=5.7in]{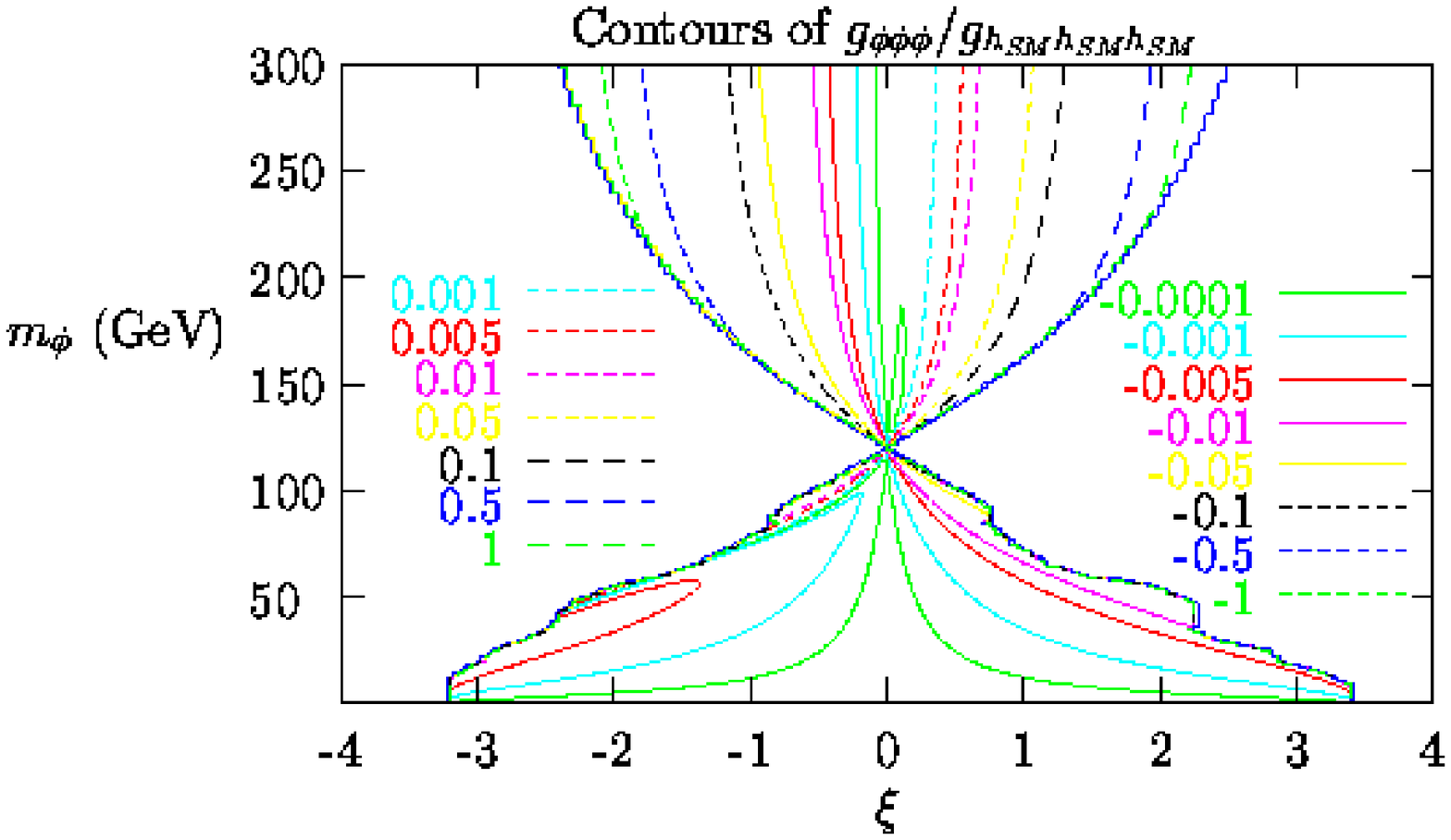}
\vspace*{.2in}

\includegraphics[width=4in,angle=90]{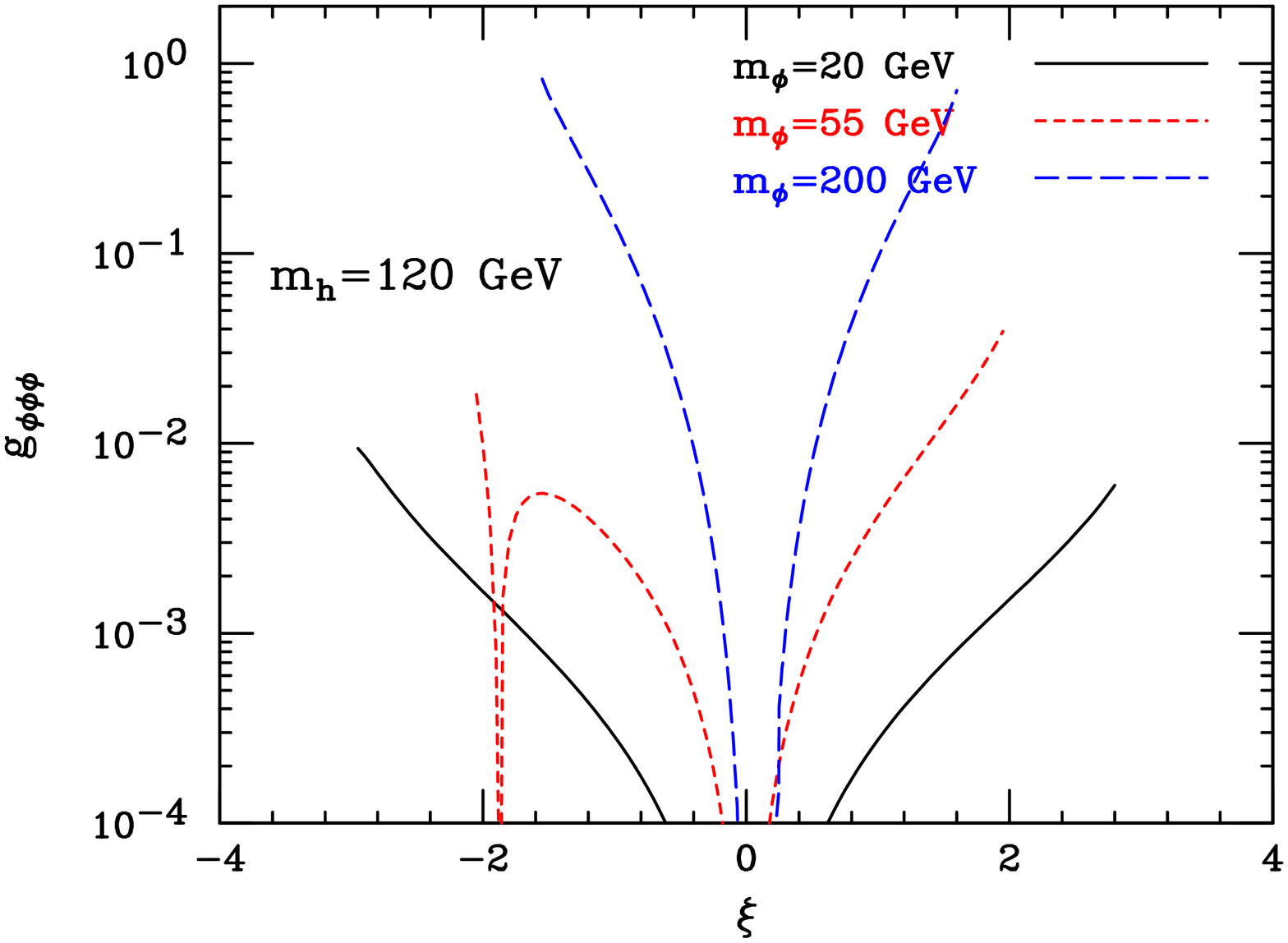}
\vspace*{-.1in}
\caption{For $\mh=120\gev$ and $\lphi=5\tev$, we plot the
ratio, $g_{\phi\phi\phi}=\overline g_{\phi\phi\phi}/\overline 
g_{\hsm\hsm\hsm}$, of the $\phi^3$ self coupling
to the SM prediction for the $\hsm^3$ coupling
taking  $\mhsm=\mphi$. The first plot gives contours,
while the 2nd plot gives results for  $|g_{\phi\phi\phi}|$
at $\mphi=20$, $55$ and $200\gev$. In the latter plot, $g_{\phi\phi\phi}$
is: $<0$ for all $\xi$ at $\mphi=20\gev$; $>0$ for $\xi<-1.87$ and $<0$
otherwise at $\mphi=55\gev$; and 
$<0$ for $\xi<0.2$ and $>0$ for $\xi>0.2$ at $\mphi=200\gev$.
}
\label{gphiphiphi_mh120}
\end{figure}

A coupling of particular interest in testing the nature of
electroweak symmetry breaking is the $h^3$ self coupling.
The algebraic form of the $h^3$ coupling
appears in the Appendix. The SM limit for this coupling corresponds
to $d=-a=1$, $c=b=0$. In addition, one employs $\gam^{-1}/\lphi=1/\vo$.
In Fig.~\ref{ghhh_mh120}, we plot the ratio $g_{hhh}\equiv {\anti g_{hhh}
/ \anti g_{\hsm\hsm\hsm}}$ of the $h^3$ coupling relative
to the corresponding SM value computed for a SM Higgs boson
mass equal to $\mh$.
(Recall that, in our notation, a bar indicates the full coupling as opposed
to the value relative to the SM, which ratio is indicated by a $g$
without a bar.)
We see that there are typically rather substantial
deviations that one could easily probe at a linear collider.

It is also interesting to examine the $\phi^3$ self coupling.
Taking $\mhsm=\mphi$, we plot 
in the upper figure of Fig.~\ref{gphiphiphi_mh120} contours of 
$g_{\phi\phi\phi}\equiv 
\anti g_{\phi\phi\phi}/\anti g_{\hsm\hsm\hsm}$; in the lower figure,
we plot $|g_{\phi\phi\phi}|$. 
As expected, $g_{\phi\phi\phi}$
vanishes for $\xi=0$. It is often negative (\ie\ $\anti g_{\phi\phi\phi}$
has the opposite sign compared to
$\anti g_{\hsm\hsm\hsm}$) and is generally $<1$, implying suppression
relative to the SM case.
Only for $\mphi=200\gev$ and the largest allowed $|\xi|$ values
can the $\phi^3$ coupling 
take values comparable to the SM strength.
Thus, in general its measurement may be quite difficult.
Of course, the  `background
diagrams' contributing to the same final state
($\epem\to t\anti t\phi\phi$ or $\epem\to Z\phi\phi$)
will also be suppressed in comparison
to the SM case; they 
have two $f\anti f \phi$ or $VV\phi$ vertices proportional
to $(c+\gam a)^2$, and $(c+\gam a)^2$ is substantially smaller
than 1 for much of the parameter space being considered.

A particularly important feature of the above plots is that
once $\mh$ is large enough ($\mh\gsim 115\gev$
is sufficient) there is
a substantial range of $\xi$ values for any $\mphi<\mh/2$ 
(so that $\h\to\phi\phi$ decays are possible) that
cannot be excluded by LEP/LEP2 constraints. 
The reverse is also true;
allowed parameter regions exist for which $\phi\to hh$ decays
are possible once $\mphi\gsim 230\gev$. 
We now turn to a discussion of branching ratios,
including the $h\to \phi\phi$ final mode.

The partial width of the $\h$ in which we are most interested is
that for $\h\to \phi\phi$:
\beq \Gamma(h\to \phi\phi )={\anti g_{\phi\phi h}^2\over
32\pi\mh}(1-4r_\phi)^{1/2}\,.
\label{htophiphiwdth}
\eeq
In the above equation,
$\lam(1,r_1,r_2)\equiv 1+r_1^2+r_2^2-2r_1-2r_2-2r_1r_2$,
$r_\phi=\mphi^2/\mh^2$.
We also give the expression for $\h\to h^n\phi$:
 \beq
\Gamma(h\to \h^n \phi)=\frac{\anti g_{n \phi\h}^2  }{192 \pi}
{\mh^3} {\lam^{5/2}(1,r_\phi,r_n) \over r_n^2} \,,
\label{htohnphiwdth}
\eeq
where $r_n=m_n^2/\mh^2$
(Corresponding results apply for $\phi \to hh$ and $\phi \to h^n h$.)
 Expanding in powers of
$\gamma=\vo/\lphi$ using Eq.~(\ref{gnphlim}),
we find that $\Gamma(h\to \h^n \phi) \sim \mh^3/r_n^2 \sim \mh^7$.

It is interesting to investigate Higgs-boson 
branching ratios for various decay channels
in the presence of the $\xi$-mixing. 
If we neglected the $gg$ and $\gam\gam$ anomalous
couplings, we would have
\beq
\Gamma(h\to all)=(d+\gamma b)^2\Gamma_{SM}(h\to all)+
\Gamma(h\to \h^n \phi)+\Gamma(h\to \phi\phi )\,,
\label{width}
\eeq
where $\Gamma_{SM}(h\to all)$ is the SM total width.
However, the $gg$ width can be quite enhanced and this must be included.
We have done this in the context of a modified version of HDECAY \cite{HDECAY},
which includes all relevant radiative corrections to couplings
and branching ratios. In particular, the running $b$ mass decreases
the $b\anti b$ branching ratio of the $h$, resulting in some
increase in $BR(h\to \phi\phi)$.

\begin{figure}[p]
\begin{center}
\includegraphics[height=4in,angle=90]{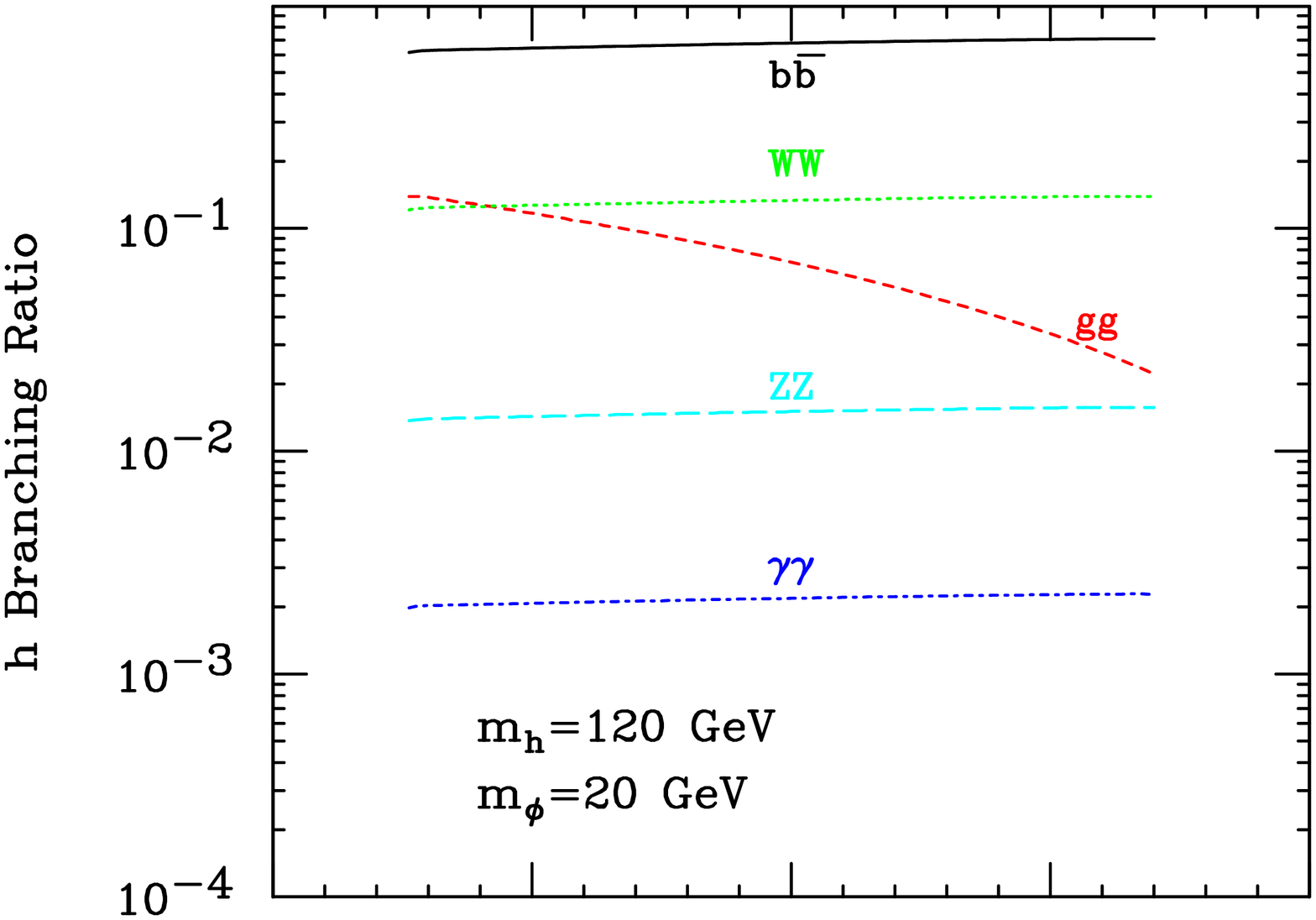}
\includegraphics[height=4in,angle=90]{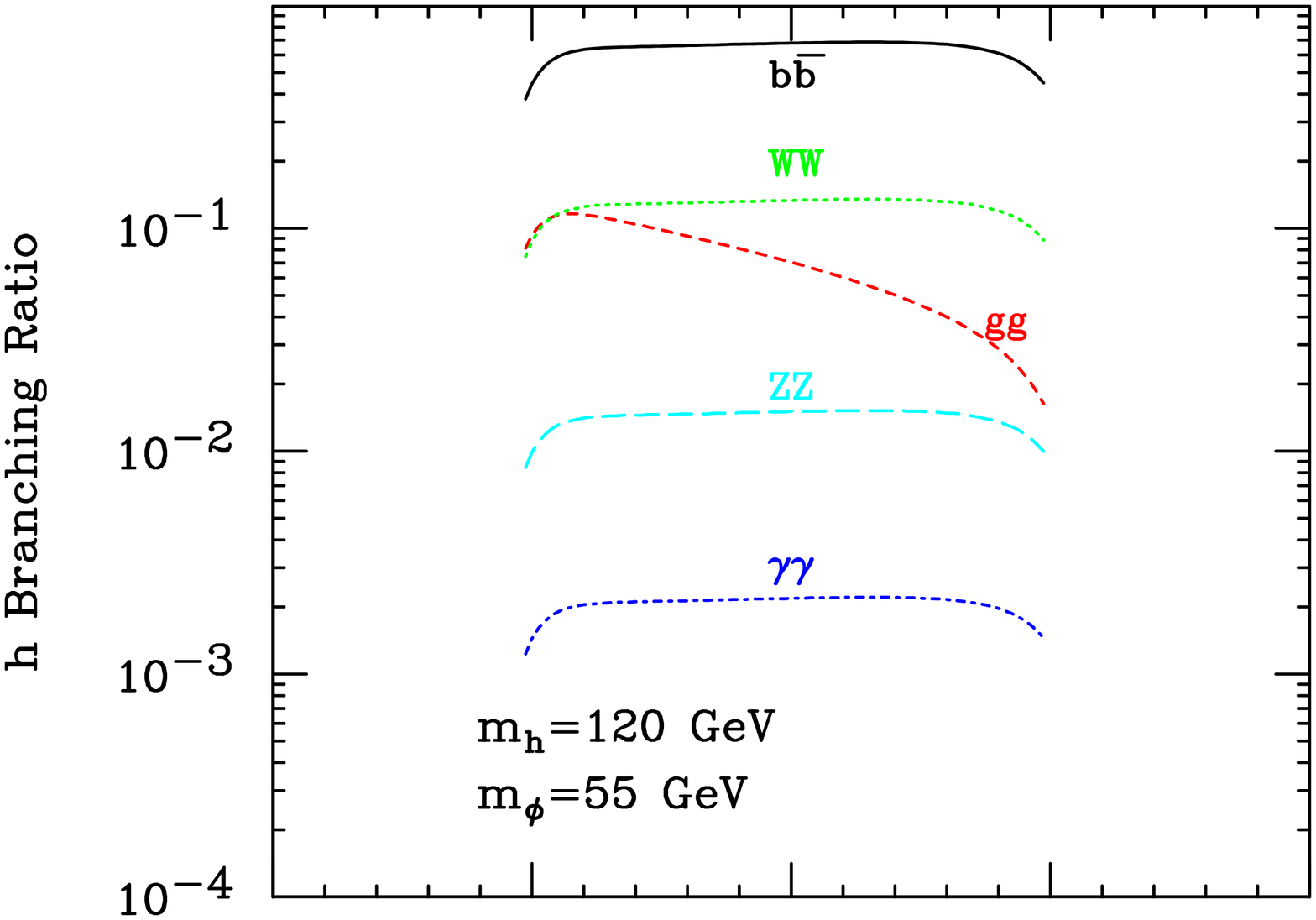}
\includegraphics[height=4in,angle=90]{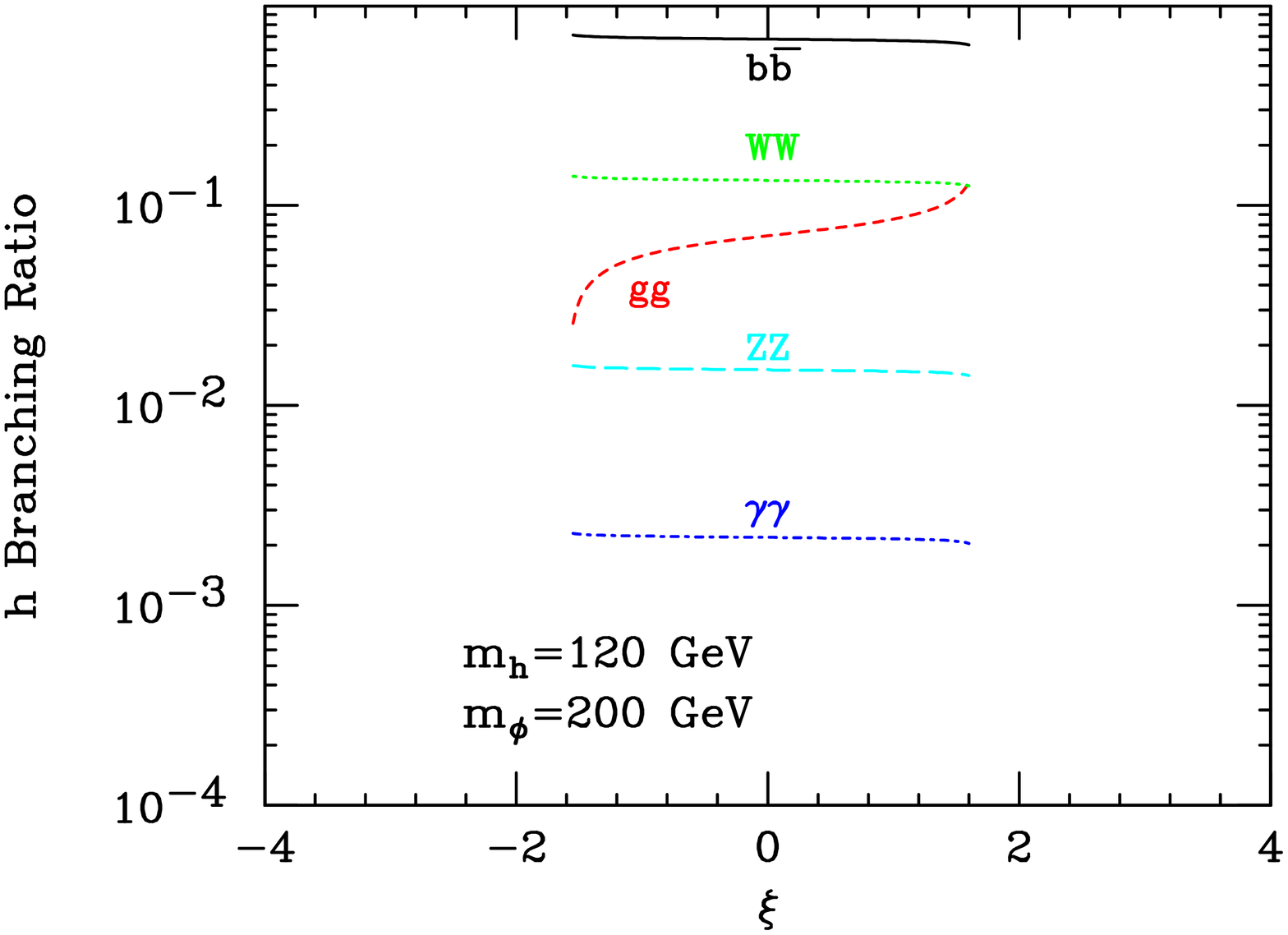}
\end{center}
\vspace*{-.2in}
\caption{The branching ratios for $h$ decays to $b\anti b$,
$gg$, $WW^*$, $ZZ^*$ and $\gam\gam$ 
for $m_h=120\gev$  and $\lphi=5\tev$ as functions of $\xi$
for $\mphi=20$, $55$ and $200\gev$.
}
\label{brhall_mh120}
\end{figure}
\begin{figure}[p]
\begin{center}
\includegraphics[height=4in,angle=90]{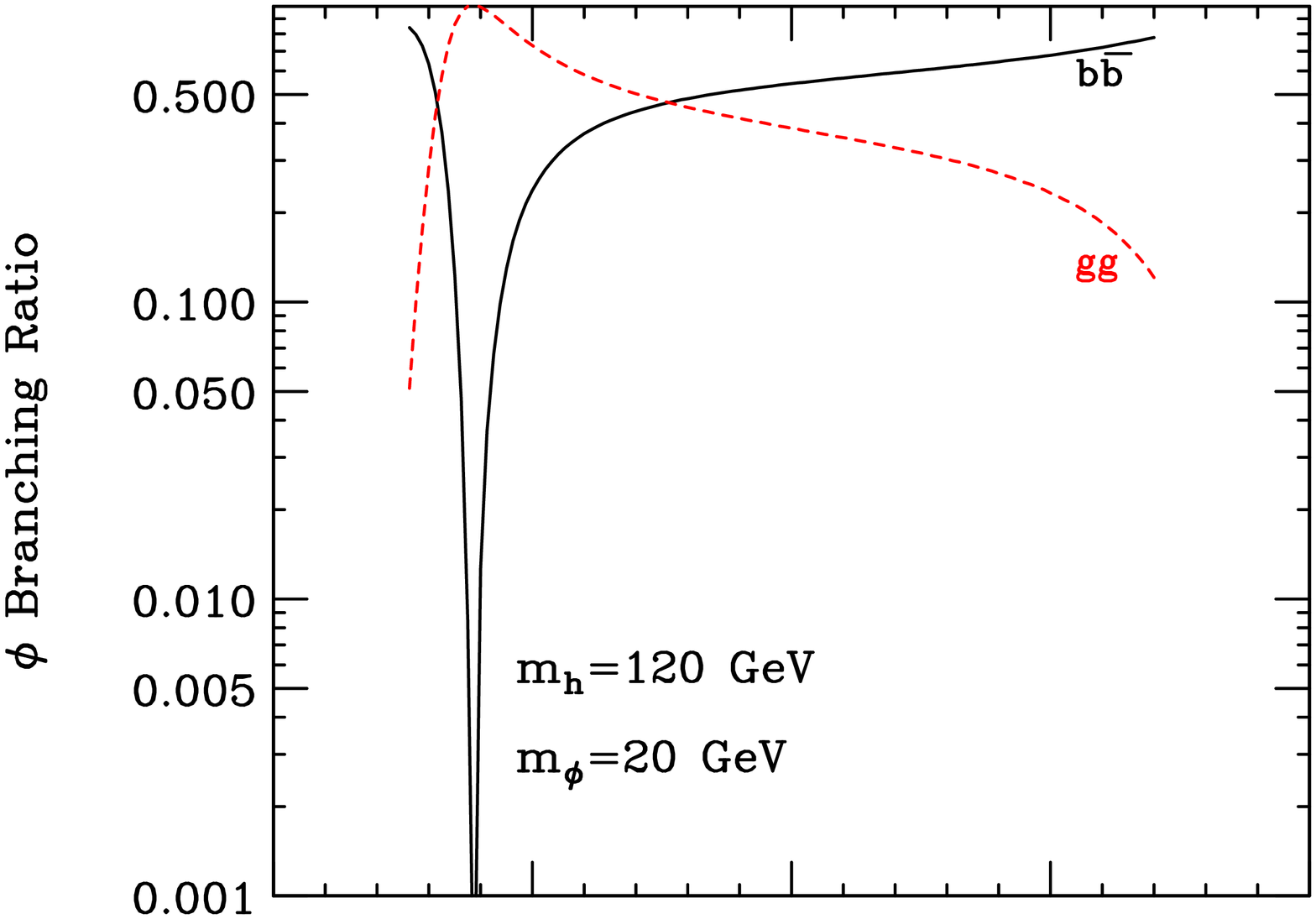}
\includegraphics[height=4in,angle=90]{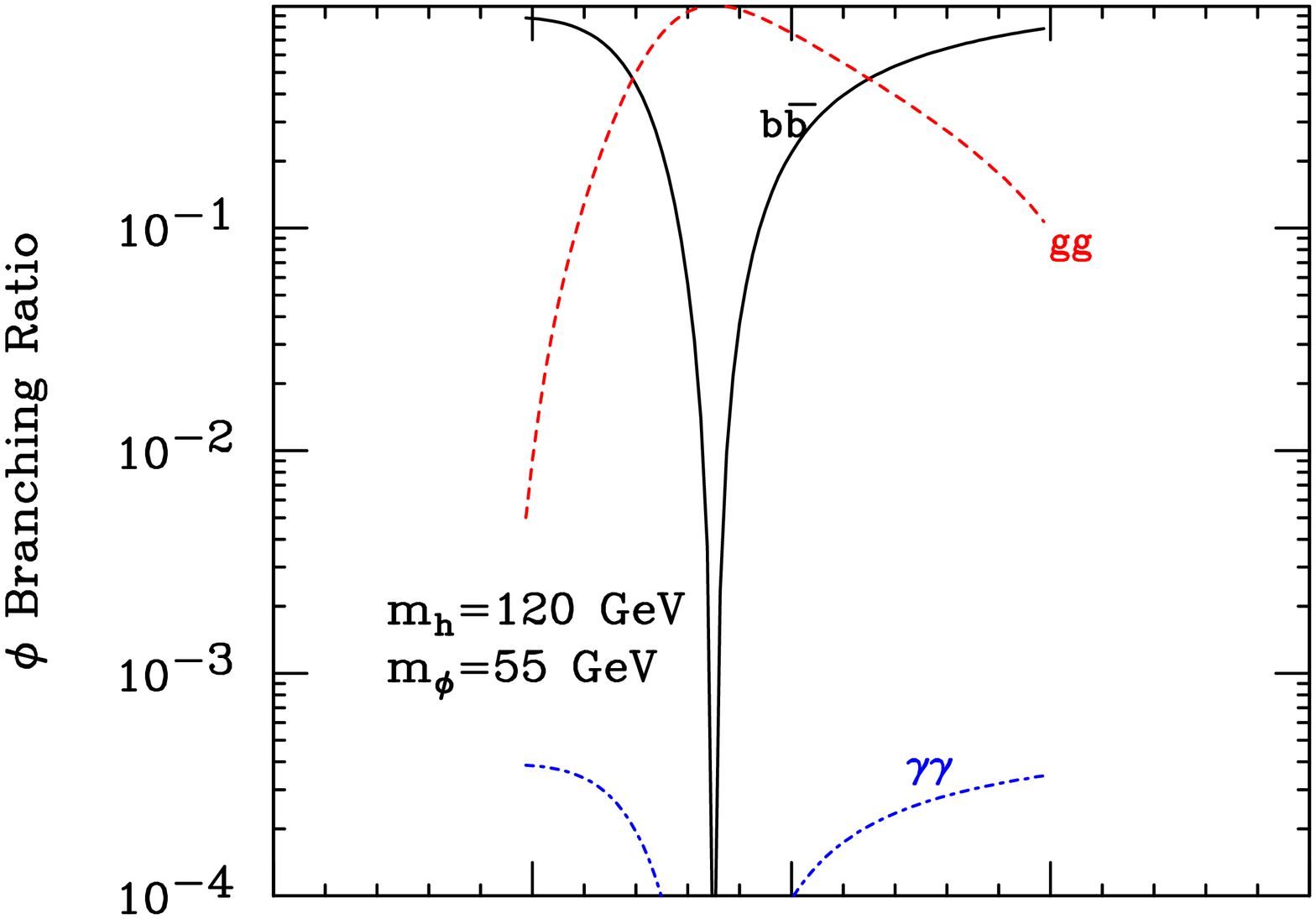}
\includegraphics[height=4in,angle=90]{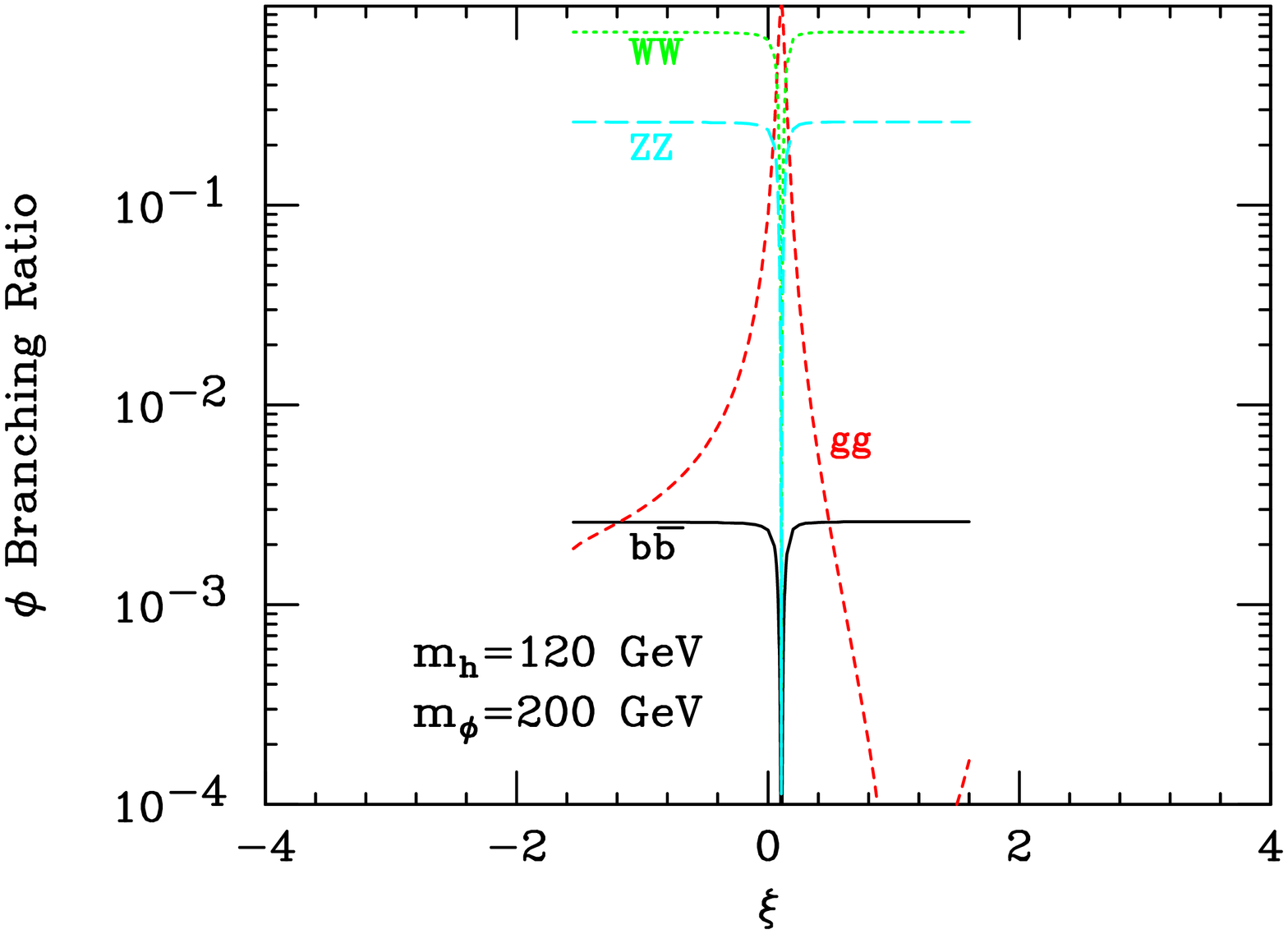}
\end{center}
\vspace*{-.2in}
\caption{The branching ratios for $\phi$ decays to $b\anti b$,
$gg$, $WW^{(*)}$, $ZZ^{(*)}$ and $\gam\gam$ 
for $m_h=120\gev$  and $\lphi=5\tev$ as functions of $\xi$
for $\mphi=20$, $55$ and $200\gev$.
}
\label{brphiall_mh120}
\end{figure}

In Fig.~\ref{brhall_mh120}, we plot the branching ratios for $h\to b\anti b$,
$gg$, $WW^*$, $ZZ^*$ and $\gam\gam$ 
as a function of the mixing parameter $\xi$,
taking $\mh=120\gev$ and $\lphi=5\tev$. Results are shown
for three different $\mphi$ values: $20$, $55$ and $200\gev$.
(The case of $\mphi=55\gev$ is one for which $BR(\h\to\phi\phi)$
can be quite large when $\mh=120\gev$.)
These plots are limited to $\xi$ values allowed by the theoretical
constraints discussed earlier. We have chosen
to not include a more restrictive
bound on $g_{ZZh}^2$ as might be appropriate 
in order to guarantee that the $h$
contributions to precision electroweak observables be
consistent with $S$ and $T$ remaining within the usual $95\%$ 
CL ellipse.\footnote{Contributions
of the $\phi$ to $S$ and $T$ are small.
Indeed, referring to Fig.~\ref{couplingsphi}, we see that $g_{ZZ\phi}^2<1$
almost everywhere when $\lphi=5\tev$.} Fig.~\ref{couplingsh} shows
that $g_{ZZ\h}^2>2$ (a rough estimate of the needed bound) only
at the very largest $|\xi|$ values when $\mphi$ is small.

The most important features of Fig.~\ref{brhall_mh120} are
the following. First, large values for the $gg$ branching ratio
(due to the anomalous contribution to the $h gg$ coupling)
are the norm. This suppresses the other branching ratios to some extent.  
(The anomalous contribution to the $h\gam\gam$
coupling is less important due to presence of the large $W$ loop contribution
in this latter case.) Second, for $\mphi=55\gev$,
$BR(\h\to \phi\phi)$ is large at large $|\xi|$ and suppresses the conventional
branching ratios.  In general, changes in the branching ratio
of the $\h$ with respect to the SM
are modest, but nonetheless they are 
at an observable level, at least at the LC. Note, however, that
the modest $BR$ changes belie the fact that the 
$f\anti f$ and $VV$ coupling-squared factor $(c+\gam d)^2$
is often changing dramatically, implying dramatic changes in $\h$
production rates with respect to expectations for a SM $\hsm$.

Results for the $\phi$ branching ratios are plotted in 
Fig.~\ref{brphiall_mh120}. We observe that the $gg$ decay
is generally dominant over the $b\anti b$ mode
and that it has the largest branching ratio
until  the $WW^{(*)},ZZ^{(*)}$
modes increase in importance at larger $\mphi$.  Of course,
the zero in $g_{ZZ\phi}^2=(c+\gamma a)^2$ has a very
large impact.

\begin{figure}[t]
\begin{center}
\includegraphics[width=4in,angle=90]{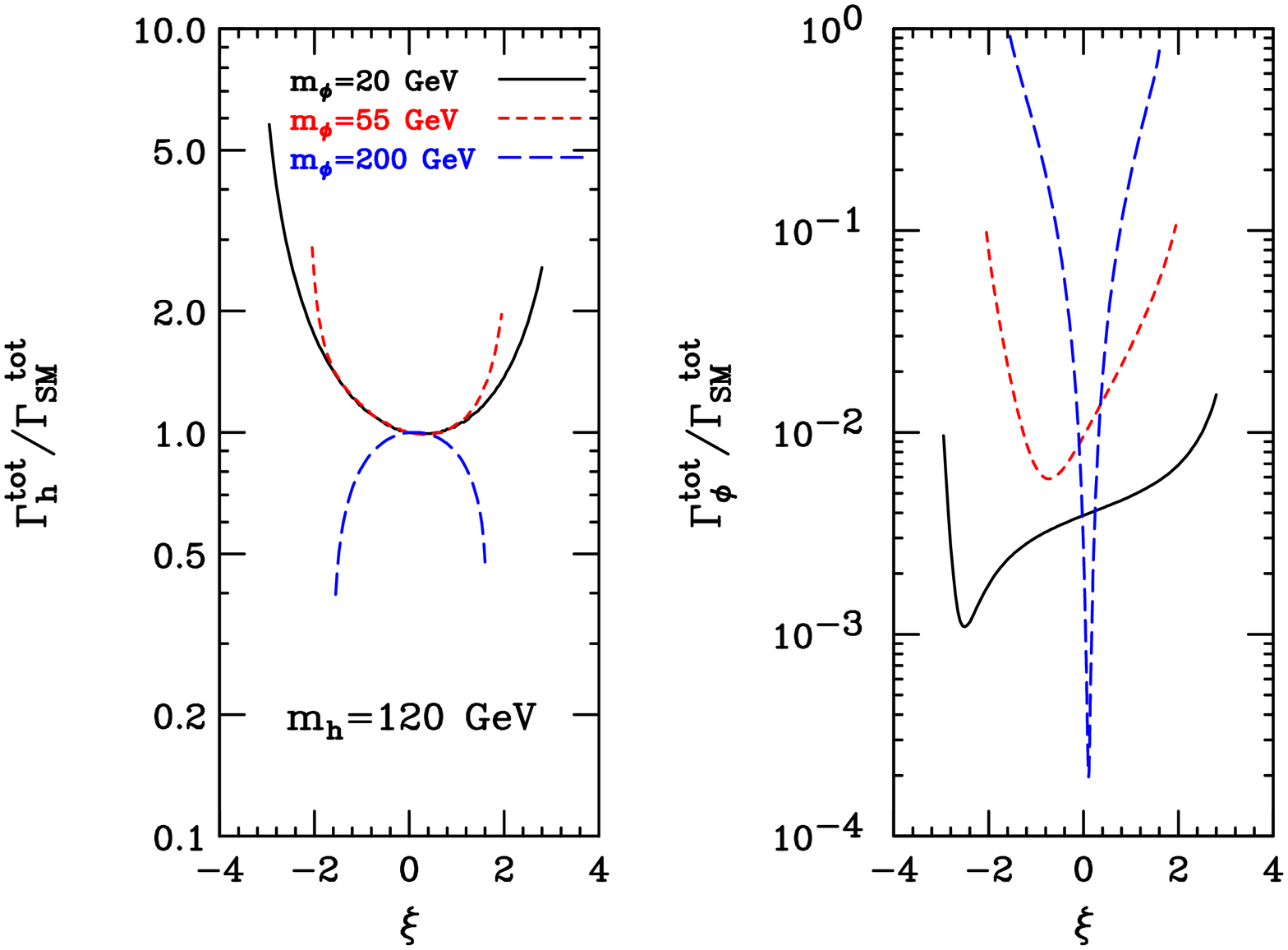}
\end{center}
\vspace*{-.2in}
\caption{The total widths for the $h$ and $\phi$ relative to the
value for a SM Higgs boson of the same mass are plotted as functions
of $\xi$ for $\mh=120\gev$  and $\lphi=5\tev$ taking
$\mphi=20$, $55$ and $200\gev$.
}
\label{widthsall_mh120}
\end{figure}

Also important for $h$ discovery is its total width. 
In the left-hand window of
Fig.~\ref{widthsall_mh120}, we plot the ratio of the total $h$
width to the corresponding width of a SM Higgs boson of the same mass,
$\Gamma_h^{\rm tot}/\Gamma_{SM}^{\rm tot}$, as a function of $\xi$
for $\mphi=20$, $55$ and $200\gev$.
Note that a substantially larger total width
for the $h$ is possible if $\mphi$ is small.
In the right-hand window, we plot the ratio 
$\Gamma_\phi^{\rm tot}/\Gamma_{SM}^{\rm tot}$
(for $\mhsm=\mphi$) as a function of $\xi$. The $\phi$ is generally
quite narrow. This is true even for $\mphi=200\gev$, for which $WW/ZZ$ decays
are allowed, near the zero in $(c+\gamma a)^2$.

\begin{figure}[h!]
\begin{center}
\includegraphics[width=4in,angle=90]{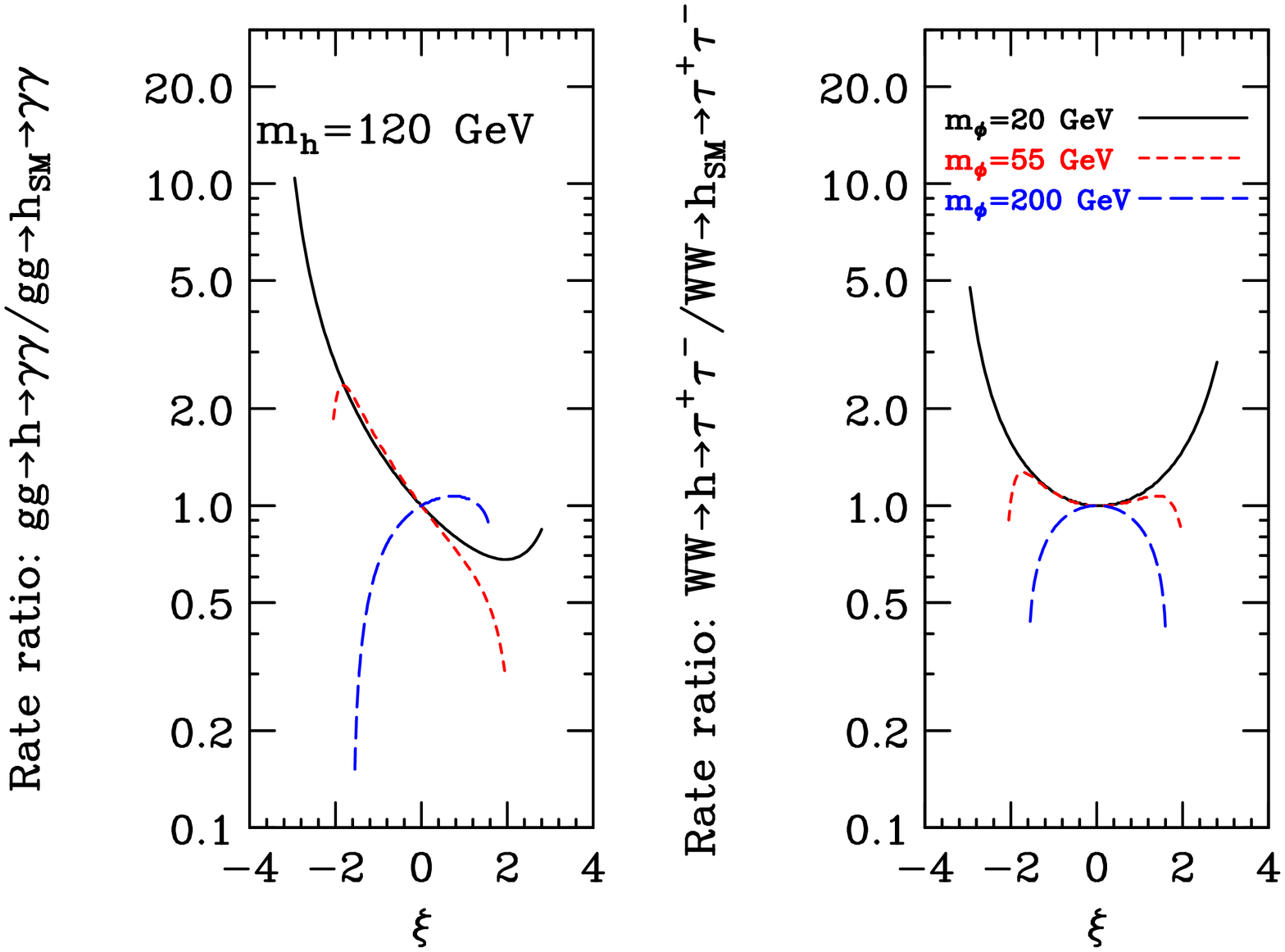}
\end{center}
\vspace*{-.2in}
\caption{The ratio of the rates for $gg\to h \to \gam\gam$
and $WW\to h \to \tau^+\tau^-$ (the latter is the same as
that for $gg\to t\anti t h\to t\anti t b\anti b$)
to the corresponding rates for the SM Higgs boson.
Results are shown
for $m_h=120\gev$  and $\lphi=5\tev$ as functions of $\xi$
for $\mphi=20$, $55$ and $200\gev$.
}
\label{prodh_mh120}
\end{figure}

Experimentally, the above results imply that detection of the $h$
at the LHC could be significantly impacted if $|\xi|$ is large.
To illustrate this, we plot in Fig.~\ref{prodh_mh120} the
ratio of the rates for $gg\to h \to \gam\gam$, $WW\to h \to \tau^+\tau^-$
and $gg\to t\anti t h \to t\anti t b\anti b$ (the latter
two ratios being equal) to the corresponding rates for
the SM Higgs boson.  For this figure, we take $\mh=120\gev$
and $\lphi=5\tev$ and show results for $\mphi=20$, $55$ and $200\gev$.
In the case of $\mphi=55\gev$, the $h\to \phi\phi$ decay, discussed
in more detail later, is substantial for large $|\xi|$.  
The resulting suppression of the standard LHC modes at the largest allowed
$|\xi|$ values is most evident in the $\wp\wm\to h\to \tau^+\tau^-$
curves. Another important impact of mixing is through 
communication of the anomalous $gg$ coupling
of the $\phi_0$ to the $h$ mass eigenstate. The result is
that prospects for $h$ discovery in the 
$gg\to h \to \gam\gam$ mode could be either substantially poorer 
or substantially better than for a SM
Higgs boson of the same mass, depending on $\xi$ and $\mphi$.
\footnote{We note that even for parameters such that
$\Gamma_h^{\rm tot}$ is enhanced relative to $\Gamma_{\hsm}^{\rm tot}$,
the very tiny SM Higgs
width at $\mh=120\gev$ implies that $\Gamma_h^{\rm tot}$
will remain much smaller than the experimental resolution,
even in the important $\gam\gam$ final state.}

At the LC, the potential for $h$ discovery is primarily determined
by $g_{ZZh}^2$. As shown in Fig.~\ref{couplingsh}, this 
reduced coupling-squared (defined relative to the SM value)
is often $>1$ (and can be as large as $\sim 5$), but can also
fall to values as low as $\sim 0.4$, implying signficant
suppression relative to SM expectations.
The latter suppression is well within the reach of 
the $e^+e^-\to Z h$ recoil mass discovery technique at a LC with
$\sqrt s=500\gev$ and $L=500\fbi$.  The techniques that have
been developed for measuring the total width of a Higgs boson
at the LC indirectly would remain applicable and could reveal
the presence of $\xi\neq 0$ mixing through a sizable deviation
with respect to the SM prediction.

\begin{figure}[h!]
\begin{center}
\includegraphics[width=4in,angle=90]{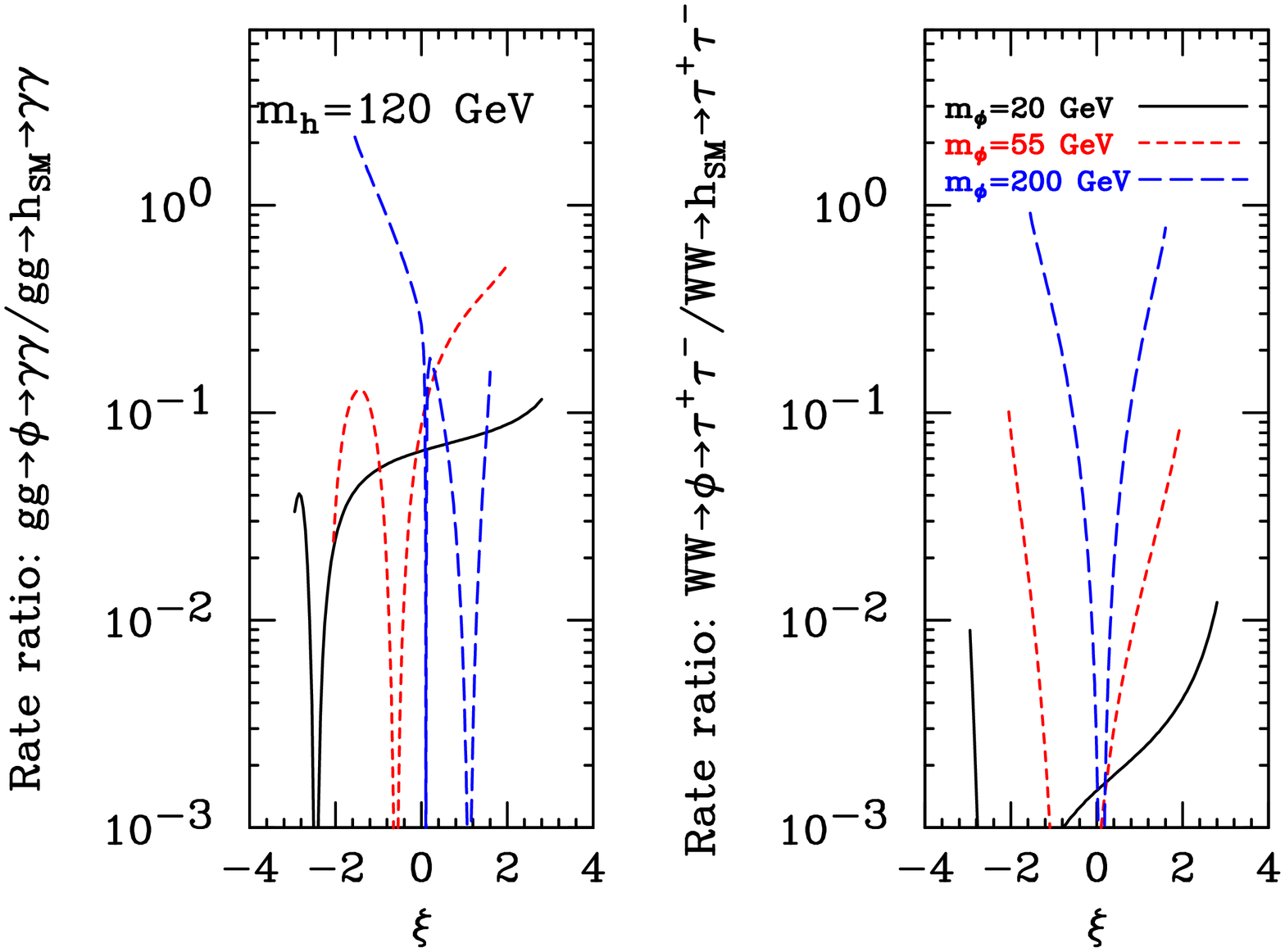}
\end{center}
\vspace*{-.2in}
\caption{The ratio of the rates for $gg\to \phi\to \gam\gam$ 
and for $WW\to \phi \to \tau^+\tau^-$ (the latter being the same as
that for $gg\to t\anti t \phi \to t\anti t b\anti b$)
to the corresponding rates for the SM Higgs boson.
Results are shown
for $m_h=120\gev$  and $\lphi=5\tev$ as functions of $\xi$
for $\mphi=20$, $55$ and $200\gev$. 
}
\label{prodphi_mh120}
\end{figure}

\begin{figure}[h!]
\begin{center}
\includegraphics[width=4in,angle=90]{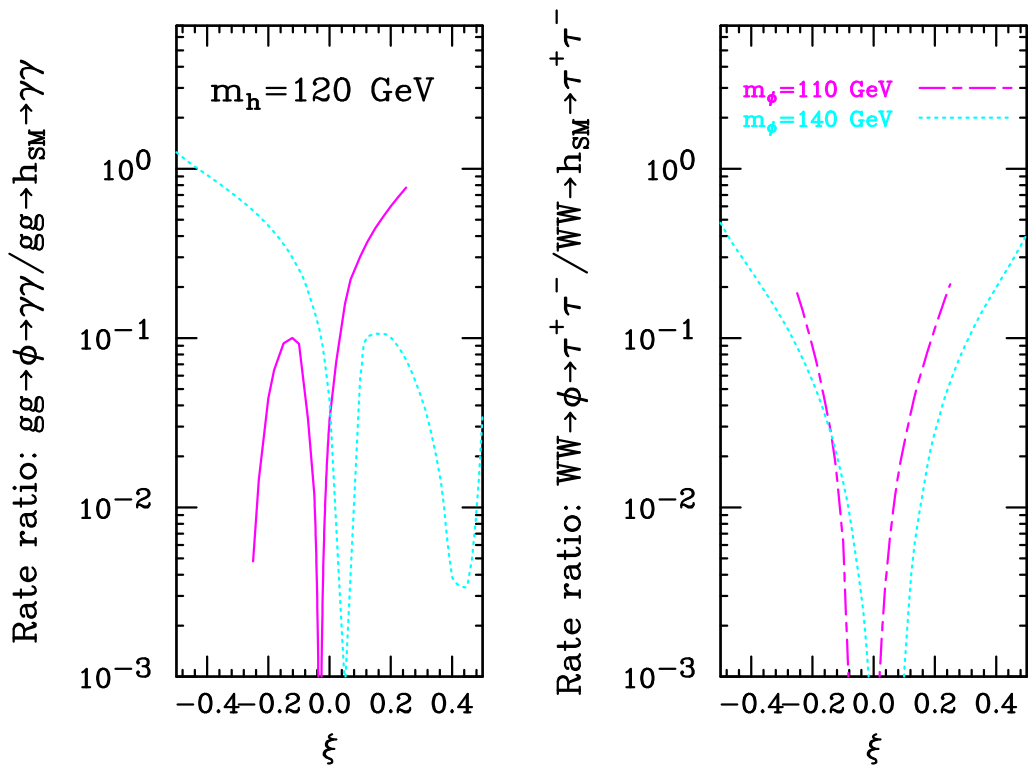}
\end{center}
\vspace*{-.2in}
\caption{As in Fig.~\ref{prodphi_mh120}, but
for $\mphi=110$ and $140\gev$. 
}
\label{prodphi_mh120_2}
\end{figure}
\begin{figure}[h!]
\begin{center}
\includegraphics[width=4in,angle=90]{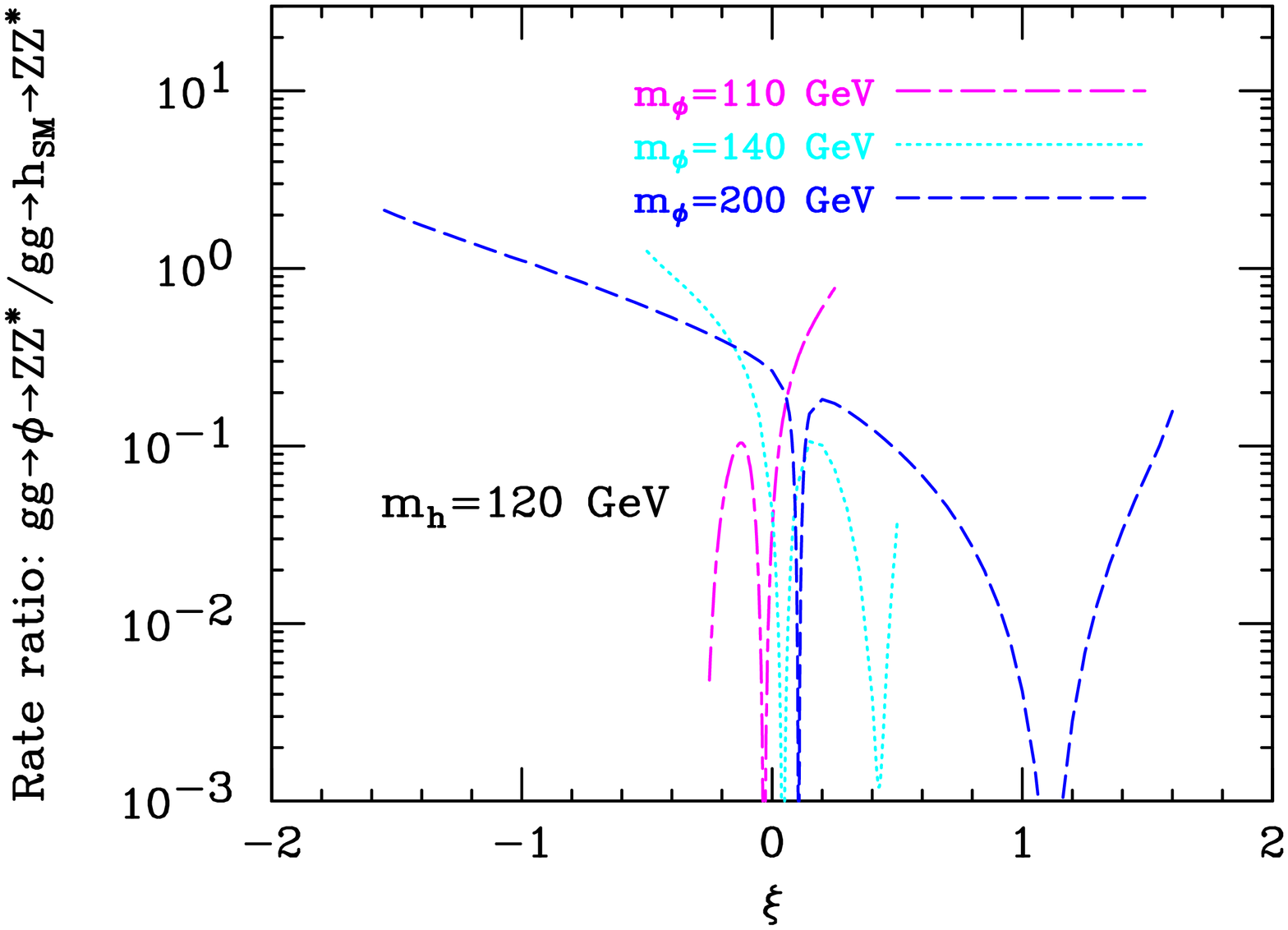}
\end{center}
\vspace*{-.2in}
\caption{The ratio of the rate for $gg\to \phi\to ZZ$
to the corresponding rate for a SM Higgs boson with mass $\mphi$
assuming $\mh=120\gev$ and $\lphi=5\tev$ as a function of $\xi$
for $\mphi=110$, $140$ and $200\gev$. Recall that the $\xi$ range
is increasingly restricted as $\mphi$ becomes more degenerate
with $\mh$.
}
\label{prodphiggtozz_mh120}
\end{figure}

What about prospects for $\phi$ detection at the LHC?
In Figs.~\ref{prodphi_mh120} and \ref{prodphi_mh120_2}, 
we plot the same ratios for the $\phi$
as we did for the $h$
in Fig.~\ref{prodh_mh120}.
For this figure, we take $\mh=120\gev$
and $\lphi=5\tev$ and show results for $\mphi=20$, $55$ and $200\gev$
in Fig.~\ref{prodphi_mh120} and for $\mphi=110$ and $140\gev$
in Fig.~\ref{prodphi_mh120_2}.
For all masses, the $gg\to\phi\to\gam\gam$ 
rate is generally significantly suppressed relative
to the prediction for a SM Higgs boson, depending upon $\xi$ and $\mphi$.
The dip in the $\phi$ rates 
is due to a cancellation that zeroes the $\gam\gam$ coupling,
and occurs very close to the point at which the $\phi$'s couplings
to vector bosons and fermions, $(c+\gam a)^2$, vanishes.
Detection of the $\phi$ in $gg\to\phi\to\gam\gam$ will generally be
quite difficult.
 The $WW\to \phi \to \tau^+\tau^-$ and
$gg\to t\anti t \phi\to t\anti t b\anti b$ modes are generally
also quite suppressed relative to SM rates and would 
probably not be visible.
For $\mphi\gsim 110\gev$, in addition to the above three modes
one can consider the standard $gg\to \phi \to ZZ^{(*)}\to 4\ell$ signal.
The ratio for this rate relative to the SM prediction is plotted
in Fig.~\ref{prodphiggtozz_mh120} for $\mphi=110$, $140$ and $200\gev$.
For $\mphi=200\gev$, the very high level
of statistical significance predicted for the SM $ZZ$ final state signal
at this mass implies that $\phi$ detection in this mode should be
possible except near the zeroes in the $ZZ\phi$ coupling.
For the $\mphi=110$ and $140\gev$ cases,  the dip region occupies
a lot of the allowed $\xi$ range and suppression is generally
present even away from the dip regions.  Detection in the $4\ell$
mode would be unlikely.

At the LC, the potential for $\phi$ discovery is primarily determined
by $g_{ZZ\phi}^2$. As shown in Fig.~\ref{couplingsphi}, 
this reduced coupling-squared
is typically substantially suppressed
relative to the SM value of 1. Still, because of the very high statistical
significance associated with a SM Higgs signal in the $e^+e^-\to Z$+Higgs
mode for $\sqrt s=500\gev$ and $L=500\fbi$, 
detection of the $\phi$ will be possible except near the zero
in the $ZZ\phi$ coupling.
As discussed earlier, the width of the $\phi$ would be
much smaller than anticipated.  
This could be checked using the techniques that have
been developed for measuring the total width of a narrow Higgs boson
at the LC indirectly.

\begin{figure}[h!]
\begin{center}
\includegraphics[width=4in,angle=90]{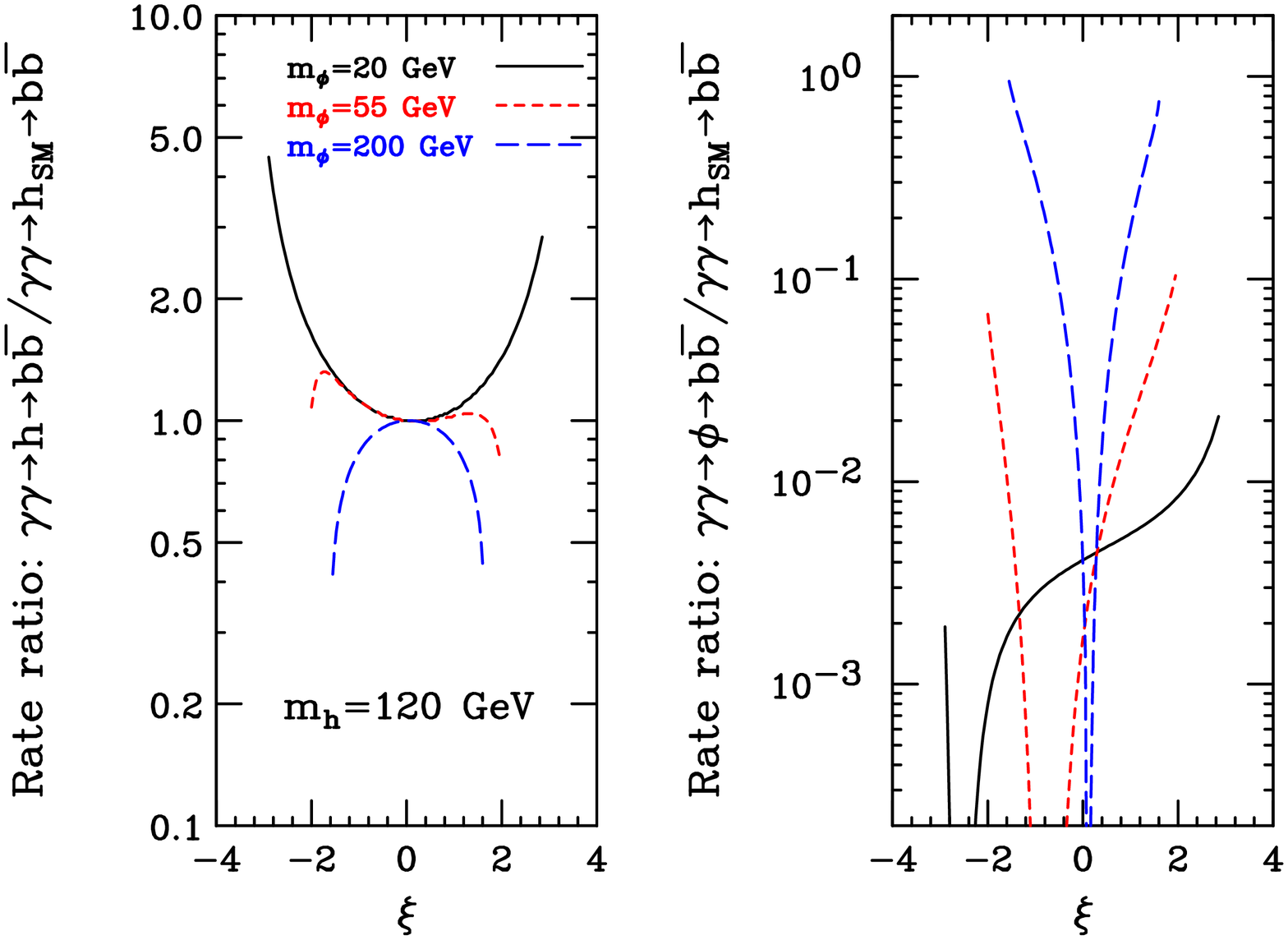}
\end{center}
\vspace*{-.2in}
\caption{The rates for $\gam\gam\to h \to b\anti b$
and $\gam\gam\to\phi\to b\anti b$ relative
to the corresponding rate for a SM Higgs boson of the same mass.
Results are shown
for $m_h=120\gev$  and $\lphi=5\tev$ as functions of $\xi$
for $\mphi=20$, $55$ and $200\gev$.
}
\label{gagatobb_mh120}
\end{figure}

Also of considerable interest is how $\xi\neq0$ would affect prospects
for $h$ and $\phi$ detection at a $\gam\gam$ collider. To assess this,
we plot in Fig.~\ref{gagatobb_mh120}
the $\gam\gam\to h\to b\anti b$ and $\gam\gam\to \phi\to b\anti b$
rates relative to the SM rates evaluated for Higgs mass equal
to $\mh$ or $\mphi$, respectively. 
In the case of the $\h$, the plot differs 
only slightly from Fig.~\ref{prodh_mh120}
for the $WW\to\h\to \tau^+\tau^-$
LHC discovery mode. This means that
the anomaly contribution to the $\gam\gam\h$ 
coupling is much smaller than that from the standard fermion and
$W$ boson loops. In the case of the $\phi$, differences between 
these $\gam\gam\to \phi \to b\anti b$ curves and the corresponding
$WW\to\phi\to\tau^+\tau^-$ curves of Fig.~\ref{prodphi_mh120}
are somewhat larger, especially in the vicinity of the zeroes.

To summarize the results, in the case of the $\h$, for the parameters
considered, the rate is suppressed by at most a factor of 0.5
and would thus be quite sufficient to yield a highly detectable
and accurately measurable signal.  
The $\phi$ would typically be much more difficult to discover
in $\gam\gam$ collisions.  Large dips in the rate occur
in the vicinity of the zero in the $\phi\to b\anti b$
coupling and branching ratio, which is at the same location
as the zero in $g_{ZZ\phi}$.  The 
$\gam\gam\to \phi\to gg$ channel is somewhat less suppressed
in the dip region due to the anomalous contribution
to the $\phi gg$ coupling. However, the signal is still small
in the dip regions and this channel would 
have large backgrounds. Although it would 
be difficult to isolate, further study might be warranted.

\begin{figure}[p!]
\begin{center}
\includegraphics[height=4.8in,angle=90]{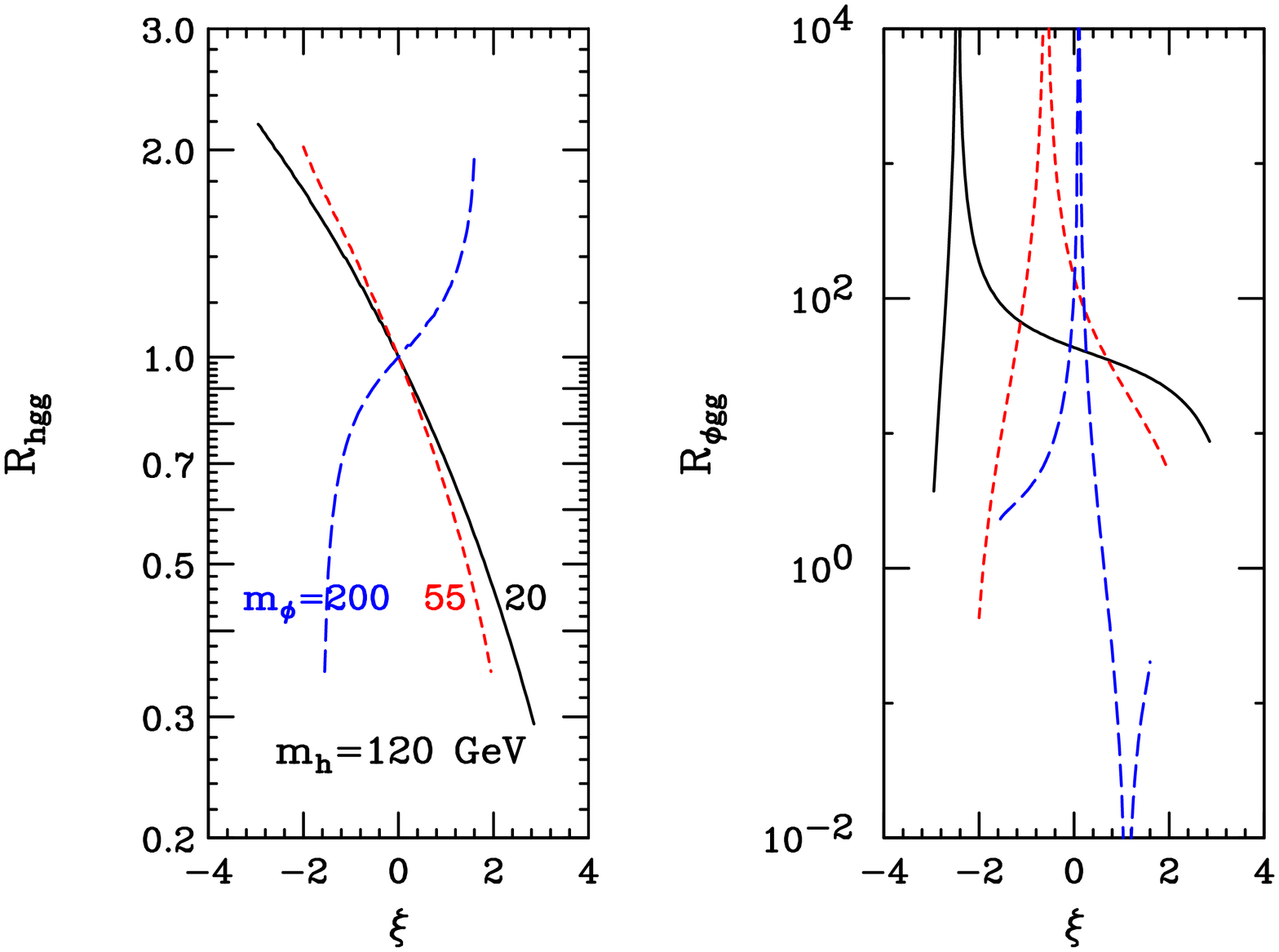}
\includegraphics[height=4.8in,angle=90]{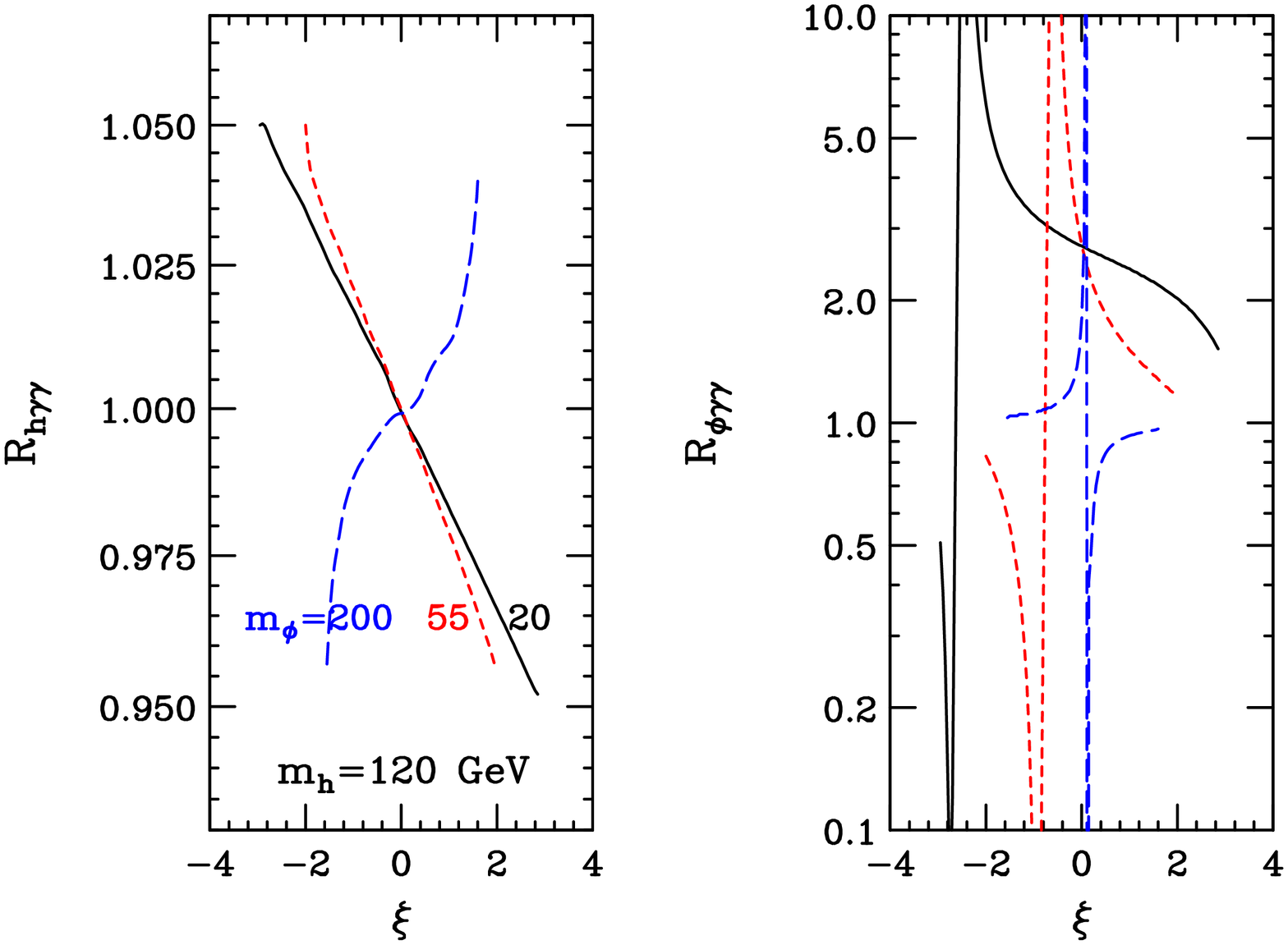}
\end{center}
\vspace*{-.2in}
\caption{In the upper plots, we give the ratios $R_{\h gg}$ and $R_{\phi gg}$
of the $\h gg$ and $\phi gg$ couplings-squared
including the anomalous contribution
to the corresponding values expected in its absence.
Results for the the analogous ratios
$R_{\h\gam\gam}$ and $R_{\phi\gam\gam}$ are presented
in the lower plots. Results are shown
for $m_h=120\gev$  and $\lphi=5\tev$ as functions of $\xi$
for $\mphi=20$, $55$ and $200\gev$.
(The same type of line is used for a given $\mphi$ in 
the right-hand figure as is used in the left-hand figure.)}
\label{gggagaanomalouscoup_mh120}
\end{figure}

An important question is whether the deviations due to the anomalous
$\phi_0 gg$ and 
$\phi_0\gam\gam$ couplings are sufficiently large to be measurable.
To quantify this, we plot 
in Fig.~\ref{gggagaanomalouscoup_mh120} the ratios~\footnote{Once again,
we remind the reader that the $\anti g$ notation refers to the
full coupling strength as normally defined, whereas $g$'s without
a bar are reserved for certain coupling ratios.} 
\beq
R_{sgg}\equiv{\anti g_{s gg}^2(\mbox{with anomaly})\over \anti g_{s gg}^2(\mbox{without anomaly})}\,,
\quad\mbox{and}\quad
R_{s\gam\gam}\equiv {\anti g_{s \gam\gam}^2(\mbox{with anomaly})\over 
\anti g_{s \gam\gam}^2(\mbox{without anomaly})}\,,
\label{ratiodefs}
\eeq 
for $s=\h$ and $s=\phi$. 
These ratios can be determined experimentally.  
First, (model-independent) measurements of the
$g_{ZZ\h}^2=(d+\gam b)^2$ and $g_{ZZ\phi}^2=(c+\gam a)^2$ coupling
factors for the $\h$ and $\phi$ are obtained
using $\epem\to Z\h$ and $\epem\to Z\phi$
production at the linear collider. The $sgg$ and $s\gam\gam$ couplings
($s=\h$ or $\phi$) expected 
from the standard fermion and $W$-boson loops
in the absence of the anomalous contribution can then
be computed. Meanwhile, the actual couplings-squared,
$\anti g_{s gg}^2$ and $\anti g_{s\gam\gam}^2$, 
including any anomalous contribution,
can be directly measured using a combination of $\gam\gam\to s \to b\anti b$
and $gg\to s\to \gam\gam$ data. 
  
In more detail, we employ the following procedures.
\bit
\item First, obtain
$g_{ZZs}^2$ (defined relative to the SM prediction
at $\mhsm=m_s$) 
from $\sigma(\epem\to Z s)$ (inclusive recoil technique).
\item
Next, determine $BR(s\to b\anti b)=\sigma(\epem\to Zs\to Zb\anti b)/
\sigma(\epem\to Zs)$. 
\item
Then, compute $\anti g_{s\gam\gam}^2$
from $\sigma(\gam\gam\to s\to b\anti b)/BR(s\to b\anti b)$.
\item 
To display the contribution to the $s\gam\gam$ coupling-squared
from the anomaly one would then compute
\beq
R_{s\gam\gam}\equiv
{\anti g_{s\gam\gam}^2(\mbox{from experiment})
\over \anti g_{\hsm \gam\gam}^2(\mbox{as computed for $\mhsm=m_s$})
\times g_{ZZs}^2(\mbox{from experiment})}
\label{rgamgamdef}
\eeq
\item To determine $\anti g_{sgg}^2$ experimentally requires one more step.
We must compute $\sigma(gg\to s\to \gam\gam)/BR(s\to \gam\gam)$.
To obtain $BR(s\to \gam\gam)$,
we need a measurement of $\Gamma_s^{\rm tot}$.

Given such a measurement, we then compute
\beq
BR(s\to\gam\gam)={\Gamma(s\to\gam\gam)(\mbox{computed from 
$\anti g_{s\gam\gam}^2$})\over \Gamma_s^{\rm tot}(\mbox{from experiment})}\,,
\eeq 
where the above experimental determination of $\anti g_{s\gam\gam}^2$
is employed and the experimental techniques outlined
in \cite{lcvol2} are employed for $\Gamma_s^{\rm tot}$.
\item The ratio analogous to Eq.~(\ref{rgamgamdef})
for the $gg$ coupling is then 
\beq
R_{sgg}\equiv
{\anti g_{sgg}^2(\mbox{from experiment})
\over \anti g_{\hsm gg}^2(\mbox{as computed for $\mhsm=m_s$})
\times g_{ZZs}^2(\mbox{from experiment})}\,.
\label{rggdef}
\eeq

\eit
For a light SM Higgs boson, the various cross sections
and branching ratios needed for the $s\gam\gam$ coupling
can be determined with errors of order a few percent \cite{lcvol2}. We see
from Fig.~\ref{gggagaanomalouscoup_mh120}
that for large $\xi$ this level of accuracy is
on the edge of being sufficient to detect
the deviation in the case of the $\h$.  In the case of the $\phi$,
the expected deviation is typically much larger, especially
near the zeroes in the rates. Indeed, 
the size of the deviation is largest when the 
$\phi$ rate is smallest.
A careful study is needed to assess the prospects.
For the $gg$ coupling, errors might be dominated by the accuracy
with which the total width can be determined.  Estimates 
for this error in
the case of the SM $\hsm$ are in the neighborhood of $10\%$
for $\mhsm=120\gev$ \cite{lcvol2}, decreasing for higher $\mhsm$.
Thus, the factor of two deviations expected in the case
of the $\h gg$ coupling-squared at the higher $\xi$
values might well be discernable experimentally. Since the $\phi$
may prove difficult to detect at the LHC, a much more detailed
study is required to see if deviations in the $\phi gg$
coupling due to the anomalous contribution could be detected.

\begin{figure}[p]
\begin{center}
\includegraphics[width=4in,angle=90]{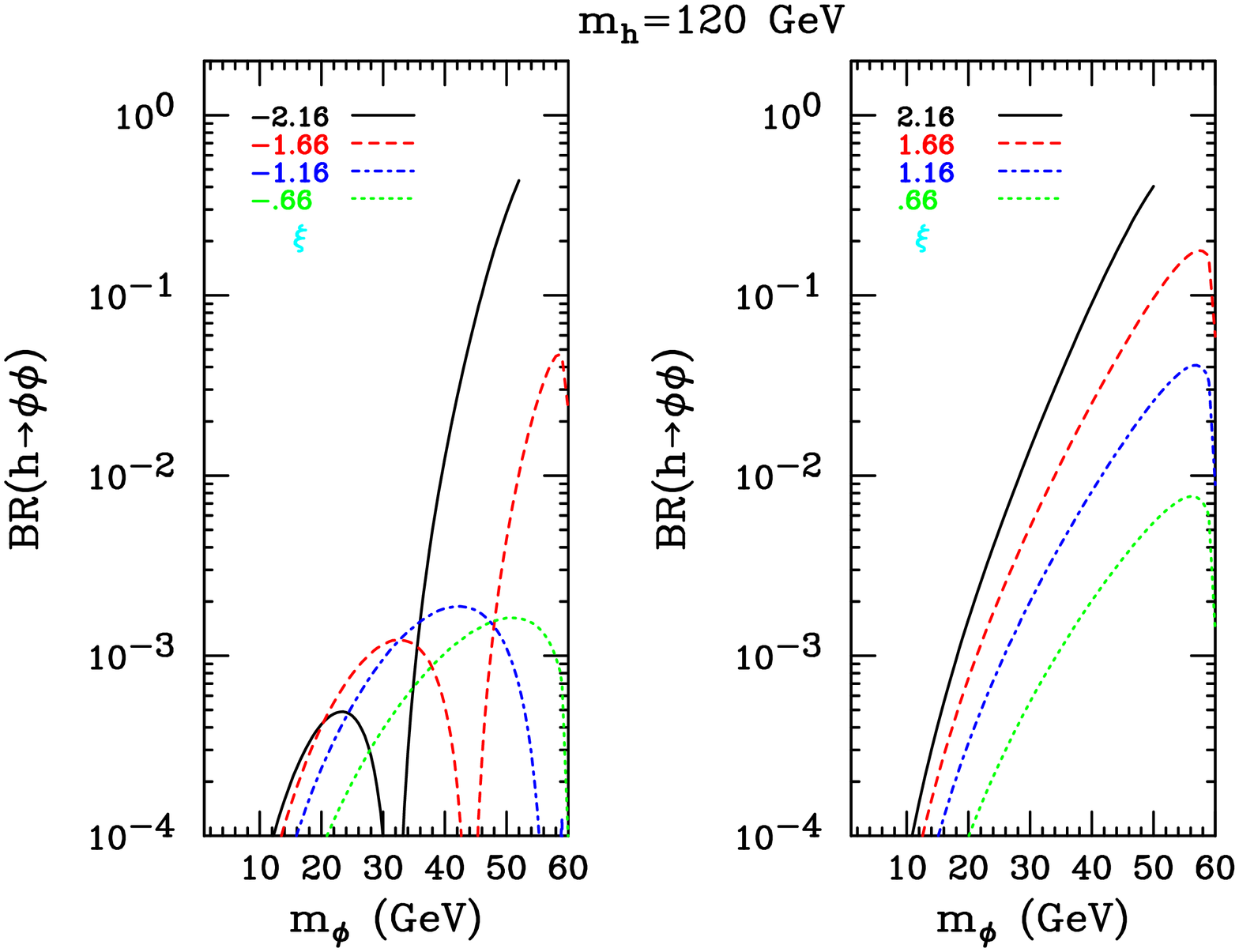}
\vspace*{-.2cm}
\includegraphics[width=4in,angle=90]{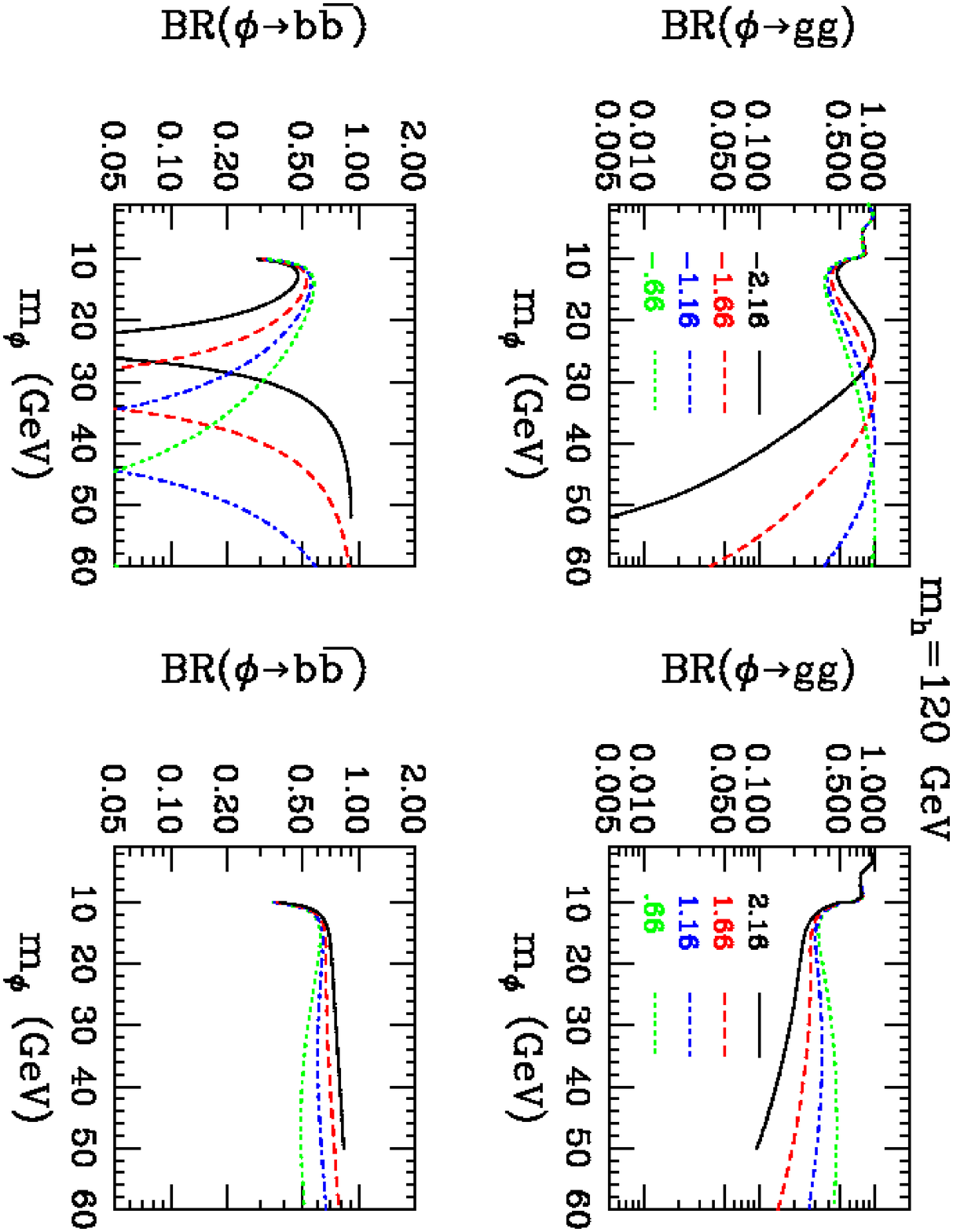} 
\end{center}
\caption{The branching ratios for $h\to \phi\phi$, $\phi\to gg$
and $\phi\to b\anti b$ for $m_h=120\gev$  
and $\lphi=5\tev$ as a function of $\mphi$ for $\xi=-2.16$, $-1.66$,
$-1.16$ and $-0.66$ (left-hand graphs) and for $\xi=0.66$, $1.16$,
$1.66$, and $2.16$ (right-hand graphs).
}
\label{br_mh120}
\end{figure}

We will now turn to a more thorough exploration of the parameter
regions in which $h\to \phi\phi$ decays are large.
The branching ratios for $h\to \phi\phi$ in the case of $\mh=120\gev$
and $\lphi=5\tev$ are shown in Fig.~\ref{br_mh120} for various $\xi$
choices within the allowed region. The plots show that $h\to\phi\phi$
decays can be quite important at the largest $|\xi|$ values
when $\mphi$ is close to $\mh/2$.
Detection of the $h\to \phi\phi$
decay mode could easily provide the most striking evidence
for the presence of $\xi\neq 0$ mixing. 
In order to understand how to search for the $h\to\phi\phi$ decay
mode, it is useful to know how the $\phi$ decays.  
In Fig.~\ref{br_mh120} we give detailed results
for $BR(\phi\to gg)$ and $BR(\phi\to b\anti b)$
for the same $\mphi$ and $\xi$ values
for which $BR(h\to \phi\phi)$ is plotted.
(The $c\anti c$ and $\tau^+\tau^-$ channels supply the remainder.)
For $\xi>0$, $BR(\phi\to b\anti b)$ is always substantial and
might make detection of the $\h\to \phi\phi\to 4b$
and $\h\to \phi\phi\to 2g 2b$ final
states possible.
The $\phi\to\gam\gam$ decay mode always has a 
very tiny branching ratio and the related detection channels
would not be useful.

One will probably first search for the $h$ in the modes that
have been shown to be viable for the SM Higgs boson.
We have given in Fig.~\ref{prodh_mh120} the rates for important
LHC discovery modes relative to the corresponding SM values
in the case of $\mphi=55\gev$. Results for other $\mphi<\mh/2$ values
are similar in nature. We observe that
the $WW\to h\to \tau^+\tau^-$ and $gg\to t\anti t h\to t\anti t b\anti b$
detection modes are generally sufficiently mildly suppressed
that detection of the $h$ in these modes should be possible
(assuming full $L=300\fbi$ luminosity per detector).
The $gg\to h\to \gam\gam$
detection mode could either be enhanced or significantly suppressed
relative to the SM expectation.
Once the $h$ has been detected in one of the SM modes, 
a dedicated search for
the $h\to \phi\phi\to b\anti b b\anti b$
and $h\to \phi\phi\to b\anti b gg$ decay modes will be important.
At the LHC, backgrounds for these modes will be substantial
and a thorough Monte Carlo assessment is needed.

At the LC, since $g_{ZZh}^2$ is close to 1 (relative to the SM Higgs
value), the $h$ will be readily detectable using the recoil
mass procedure in $e^+e^-\to Z h$ events.  Once the $h$
mass peak is detected, it should be possible to delineate in detail
the $h$ and $\phi$ branching ratios.

As for detection of the $\phi$ at the LC, the most relevant quantity
is $g_{ZZ\phi}^2$. Detailed plots of this quantity 
appear in Fig.~\ref{couplingsphi}.
These plots indicate that LC detection of $e^+e^-\to Z\phi$
using the recoil mass method will require being far from
the zero in $g_{ZZ\phi}^2$. For a significant portion
of parameter space, it seems quite apparent that the only
way to detect the $\phi$ would be through the $\h\to \phi\phi$ decays.

In order to have substantial $BR(\h\to \phi\phi)$ it is necessary
that $\mh<2\mw$.  As $\mh$ is increased above $2\mw$, 
the $WW$ and then $ZZ$ modes become strong and overwhelm the $\phi\phi$
decay mode. For example, for $\mh=200\gev$, 
the largest value found for $BR(\h\to\phi\phi)$ is of order $1\div 2\,\%$,
and such values are again achieved when $|\xi|$ is as large as possible
and $\mphi$ is just below $\mh/2$.

\begin{figure}[h!]
\begin{center}
\includegraphics[width=4in,angle=90]{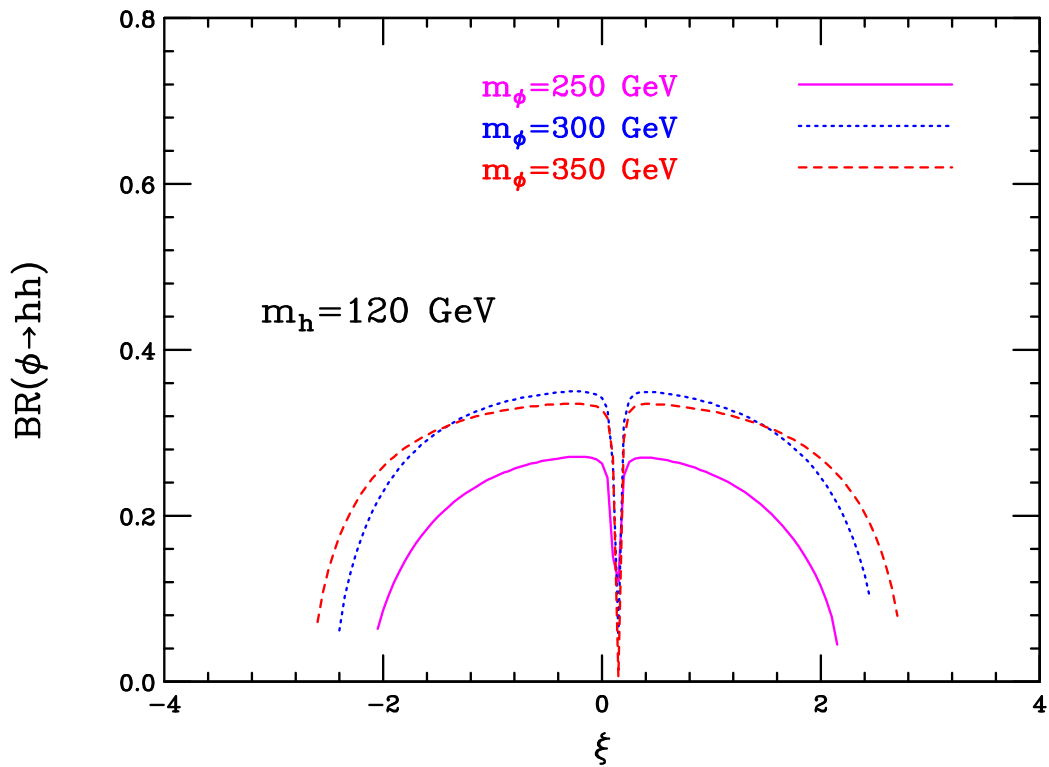}
\end{center}
\vspace*{-.2in}
\caption{The $\phi\to hh$ 
branching ratio is plotted as a function of $\xi$ for
$\mh=120\gev$ and $\mphi=250$, $300$ and $350\gev$.
We have taken $\lphi=5\tev$
and assumed $m_1>\mphi$. 
}
\label{brphihh}
\end{figure}

Let us now discuss $\phi\to hh$ decays. For $\mh=120\gev$,
these are present once $\mphi\gsim 240\gev$.
When allowed, these decays will be quite strong since
the $\phi hh$ coupling
is typically larger than the $h\phi\phi$ coupling away from 
zeroes in the coupling. In addition,
the $\phi$ decays to $f\anti f$ and $VV$ are typically suppressed
compared to those of the $\h$ because of the smaller size
of $(c+\gam a)^2$ compared to $(d+\gam b)^2$ 
when $\mphi>\mh$ (see Figs.~\ref{couplingsh} and \ref{couplingsphi}) 
for all but the
largest $|\xi|$ values. The importance
of the $\phi\to hh$ decays is illustrated for $\mh=120\gev$
and $\mphi=250$ $300$ and $350\gev$ in Fig.~\ref{brphihh}.
Even though $\mphi>2\mw$ in all these cases, 
$BR(\phi\to hh)$ is
still of order $0.3\div 0.4$ for most of the allowed $\xi$ range
not near a zero in the $\phi hh$ coupling.

\section{Phenomenology for {\boldmath $\lphi=1\tev$}}
\label{secpheno1}

In this section, we consider the more marginal  choice
of $\lphi=1\tev$. For this case, we will consider
first results obtained assuming $m_1$ is large ($>1\tev$).
As discussed earlier, such large $m_1$ requires large curvature,
$m_0/\mpl \sim \mathcal{O}(1)$, that would presumably imply
significant corrections to the RS ansatz.  Nonetheless,
the large-$m_1$ results provide a useful benchmark that might
provide a reasonable first approximation in such a case.
We also noted that for $\lphi=1\tev$ and $m_0/\mpl\sim 1$ 
large $m_1$ is needed to clearly avoid any constraints from
RunI Tevatron data.  We next consider results obtained
in two small-curvature cases:
$m_0/\mpl\sim 0.065$ and $\sim 0.195$, corresponding to
$m_1=100$ and $300\gev$, respectively.
As discussed, such small $m_1$ values might or might not be inconsistent
with constraints from current RunI Tevatron data 
and from the $S$ and $T$ electroweak observables.
However, the very interesting physics associated with Higgs decays
to KK excitations that emerges deserves attention just
in case this scenario should arise.  In our presentation for
$\lphi=1\tev$, 
we focus only on the significant changes as compared to $\lphi=5\tev$.

\begin{figure}[h!]
\begin{center}
\includegraphics[width=4in]{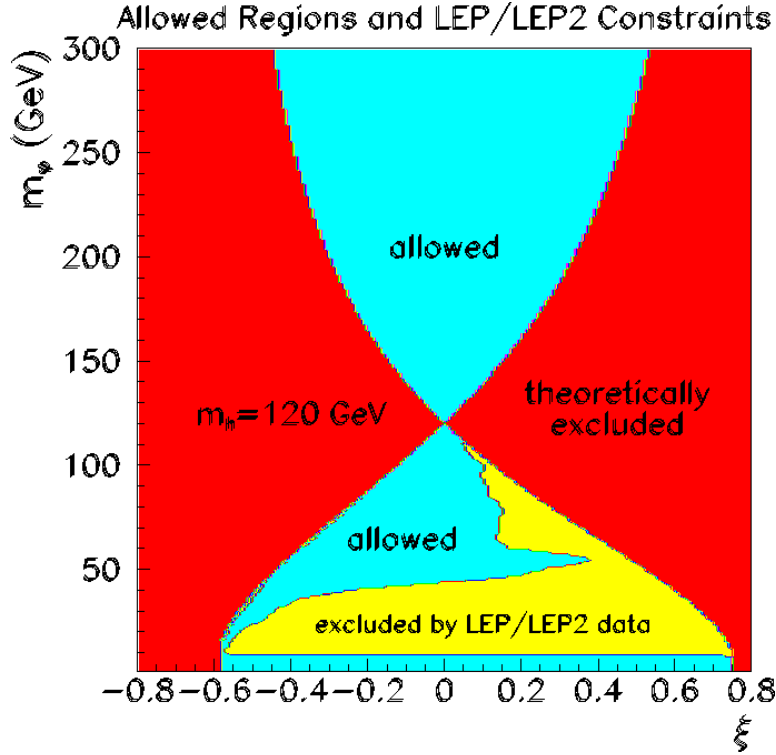}
\end{center}
\vspace*{-.3in}
\caption{As in Fig.~\ref{allowed_mh120} but for
$\lphi=1\tev$. The region  of theoretically allowed 
$\xi>0$ values below $\sim 0.45$ with 
$55\gev \lsim\mphi\lsim 115\gev$ that are in the yellow LEP/LEP2-excluded
region will be referred to as the `LE' region.
}
\label{allowed_mh120_1tev}
\end{figure}

\begin{figure}[h!]
\begin{center}
\includegraphics[width=5.5in]{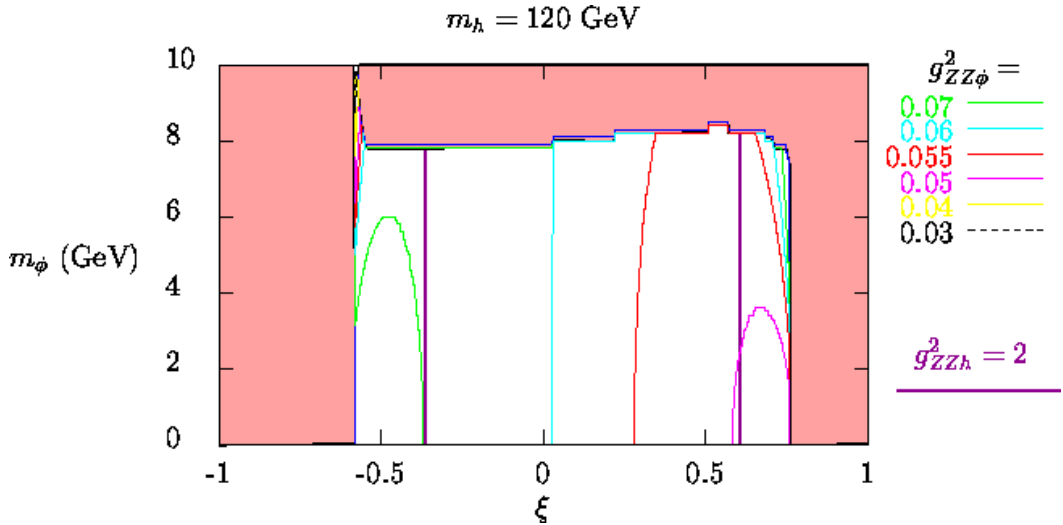}
\end{center}
\vspace*{-.2in}
\caption{For $\mh=120\gev$ and $\lphi=1\tev$, we plot contours for 
$g_{ZZ\phi}^2=(c+\gam a)^2$ 
inside the region that can only be excluded
if we assume the limits of \cite{Buskulic:1993gi} apply. 
The two thick (magenta) lines are the $\xi$ values
such that $g_{ZZh}^2=2$.  The region between these lines
has $g_{ZZh}^2<2$ and is that most likely to 
be consistent with precision electroweak data.
}
\label{mphi_0}
\end{figure}

First, we present the allowed region in $(\xi,\mphi)$ parameter
space for $\mh=120\gev$ in Fig.~\ref{allowed_mh120_1tev}.
Of course, the allowed $\xi$ range is
very much reduced compared to $\lphi=5\tev$ 
since $\gam$ is five times larger; see Eq.~(\ref{xilim}).
As compared to Fig.~\ref{allowed_mh120}, we see that there
is a significant region with lower $\mphi$, but with $\mphi\gsim 8\gev$,
that is excluded by
the LEP/LEP2 limits coming from untagged hadronic
events and/or from $b$-tagged final states.
A similar region is not excluded in the $\lphi=5\tev$ case because
the $g_{ZZ\phi}^2$ coupling
for $\lphi=5\tev$ is substantially smaller than for $\lphi=1\tev$.
Returning to the $\lphi=1\tev$ case, points with $\mphi\lsim 8\gev$
are not exluded because the upper bound on $g_{ZZ\phi}^2$
 coming from untagged hadronic
final states rises very rapidly as one
moves to lower masses and there are no limits from $b$-tagged final states.
However, we should note that
if the limits of \cite{Buskulic:1993gi} apply
(we have assumed they do not because of the dominance of $\phi\to gg$ decays),
for $\lphi=1\tev$ the $\mphi\lsim 8\gev$ region would
be excluded as well as the $\mphi\gsim 8\gev$ regions shown.
As illustrated in Fig.~\ref{mphi_0}, for $\lphi=1\tev$ the magnitude of
$g_{ZZ\phi}^2$ in this low-$\mphi$ region is not so very small.

In what follows, it will be convenient to include in some of our plots some 
$(\xi,\mphi)$ values that are marked as LEP/LEP2-excluded 
in Fig.~\ref{allowed_mh120_1tev}:
namely, we include all those theoretically allowed 
$\xi>0$ values below $\sim 0.45$ with 
$55\gev \lsim\mphi\lsim 115\gev$ that are marked in yellow.
We will refer to this region as region `LE' in what follows.

\begin{figure}[h!]
\begin{center}
\includegraphics[width=5.8in]{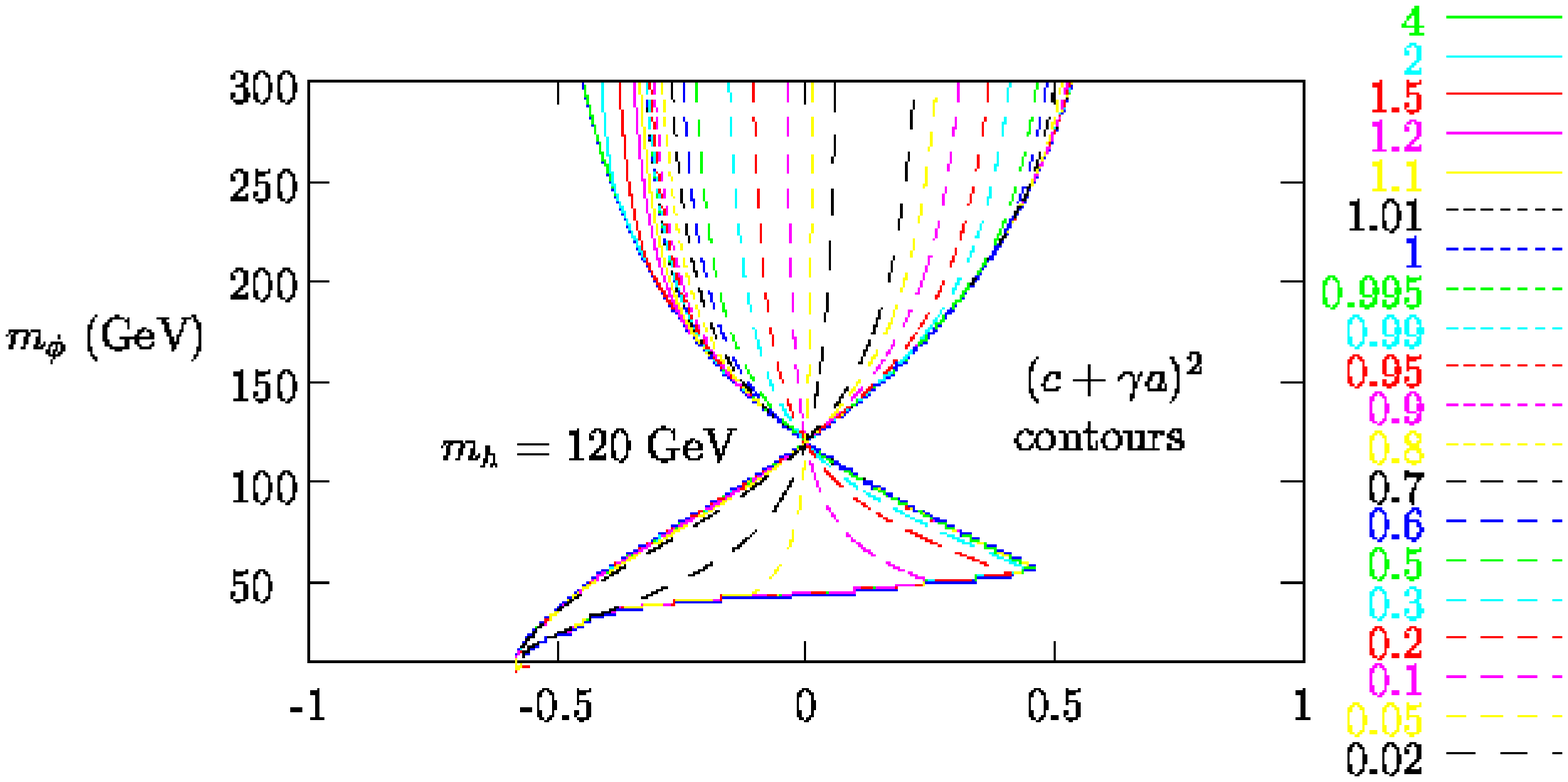}
\vspace*{.2in}

\includegraphics[width=5.8in]{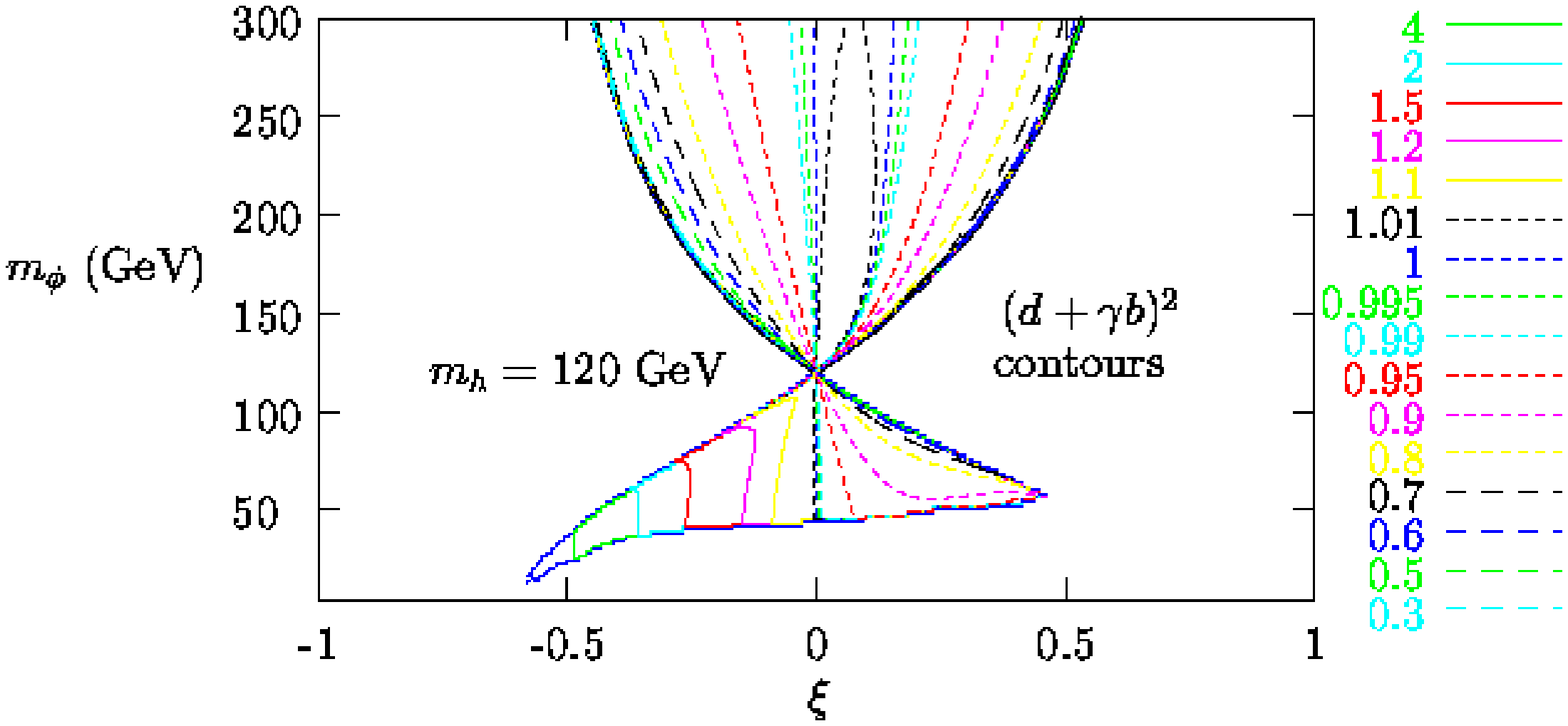}
\end{center}
\caption{For $\mh=120\gev$ and $\lphi=1\tev$, we plot contours for 
the quantities $(d+\gam b)^2$ and $(c+\gam a)^2$.
For $(d+\gam b)^2$, 
only the region $\mphi\geq 15\gev$ is shown. For $(c+\gam a)^2$,
we show the narrow pipe that connects to the allowed region 
of very small $\mphi$; see Fig.~\ref{mphi_0}.
}
\label{couplings_1tev}
\end{figure}

Contours of $g_{ZZh}^2=(d+\gam b)^2$ and $g_{ZZ\phi}^2=(c+\gam a)^2$
are presented in Fig.~\ref{couplings_1tev}.  There, we see that region
LE is excluded by LEP/LEP2 data because in this region the $g_{ZZ\phi}^2$
value gets to be a reasonable fraction of one, the SM value.
Globally speaking, the main difference between the couplings for $\lphi=1\tev$
versus those for $\lphi=5\tev$ of Figs.~\ref{couplingsh} and \ref{couplingsphi}
is that $g_{ZZ\phi}^2$ is overall much larger in the $\lphi=1\tev$ case.

\begin{figure}[p!]
\begin{center}
\includegraphics[width=3.5in,angle=90]{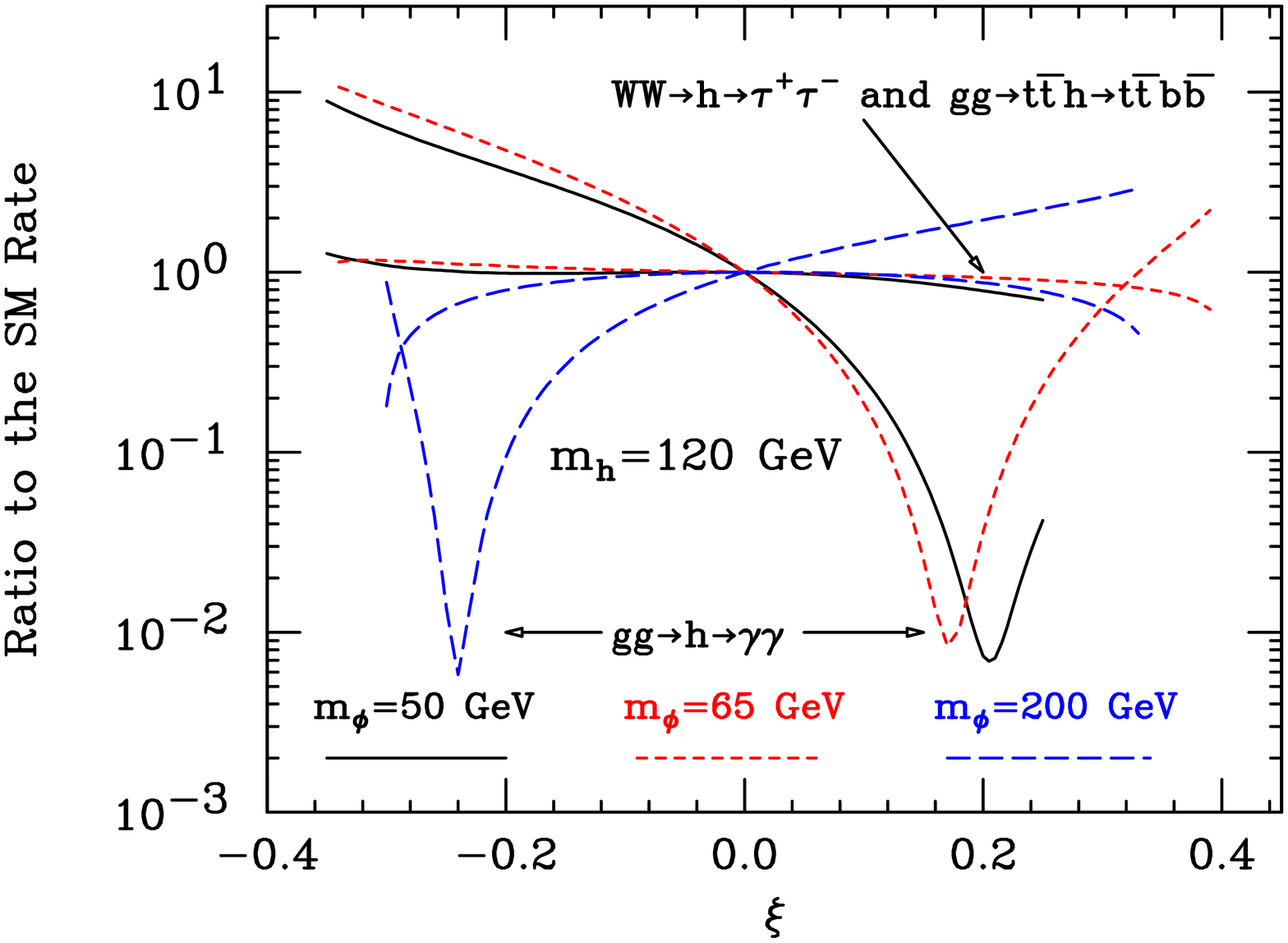}
\end{center}
\vspace*{-.2in}
\caption{The ratio of the rates for $gg\to h \to \gam\gam$
and $WW\to h \to \tau^+\tau^-$ (the latter being the same as
that for $gg\to t\anti t h\to t\anti t b\anti b$)
to the corresponding rates for the SM Higgs boson.
Results are shown
for $m_h=120\gev$  and $\lphi=1\tev$ as functions of $\xi$
for $\mphi=50$, $65$ and $200\gev$.
}
\label{prodh_mh120_1tev}
%
\begin{center}
\includegraphics[width=3.5in,angle=90]{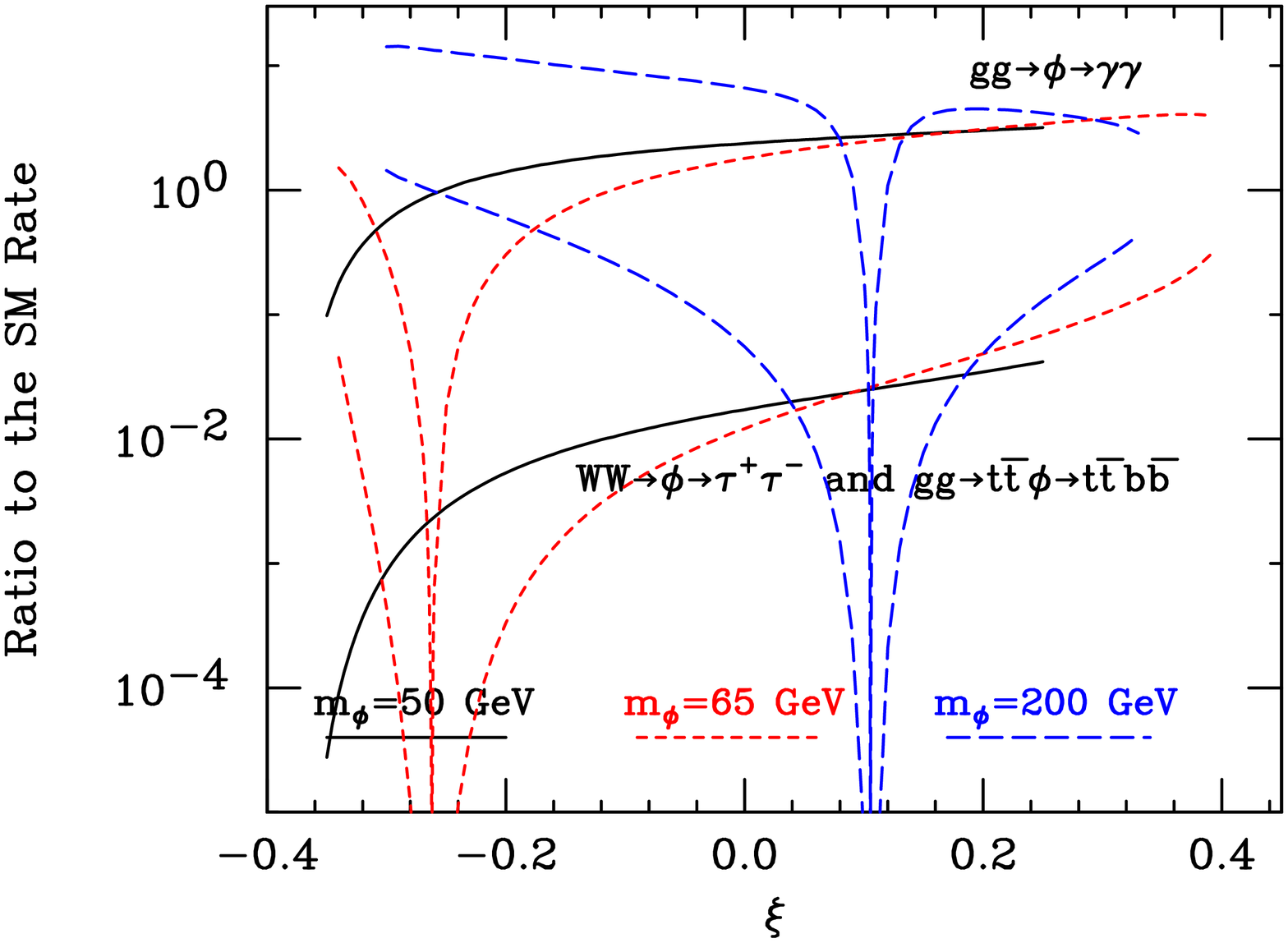}
\end{center}
\vspace*{-.2in}
\caption{The ratio of the rates for $gg\to \phi\to \gam\gam$ (the higher
curves for a given $\mphi$) 
and for $WW\to \phi \to \tau^+\tau^-$ (the latter being the same as
that for $gg\to t\anti t \phi \to t\anti t b\anti b$)
to the corresponding rates for the SM Higgs boson.
Results are shown
for $m_h=120\gev$  and $\lphi=1\tev$ as functions of $\xi$
for $\mphi=50$, $65$ and $200\gev$. 
}
\label{prodphi_mh120_1tev}
\end{figure}

\begin{figure}[p!]
\begin{center}
\includegraphics[width=3.5in,angle=90]{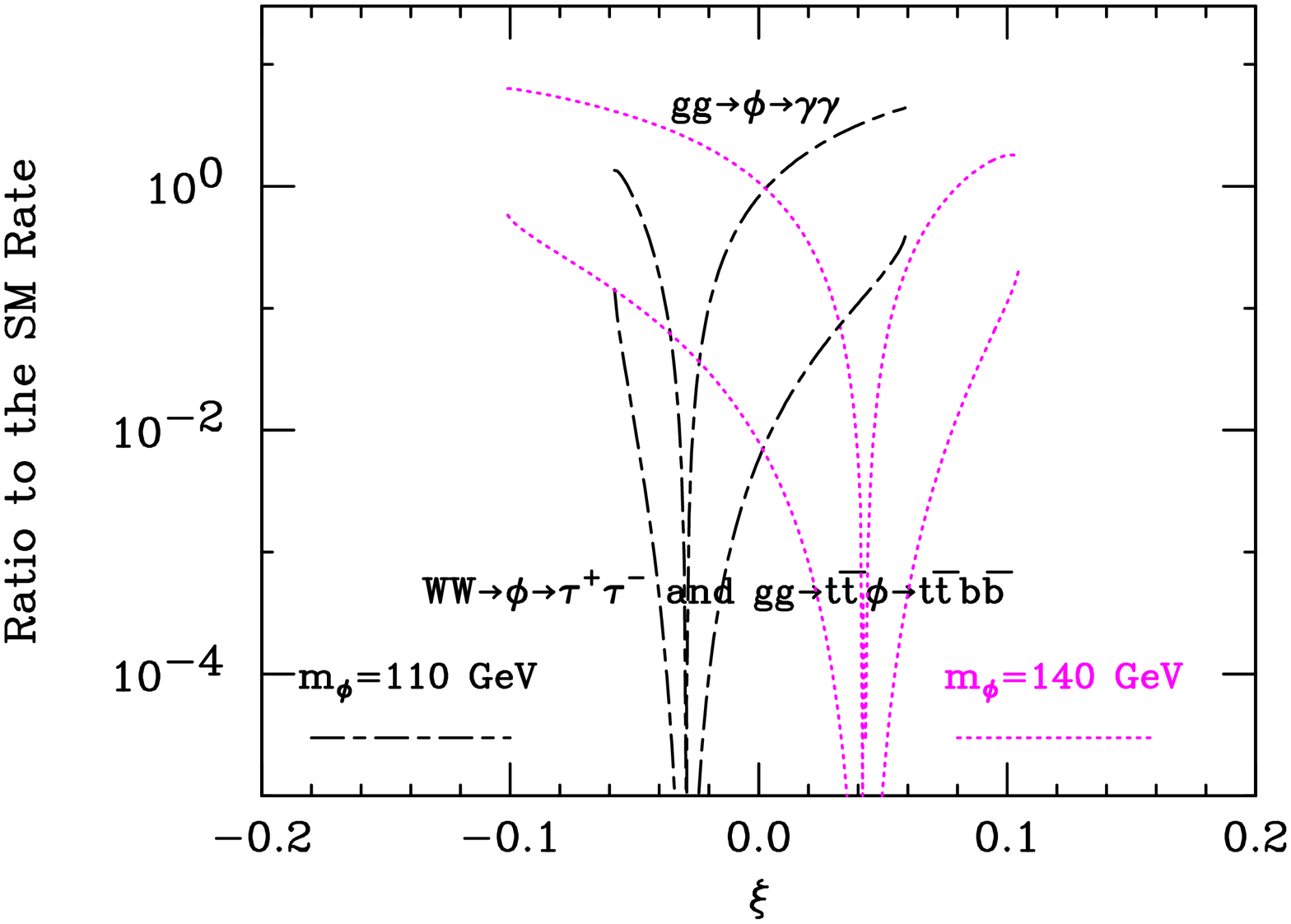}
\end{center}
\vspace*{-.2in}
\caption{The ratio of the rates for $gg\to \phi\to \gam\gam$ (the higher
curves for a given $\mphi$) 
and for $WW\to \phi \to \tau^+\tau^-$ (the latter being the same as
that for $gg\to t\anti t \phi \to t\anti t b\anti b$)
to the corresponding rates for the SM Higgs boson.
Results are shown
for $m_h=120\gev$  and $\lphi=1\tev$ as functions of $\xi$
for $\mphi=110$ and $140\gev$.
}
\label{prodphi_mh120_2_1tev}
%

%
\begin{center}
\includegraphics[width=3.5in,angle=90]{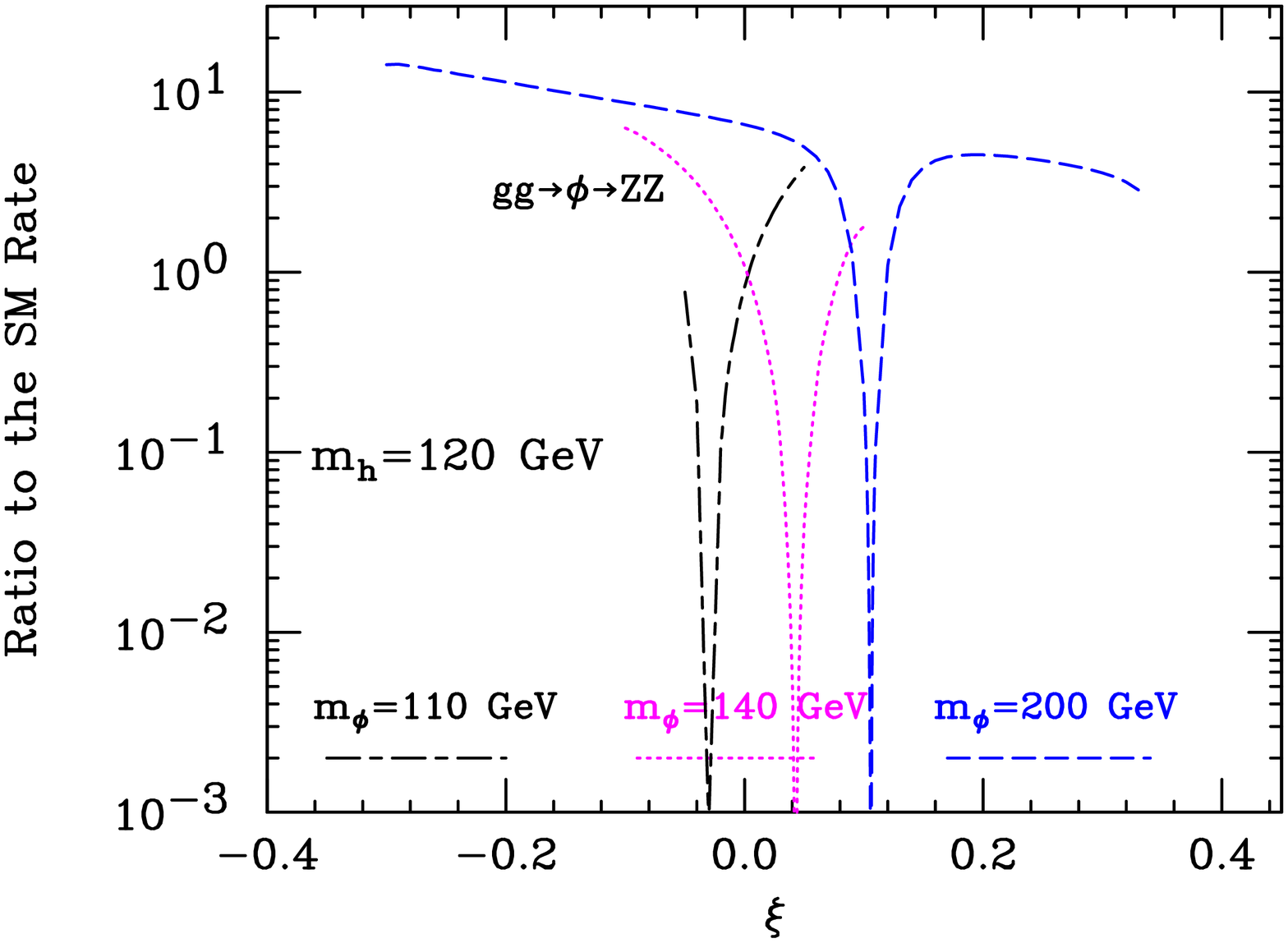}
\end{center}
\vspace*{-.2in}
\caption{The ratio of the rate for $gg\to \phi\to ZZ$
to the corresponding rate for a SM Higgs boson with mass $\mphi$
assuming $\mh=120\gev$ and $\lphi=1\tev$ as a function of $\xi$
for $\mphi=110$, $140$ and $200\gev$. Recall that the $\xi$ range
is increasingly restricted as $\mphi$ becomes more degenerate
with $\mh$.
}
\label{prodphiggtozz_mh120_1tev}
\end{figure}
\begin{figure}[h!]
\begin{center}
\includegraphics[width=3.5in,angle=90]{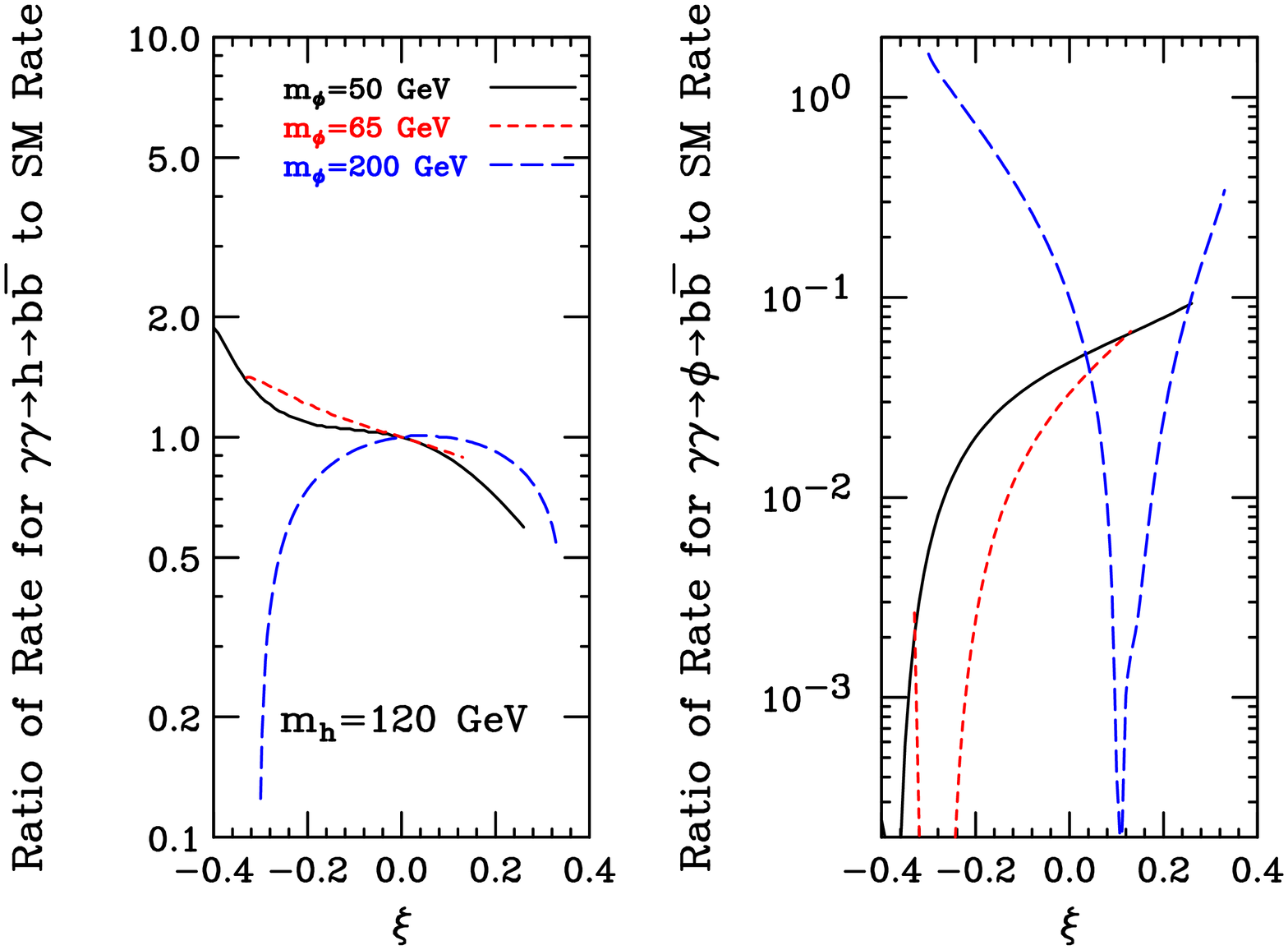}
\end{center}
\vspace*{-.2in}
\caption{The rates for $\gam\gam\to h \to b\anti b$
and $\gam\gam\to\phi\to b\anti b$ relative
to the corresponding rate for a SM Higgs boson of the same mass.
Results are shown
for $m_h=120\gev$  and $\lphi=1\tev$ as functions of $\xi$
for $\mphi=50$, $65$ and $200\gev$.
}
\label{gagatobb_mh120_1tev}
\end{figure}

Next, we present the corresponding graphs related to LHC 
and $\gam\gam$ collider discovery.
For these graphs, we have chosen to focus on $\mh=120\gev$
(as for $\lphi=5\tev$) and on the values
of $\mphi=50$, $65$ and $200\gev$. The lowest value still gives
a substantial range of allowed $\xi$ and 
will have significant $BR(h\to\phi\phi)$.  
For the middle value, these decays are forbidden.  
In all the LHC and $\gam\gam$ collider graphs, $m_1$ is assumed to be large,
in particular large enough that decays of
the Higgs or radion to $h^1$ are forbidden.
The main implication
of Figs.~\ref{prodh_mh120_1tev}, \ref{prodphi_mh120_1tev}, \ref{prodphi_mh120_2_1tev}, \ref{prodphiggtozz_mh120_1tev} and \ref{gagatobb_mh120_1tev}
is that for $\lphi=1\tev$, 
$\phi$ discovery at the LHC and in $\gam\gam$ collisions
has much better prospects (away from the usual zeroes in the $\phi gg$
and $\phi\gam\gam$ couplings) than in the case of $\lphi=5\tev$.
It is still true that the anomalous contributions to the
$\h\gam\gam$ and $\phi\gam\gam$ couplings will be hard to isolate.

\begin{figure}[p]
\begin{center}
\includegraphics[width=3.6in,angle=90]{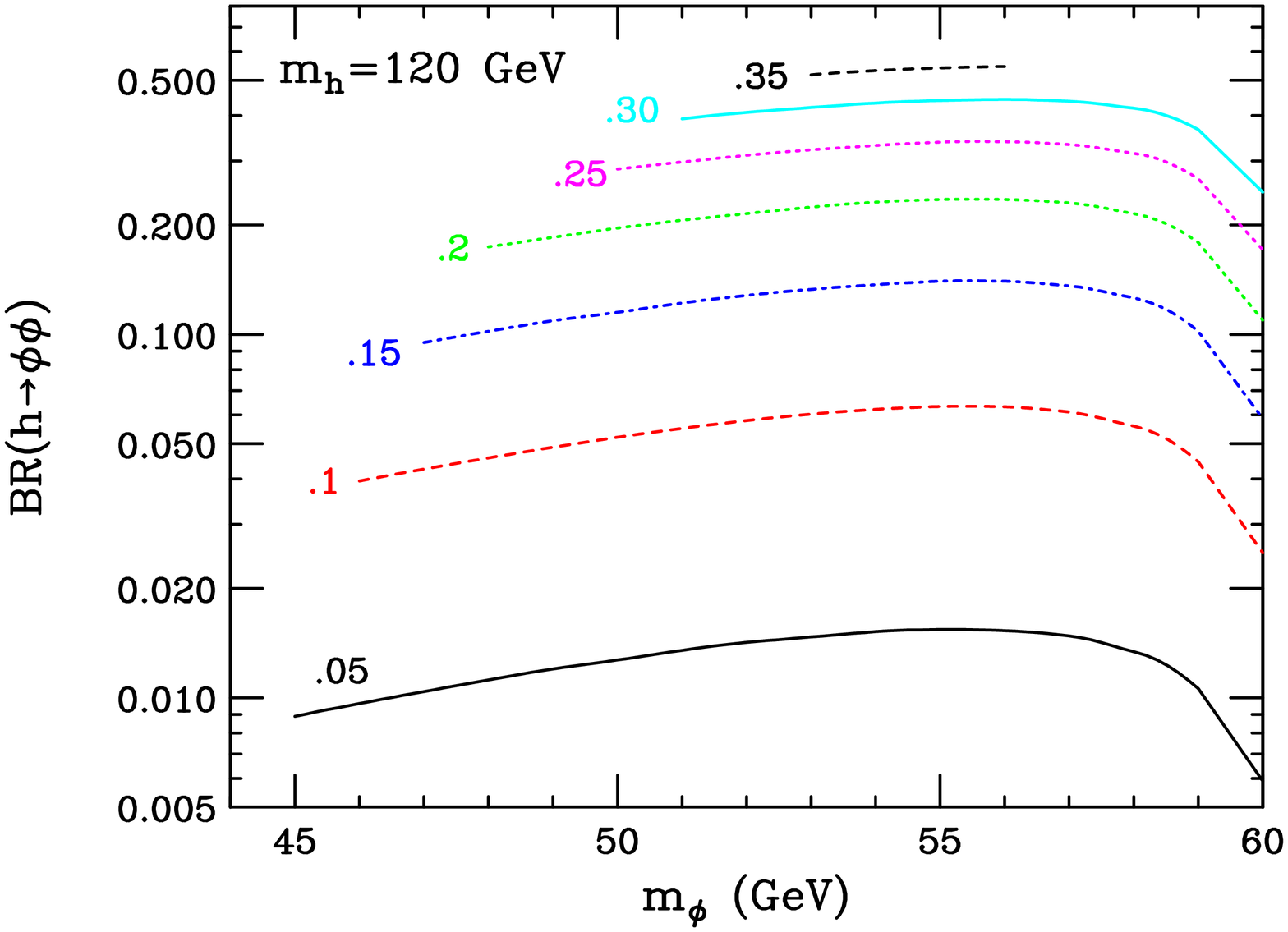}
\vspace*{.2in}
%
%
\includegraphics[width=3.7in,angle=90]{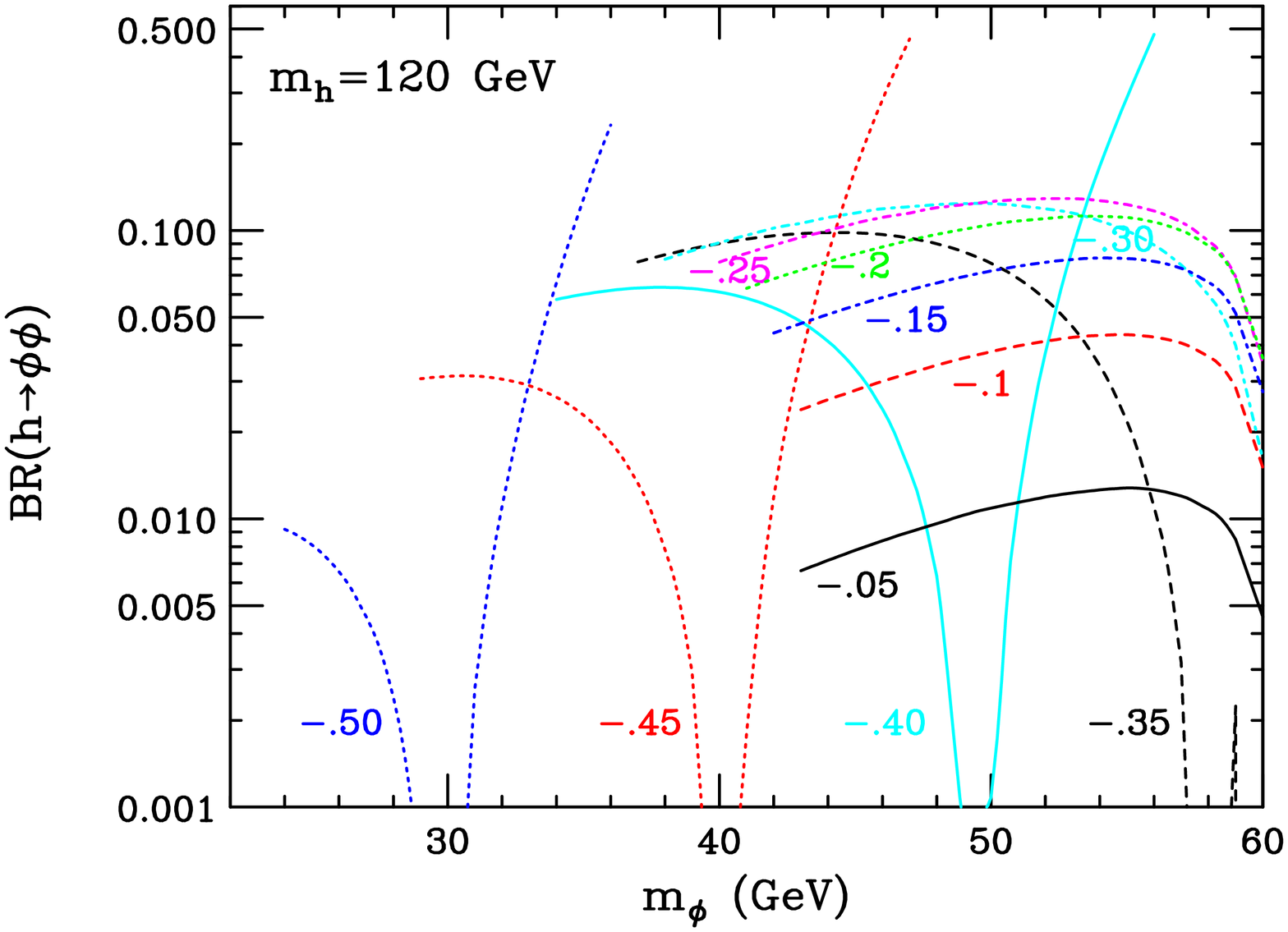}
\end{center}
\vspace*{-.2in}
\caption{For various $\xi>0$ and
$\xi<0$ values, the branching ratio for $h\to \phi\phi$ is plotted as
a function of $\mphi$, taking
$m_h=120\gev$  and $\lphi=1\tev$.  Only points not excluded by LEP/LEP2 
(the blue region of Fig.~\ref{allowed_mh120_1tev}) are plotted.
The curves terminate at low $\mphi$ when the LEP/LEP2
limits of are encountered.
}
\label{br_mh120_1tev}
\end{figure}

The greatest interest in the lower $\lphi$ value derives from
the fact that $h\to \phi\phi$ decays can be much more prominent
and that $h,\phi$ decays to final states containing the 1st KK
excitation $h^1$ become possible. The first point is illustrated in
Fig.~\ref{br_mh120_1tev}. 
$BR(\h\to\phi\phi)$ can be as large as 50\% at the highest
allowed $|\xi|$ values.

\begin{figure}[h!]
\begin{center}
\includegraphics[width=4in,angle=90]{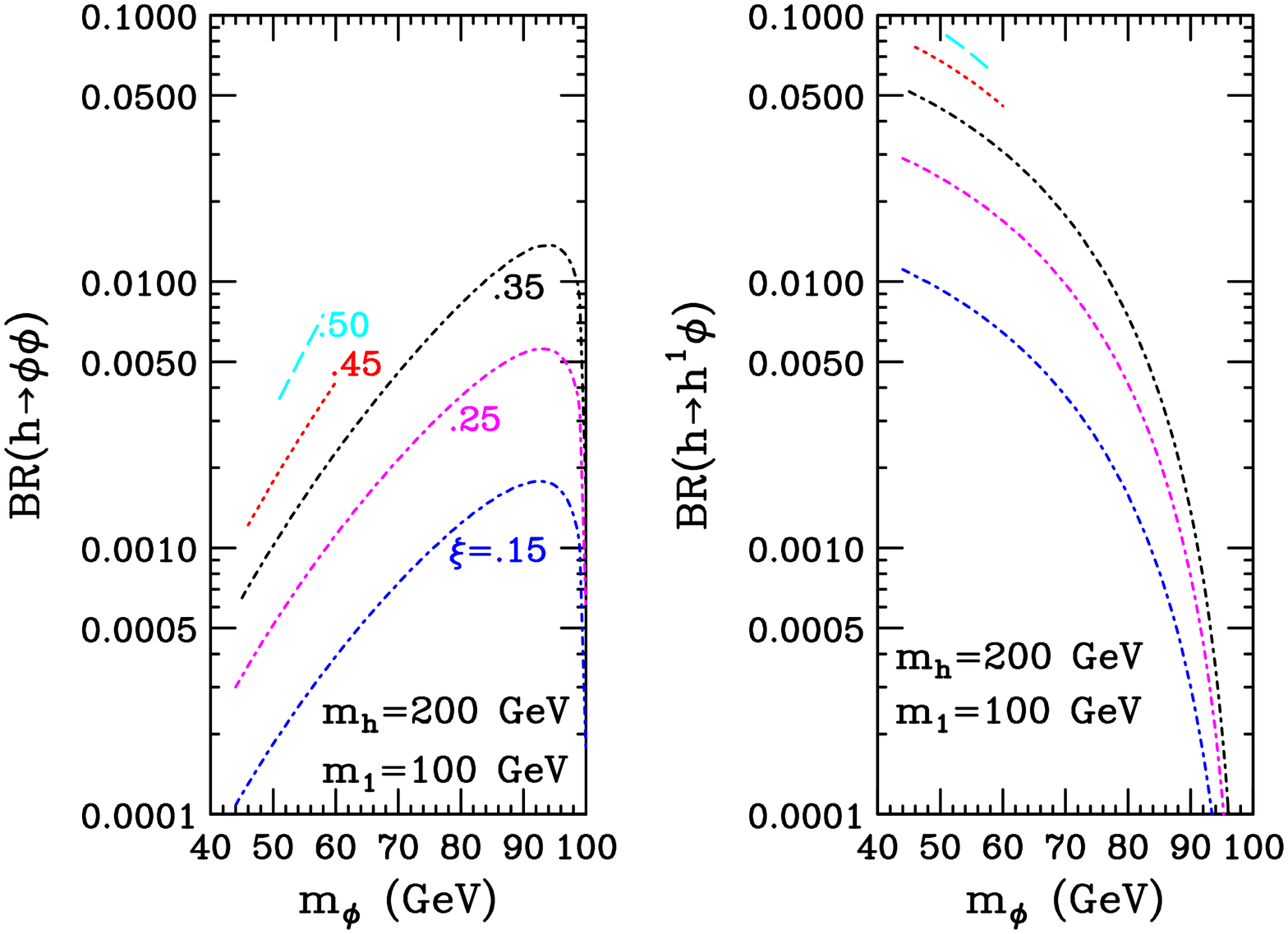}
\end{center}
\vspace*{-.2in}
\caption{The $h\to\phi\phi$ and $h\to h^1\phi$
branching ratios as a function of $\mphi$ for
$m_h=200\gev$, $m_1=100\gev$, and $\lphi=\sqrt3\lwh=1\tev$, 
for various $\xi$ choices.
Results are plotted only for $\mphi$ values 
satisfying LEP/LEP2 bounds (the blue region of Fig.~\ref{allowed_mh120_1tev}).
The curve legend for the right-hand plot is the
same as shown in the left-hand plot.
}
\label{br_mh200_1tev}
\end{figure}

Let us now discuss what happens at higher $\mh$ values.
The largest value that can be easily
consistent with precision electroweak constraints is $\mh=200\gev$.
For this value, we will require in our plots that $g_{ZZh}^2<1.2$
in order to be certain that $S,T$ lie within the 95\% CL ellipse.
In order to learn if the $h\to h^1\phi $ decay could be significant,
we retain $\lphi=1\tev$, for which Eq.~(\ref{params})
implies that $\lwh=\lphi/\sqrt3=577\gev$, and
choose $m_1=100\gev$ [corresponding to $m_0/\mpl\sim 0.065$,
see Eq.~(\ref{params})].
Results for $BR(h\to \phi\phi)$ and $BR(h\to h^1 \phi)$
are plotted in Fig.~\ref{br_mh200_1tev}
as a function of $\mphi$ for selected values of $\xi>0$.
(Only relatively small values of $|\xi|$ are not excluded
by the $g_{ZZh}^2<1.2$ requirement when $\xi<0$.)  
The $h\to \phi\phi$ branching ratio
can be significant, especially for the larger values of $\xi$
allowed by the theoretical constraint of $Z^2>0$.
Certainly, these decays should be searched for at the LC as their
presence would imply non-zero $\xi$ and would allow a measurement
of this very fundamental parameter. 
The reason for the small size of the $h\to \phi \phi$
and $h\to h^1\phi $ branching ratios is the
dominance of the $WW$ and $ZZ$ decay modes.  Once these $VV$ decays
become full strength, the $\phi\phi$ and $h^1\phi $ decays
will be rare.

\begin{figure}[h]
\begin{center}
\includegraphics[width=4in,angle=90]{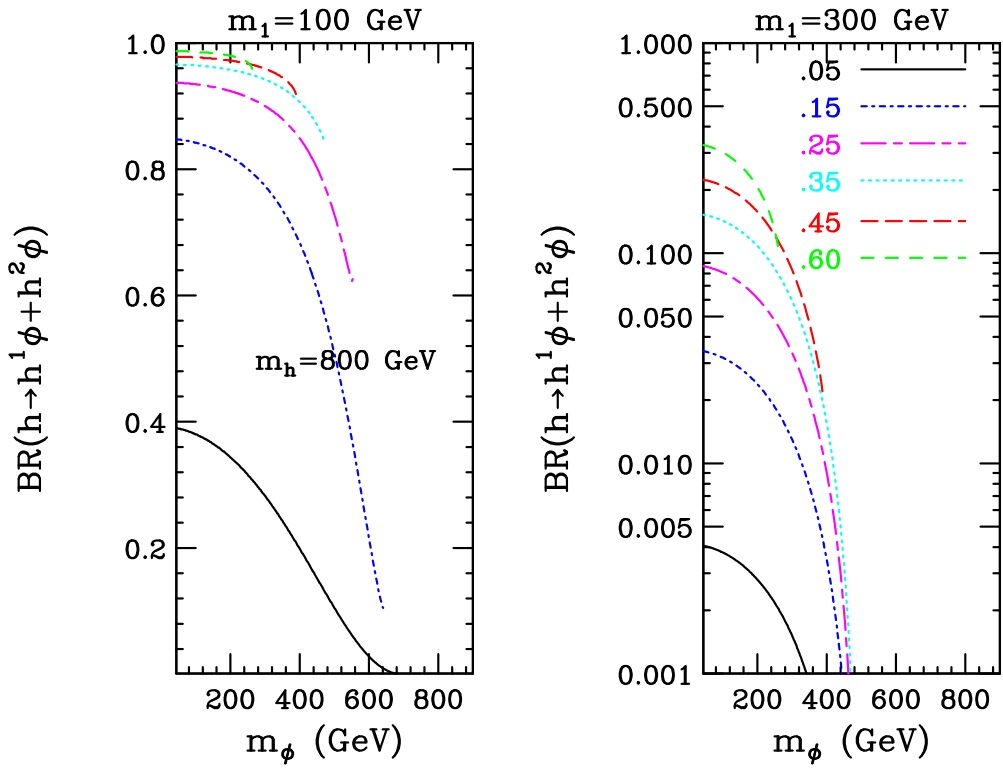}
\end{center}
\vspace*{-.2in}
\caption{We plot the $h\to h^1\phi+h^2\phi $ 
branching ratio as a function of $\mphi$ for
$m_h=800\gev$ and $\lphi=\sqrt3\lwh=1\tev$,
in the cases of $m_1=100\gev$ and $300\gev$,
for various $\xi$ choices (as indicated in the right-hand window).
Results are plotted only for $\mphi$ values 
satisfying LEP/LEP2 bounds.
In this plot, we have assumed that $h\to h^1h^1+\ldots$ 
decays (for which vertices
do exist but have not been studied in detail in this paper)
are unimportant even though they are kinematically allowed for
the $\mh$ and $m_1$ choices of this figure.
}
\label{brhphiphiphihn_mh800_1tev}
\end{figure}

As noted earlier, at still larger values of $\mh$
precision electroweak constraints become difficult to
satisfy. A future paper will explore this region in more detail.
Very roughly, the $\Gamma(h\to h^n \phi)\sim \lwh^{-2}
\mh^7/m_n^4$ behavior found earlier 
in Eq.~(\ref{htohnphiwdth}) means that 
$h\to h^n\phi$ decays can dominate over 
the $WW$ and $ZZ$ decay modes that grow only as $\mh^3$,
{\it provided}  that $m_1$ is sufficiently
small and that $\lwh$ (and hence $\lphi$) is of order a TeV.
To illustrate, in Fig.~\ref{brhphiphiphihn_mh800_1tev}
we plot $BR(h\to h^1\phi+h^2\phi)$ for $\mh=800\gev$ as a function of $\mphi$
for a number of positive $\xi$ values
and for the cases of $m_1=100\gev$ and $m_1=300\gev$. 
[$BR(h\to\phi\phi)$ is typically below or of order 0.01 
for this large a value of $\mh$.]
In obtaining the results shown, 
we have assumed that $h\to h^1h^1+\ldots$ 
decays (for which vertices
do exist but have not been studied in detail in this paper)
are unimportant even though they are kinematically allowed for
the $\mh$ and $m_1$ choices of this figure.  
We also note that $BR(h\to h^1\phi)>11 BR(h\to h^2\phi)$
 for the cases studied, as anticipated from
the $(m_1/m_2)^4\sim (3.8/7)^4\sim 0.086$ scaling noted above.
From Fig.~\ref{brhphiphiphihn_mh800_1tev},
we see that $m_1=100\gev$ yields large 
values of $BR(h\to h^1\phi+h^2\phi)$ at small $\mphi$ when $\xi$
is not small. For $m_1=300\gev$,
$BR(h\to h^1\phi+h^2\phi)$ is much smaller
than for $m_1=100\gev$, being of order
$\sim 0.04\div 0.4$ for small values of $\mphi$ and $\xi$
ranging from 0.05 to 0.60. Results for $\xi<0$ are very
similar in nature.

\section{Summary and Conclusions}
\label{secfinal}

We have discussed the scalar sector of the Randall-Sundrum model. The
effective potential (defined as a set of interaction terms that
contain no derivatives) for the Standard Model Higgs-boson ($\ho$) sector
interacting with Kaluza-Klein excitations of the graviton
($h_\mu^{\mu\, n}$) field and the radion ($\phio$) field has been
derived. Without specifying its origin, a stabilizing mass-term for
the radion has been introduced. After including this term, 
we have shown that only the
Standard Model vacuum determined by $\partial V(\ho)/\partial \ho =0$ is
allowed. Further, we find that consistency of the RS solution
requires that the Higgs potential 
vanishes at the vacuum solution. Otherwise, 
the finely tuned matching required in the RS model between the bulk and branes
would be violated. As a result, for the correct vacuum solution 
the effective potential does not
contain any terms linear in the quantum Higgs field.
The above results emerge only with a very full treatment of the effective
potential.  Truncation of its expansion in powers of the fields can
lead to erroneous conclusions.

Having confirmed that the usually assumed vacuum properties are 
correct, we pursue in more detail the phenomenology of the RS scalar
sector, focusing in particular on results found in the presence of 
a curvature-scalar mixing $\xi\, R\, \Hhat^\dagger
\Hhat$ contribution to the Lagrangian.
We delineate the somewhat tricky
`inversion' procedure for determining all the Lagrangian
parameters given the 
masses of the physical eigenstates $\h$ and $\phi$. 
A full set of Feynman rules for the resulting tri-linear interactions
among the $h$, $\phi$ and $h^n$ mass eigenstates are then derived.
We also summarize the Feynman rules for couplings to standard channels:
$b\anti b,WW,\ldots$ as well as $gg$ and $\gam\gam$ (including
the anomalous contributions to the latter).
Simple sum rules that relate Higgs-boson and radion couplings to pairs
of vector bosons and fermions are given. 
Of particular interest is the fact that non-zero $\xi$ induces interactions
linear in the Higgs field: $\phi^2 h$ and $h^n h \phi$.
The explicit forms of these interactions 
must be obtained using the above-mentioned full treatment of
the effective potential as well as a similarly full treatment
of the related derivative terms in the Lagrangian. 

We summarize the connections between the parameters of the model
and the lower bounds on the new physics scales among these parameters
required by precision electroweak and Tevatron RunI constraints.
We explore the behavior of the couplings in the range
of parameter space allowed by theoretical and existing experimental
constraints.  In particular, we derive the regions of parameter
space that are excluded by direct LEP/LEP2 limits 
on scalar particles with $ZZ$ coupling as function of scalar mass.
Of particular note is the fact that
the sum rule for $ZZh$ and $ZZ\phi$ squared-couplings noted above 
implies that it is impossible for both the $h$ and $\phi$ to be light.

We note that precision electroweak data is most naturally
satisfied if the $h$
and $\phi$ masses are modest in size, $\lsim 200\gev$.
 We focus on the case of
small to moderate $\mh$ and $\mphi$, and
discuss expectations for $h$ and $\phi$ production/detection
at the LHC and a LC in comparison to the SM Higgs boson.
In the regions of parameter space allowed by theoretical and
current experimental constraints,
we find that LHC detection of the $\phi$ is likely to be quite difficult.
In addition, LHC detection of the $h$ is not guaranteed. 
 
One particularly interesting complication for $\xi\neq 0$ is the presence of 
the non-standard decay channel $h \to \phi\phi$. 
The $h\to \phi\phi$ decay
 could easily be present since in the context of the RS model
there is a possibility (perhaps even a slight preference)
for the $\phi$ to be substantially lighter than the $h$.  
In particular, $\mphi<\mh/2$ is a distinct possibility.
We study in detail the phenomenology when $\mphi\leq 60\gev$
for $\mh=120\gev$, for which $h\to\phi\phi$ is possible.

In the main phenomenology section, Sec.~\ref{secpheno5}, we
consider the new physics scales of $\lphi=5\tev$
and $m_1\sim 750\gev$ for the first KK resonance, $h^1$.  These values
imply that constraints from precision electroweak
data and from RunI Tevatron data are clearly satisfied.
For this case, the  $h \to
h^1 \phi+\ldots$ modes,  which  are also potentially very
interesting, are forbidden for the moderate $\mh$ and $\mphi$
values explored here.

For $\mh=120\gev$, for the largest allowed $|\xi|$ values
and for $\mphi$ close to $\mh/2$, the $h\to\phi\phi$ mode will
substantially dilute the rates for the usual search channels.
In fact, we find that $BR(h \to \phi\phi)$
could easily be as large as $30 \div 40\, \%$.
Regardless of the magnitude of $BR(\h\to\phi\phi)$, detection
of this decay would be very important as it provides
 a crucial experimental signature for
non-zero $\xi$. 

Of course, it is also possible that $\mphi>2\mh$. 
Because of the typically large size of the $\phi hh$ coupling, we find
that $\phi\to hh$ decays will have a large branching
ratio even when $\mphi>2\mw$.

We give additional details regarding
direct detection of the $\phi$ for the portion of parameter
space for which $h\to \phi\phi$ decays are important.  
Prospects for direct detection at the LHC are not encouraging. 
At the LC, one
should be able to detect $e^+e^-\to Z \phi$ using the recoil
mass technique; $b$-tagging is not necessarily reliable 
due to the possibility that 
$\phi$ decays will be dominated by the $gg$ mode.

In addition to the above, we give a first assessment of whether or
not the anomalous contribution to the $\h gg$, $\h \gam\gam$,
$\phi gg$ and $\phi \gam\gam$ couplings could be observed experimentally.
Deviations in these couplings-squared due to the anomalous contribution
are plotted and compared to the errors expected from the 
outlined experimental procedures for extracting such deviations.
Prospects in the case of the $\h$ are relatively encouraging.

In a second phenomenology section, Sec.~\ref{secpheno1}, we
consider the case of much lighter new physics scales
set by $\lphi=1\tev$. In this case, we consider
both large $m_1$, which avoids precision EW and RunI constraints,
but requires large five-dimensional curvature, and
the small curvature values of $m_1=100\gev$ and $300\gev$.
It is not clear if these latter cases are ruled out by precision
electroweak and/or RunI Tevatron data. If such low scales
are allowed, the $\h\to \phi\phi$ branching ratio becomes
even more prominent for our sample choice of $\mh=120\gev$.
For $\mh=200\gev$ and above, $\h\to h^1\phi$ decays
rapidly emerge and become dominant for $\mh\gsim 500\gev$
in the case of $m_1=100\gev$. However,
it must be kept in mind that for such large $\mh$ values
other new physics must compensate the consequent large 
precision electroweak contributions from the Higgs loop graphs.

Overall, the Randall-Sundrum scenario leads to a fascinating extension
of the usual Higgs phenomenology, especially if radion-Higgs
mixing is present, as is most naturally the case.

{\it Note added.} In the course of preparing this paper, 
another article appeared dealing with the 
variations of the couplings of the $h$ and $\phi$
and their branching ratios 
due to the curvature-scalar mixing~\cite{Hewett:2002nk}.

\vspace*{0.6cm}
\centerline{ACKNOWLEDGMENTS}

\vspace*{0.3cm}

B.G. thanks Zygmunt Lalak, Krzysztof Meissner and Jacek Pawelczyk for useful discussions. J.F.G would like to thank J. Wells for useful discussions.
B.G. is supported in part by the State Committee for Scientific
Research under grant 5~P03B~121~20 (Poland). J.F.G. is supported
by the U.S. Department of Energy and by the Davis Institute for High
Energy Physics.

\clearpage

\section{Appendix: Feynman Rules}

In this Appendix, we summarize the relevant Feynman rules for the 
mass eigenstates $\phi$, $h$ and $h_{\mu\nu}^n$.
Rules for 
the $VV$ couplings of the $\phi$ and $h$, the $f\anti f$
couplings of the $\phi$ and $h$, and the tri-linear self-couplings
among the $\phi$, $h$ and $h_{\mu\nu}^n$ fields are given. 
We note that we are employing a normalization in which
$\sum_{\mu\nu}[\eps_{\mu\nu}^i]^* \eps^{j\,\mu\nu}=2\delta^{ij}$.
In this case, the
spin sum for the $h_{\mu\nu}^n$ state polarizations is
$\sum_{i=1,5}\eps_{\mu\nu}^i(k)\eps_{\rho\sigma}^{i\,*}(k)= B_{\mu\nu\rho\sigma}(k)$, where
\bea
B_{\mu\nu\rho\sig}(k)&=&\left(\eta_{\mu\rho}- {k_\mu k_\rho\over m_n^2}\right)
\left(\eta_{\nu\sig}-{k_\nu k_\sig\over m_n^2}\right)
+\left(\eta_{\mu\sig}- {k_\mu k_\sig\over m_n^2}\right)
\left(\eta_{\nu\rho}-{k_\nu k_\rho\over m_n^2}\right)\nonumber\\
&&-{2\over 3}\left(\eta_{\mu\nu}- {k_\mu k_\nu\over m_n^2}\right)
\left(\eta_{\rho\sig}-{k_\rho k_\sig\over m_n^2}\right)\,.
\eea
 The parameters
$a,b,c,d$ define the mixing between the $\xi=0$ states and the $\xi\neq 0$
mass eigenstates.  They are defined in Eqs.~(\ref{hform})
and (\ref{phiform}).  In the $\xi=0$ limit,
$d=-a=1$, $c=b=0$. The auxiliary functions for the $gg$ and $\gam\gam$
couplings of the $h$ and $\phi$ (Fig.~\ref{vvfffig}) are
given by 
\bea
F_{1/2}(\tau)&=&-2\tau[1+(1-\tau)f(\tau)]\,,\\
F_1(\tau)&=&2+3\tau+3\tau(2-\tau)f(\tau)\,,
\label{f112}
\eea
for spin-1/2 and spin-1 loop particles, respectively, with 
\bea
f(\tau)&=&-\quarter\ln\left[-{1+\sqrt{1-\tau}\over 1-\sqrt{1-\tau}}\right]^2\,
\nonumber\\
 & = & \left\{  \begin{array}{cl}
            {\rm \arcsin}^2(1/\sqrt{\tau}) \,,   &\qquad \tau \geq 1\,, \\
            -\frac{1}{4}\left[\ln \left(\frac{1+\sqrt{1-\tau}}{
                                              1-\sqrt{1-\tau}}\right)
                              -i\pi\right]^2\,, & \qquad \tau\leq 1\,.
                \end{array}
 \right.
\eea
The variable $\tau$ for a given loop is defined as $\tau\equiv 4m^2/M^2$,
where $m$ is the mass of the internal loop particle and $M$
is the mass of the scalar state, $h$ or $\phi$.
In Fig.~\ref{trilinearfig}, the $\anti g$ couplings used in the branching ratio
formulae, Eqs.~(\ref{htophiphiwdth}) and (\ref{htohnphiwdth}),
are defined. 
In the $\xi=0$ SM limit, $\anti g_{hhh}=-3\mh^2/\vo$ (where $\mh=\mho$
at $\xi=0$).
We have not derived detailed results for the $h h^n h^n$
and $\phi h^nh^n$ couplings, 
although, as explained in the text, such couplings do exist.


\begin{figure}[p]
\vspace*{.5in}
{\color{blue}
\begin{center}
\begin{picture}(330,50)(40,0)
\Text(15,45)[]{$h$}
\Text(65,65)[]{$Z,\,\mu$}
\Text(65,5)[]{$Z,\,\nu$}
\DashArrowLine(40,35)(15,35){3}
\Photon(40,35)(65,55){2}{4}
\Photon(40,35)(65,15){2}{4}
\Text(85,35)[l]{$i{g m_Z\over c_W}(d+\gam b)\eta^{\mu\nu}$}
\Text(215,45)[]{$h$}
\Text(265,65)[]{$f$}
\Text(265,5)[]{$\anti f$}
\DashLine(240,35)(215,35){3}
\ArrowLine(240,35)(265,55)
\ArrowLine(265,15)(240,35)
\Text(285,35)[l]{$-i{g\over 2}{m_f\over \mw}(d+\gam b)$}
\end{picture}\\
\end{center}
}
\smallskip
{\color{blue}
\begin{center}
\begin{picture}(330,50)(40,0)
\Text(15,45)[]{$h$}
\Text(65,65)[]{$W,\,\mu$}
\Text(65,5)[]{$W,\,\nu$}
\DashLine(40,35)(15,35){3}
\Photon(40,35)(65,55){2}{4}
\Photon(40,35)(65,15){2}{4}
\Text(85,35)[l]{$i{g m_W}(d+\gam b)\eta^{\mu\nu}$}
\end{picture}\\
\end{center}
}
\smallskip
{\color{blue}
\begin{center}
\begin{picture}(330,50)(40,0)
\Text(15,45)[]{$\phi$}
\Text(65,65)[]{$Z,\,\mu$}
\Text(65,5)[]{$Z,\,\nu$}
\DashLine(40,35)(15,35){3}
\Photon(40,35)(65,55){2}{4}
\Photon(40,35)(65,15){2}{4}
\Text(85,35)[l]{$i{g m_Z\over c_W}(c+\gam a)\eta^{\mu\nu}$}
\Text(215,45)[]{$\phi$}
\Text(265,65)[]{$f$}
\Text(265,5)[]{$\anti f$}
\DashLine(240,35)(215,35){3}
\ArrowLine(240,35)(265,55)
\ArrowLine(265,15)(240,35)
\Text(285,35)[l]{$-i{g\over 2}{m_f\over \mw}(c+\gam a)$}
\end{picture}\\
\end{center}
}
\smallskip
{\color{blue}
\begin{center}
\begin{picture}(330,50)(40,0)
\Text(15,45)[]{$\phi$}
\Text(65,65)[]{$W,\,\mu$}
\Text(65,5)[]{$W,\,\nu$}
\DashLine(40,35)(15,35){3}
\Photon(40,35)(65,55){2}{4}
\Photon(40,35)(65,15){2}{4}
\Text(85,35)[l]{$i{g m_W}(c+\gam a)\eta^{\mu\nu}$}
\end{picture}\\
\end{center}
}
\smallskip
{\color{blue}
\begin{center}
\begin{picture}(330,50)(40,0)
\Text(15,45)[]{$\phi,h$}
\Text(65,65)[]{$g,\,\mu,\,a$}
\Text(65,45)[]{$k_1$}
\Text(65,5)[]{$g,\,\nu,\,b$}
\Text(65,28)[]{$k_2$}
\Vertex(40,35){3}
\DashLine(40,35)(15,35){3}
\Gluon(40,35)(65,55){2}{4}
\Gluon(40,35)(65,15){2}{4}
\Text(85,35)[l]{$ic_g\delta^{ab}[k_1\cdot k_2\eta^{\mu\nu}-k_1^\nu k_2^\mu]:\,
c_g=-{\alpha_s\over 4\pi v}[g_{fV}\sum_i F_{1/2}(\tau_i)-2b_3g_{r}]$}
\end{picture}\\
\end{center}
}
\smallskip
{\color{blue}
\begin{center}
\begin{picture}(330,50)(40,0)
\Text(15,45)[]{$\phi,h$}
\Text(65,65)[]{$\gamma,\,\mu$}
\Text(65,45)[]{$k_1$}
\Text(65,5)[]{$\gamma,\,\nu$}
\Text(65,28)[]{$k_2$}
\Vertex(40,35){3}
\DashLine(40,35)(15,35){3}
\Photon(40,35)(65,55){2}{4}
\Photon(40,35)(65,15){2}{4}
\Text(85,35)[l]{$ic_\gam [k_1\cdot k_2\eta^{\mu\nu}-k_1^\nu k_2^\mu]:\,
c_\gam=-{\alpha \over 2\pi v}[g_{fV}
\sum_i e_i^2N_c^i F_i(\tau_i)-(b_2+b_Y)g_{r}]$}
\end{picture}\\
\end{center}
}
\vspace*{-.2in}
\caption{\it Feynman rules for the $VV$ and $f\anti f$ 
couplings of the scalars 
$h$ and $\phi$. Note: Since there are no pseudoscalars, there are
no single $V$ vertices. We have dropped the extra terms related
to gauge fixing, see Ref.~\cite{csaki_mix}, as appropriate
when considering on-shell $W$'s or $Z$'s or when working in the unitary gauge.
For $gg,\gam\gam$ final states, we give only the on-shell
rules. There, 
$\tau_i=4m_i^2/m_{h,\phi}^2$ where $m_i$ is the mass of the internal
loop particle. The auxiliary functions for spin-1/2 and spin-1 loop
particles are defined in the Appendix text. 
The SU(3)$\times$SU(2)$\times$U(1) $\beta$ function coefficients are $b_3=7$,
$b_2=19/6$ and $b_Y=-41/6$. For $c_g$, 
the $\sum_i$ is over all colored fermions (assumed to have $N_c^i=3$).
For $c_\gam$, the $\sum_i$ comprises all charged fermions (including
quarks, with $N_c^i=3$ and $e_i=2/3$ or $-1/3$, and leptons, 
with $e_i=-1$ and $N_c^i=1$) 
and the $W$ boson (with $e_i=1$ and $N_c^i=1$).
For the $h$,
$g_{fV}=g_{ZZ\h}=d+\gam b$ and $g_{r}=\gam b$. For the $\phi$,
$g_{fV}=g_{ZZ\phi}=c+\gam a$ and $g_{r}=\gam a$.}
\label{vvfffig}
\end{figure}
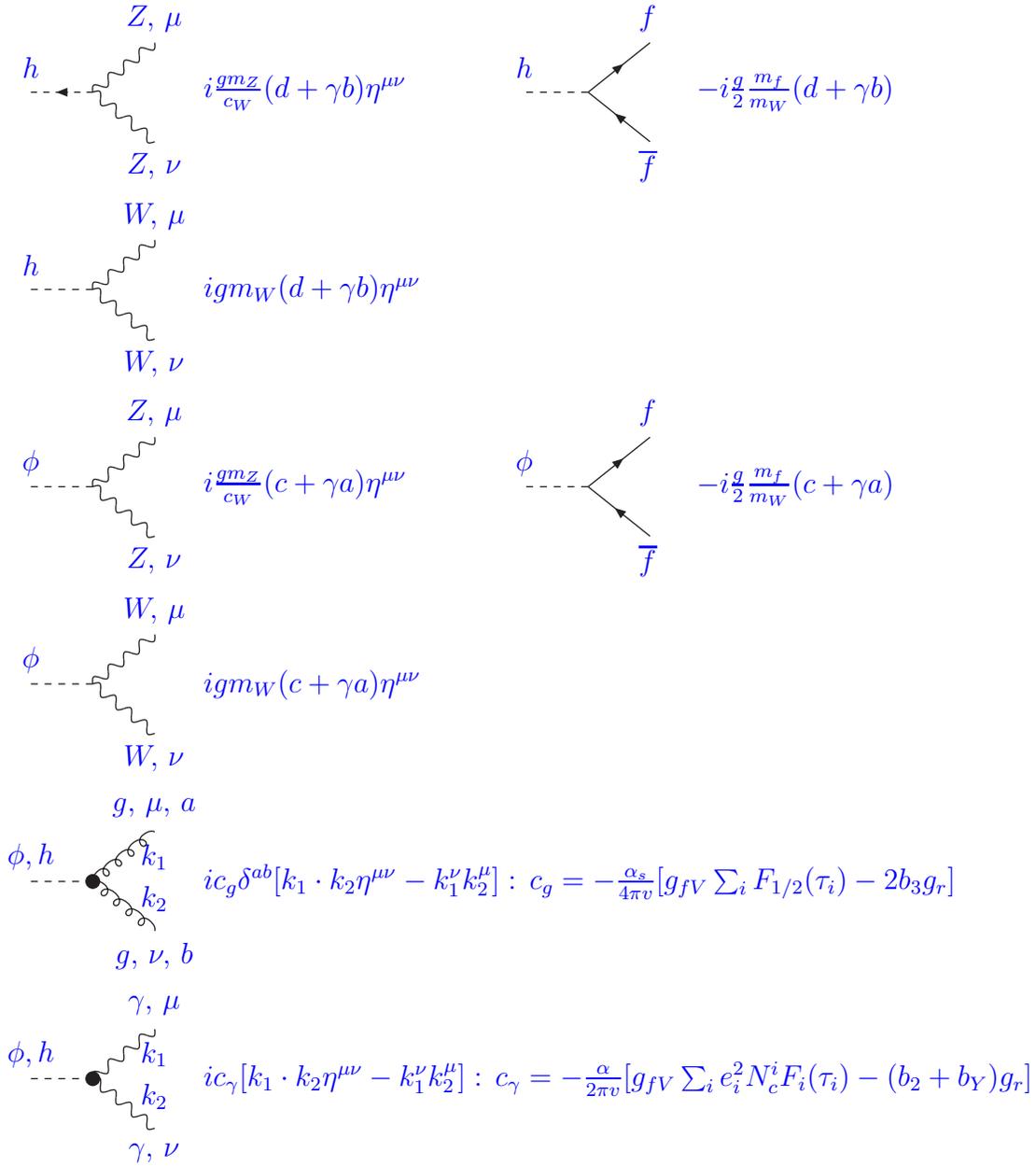

\begin{figure}[p]
{\color{blue}
\begin{center}
\begin{picture}(700,50)(10,0)
\Text(15,45)[]{$h$}
\Text(15,25)[]{$k_3$}
\Text(65,65)[]{$h$}
\Text(65,45)[]{$k_1$}
\Text(65,25)[]{$h$}
\Text(65,5)[]{$k_2$}
\DashArrowLine(40,35)(15,35){3}
\DashArrowLine(40,35)(65,55){3}
\DashArrowLine(40,35)(65,15){3}
\Text(85,35)[l]{$i{\anti g_{hhh}}\equiv$}
\Text(95,35)[l]{ $\phantom{{\anti g_{hhh}}\equiv} \hspace*{-.1in}
 {i\over \lphi}\biggl[bd\Bigl\{\left[12b\gam\xi+d(6\xi+1)\right](k_1^2+k_2^2+k_3^2)-12d\mho^2\Bigr\}-3\gam^{-1}d^3\mho^2\biggr]$}
\end{picture}\\
\end{center}
}
\smallskip
{\color{blue}
\begin{center}
\begin{picture}(700,50)(10,0)
\Text(15,45)[]{$\phi$}
\Text(15,25)[]{$k_3$}
\Text(65,65)[]{$\phi$}
\Text(65,45)[]{$k_1$}
\Text(65,25)[]{$\phi$}
\Text(65,5)[]{$k_2$}
\DashArrowLine(40,35)(15,35){3}
\DashArrowLine(40,35)(65,55){3}
\DashArrowLine(40,35)(65,15){3}
\Text(85,35)[l]{$i{\anti g_{\phi\phi\phi}}\equiv$}
\Text(95,35)[l]{$\phantom{{\anti g_{\phi\phi \phi}}\equiv} 
\hspace*{-.1in}
 {i\over \lphi}\biggl[ac\Bigl\{\left[12a\gam\xi+c(6\xi+1)\right](k_1^2+k_2^2+k_3^2)
-12c\mho^2\Bigr\}-3\gam^{-1}c^3\mho^2\biggr]$}
\end{picture}\\
\end{center}
}
\smallskip
{\color{blue}
\begin{center}
\begin{picture}(700,50)(10,0)
\Text(15,45)[]{$h$}
\Text(15,25)[]{$k_3$}
\Text(65,65)[]{$\phi$}
\Text(65,45)[]{$k_1$}
\Text(65,25)[]{$\phi$}
\Text(65,5)[]{$k_2$}
\DashArrowLine(40,35)(15,35){3}
\DashArrowLine(40,35)(65,55){3}
\DashArrowLine(40,35)(65,15){3}
\Text(85,35)[l]{$i{\anti g_{\phi\phi h}}\equiv$}
\Text(95,35)[l]{ $\phantom{{\anti g_{\phi\phi h}}\equiv} \hspace*{-.1in}
 {i\over \lphi}\biggl[\Bigl\{6a\xi(\gam(ad+bc)+cd)+bc^2\Bigr\}(k_1^2+k_2^2)$}
\Text(95,13)[l]{ $\phantom{ttttt}
+c\Bigl\{12ab\gam\xi+2ad+bc(6\xi-1)\Bigr\}k_3^2
-4c(2ad+bc)\mho^2 -3\gam^{-1}c^2 d\mho^2\biggr]$}
\end{picture}\\
\end{center}
}
\smallskip
{\color{blue}
\begin{center}
\begin{picture}(700,50)(10,0)
\Text(15,45)[]{$\phi$}
\Text(15,25)[]{$k_3$}
\Text(65,65)[]{$h$}
\Text(65,45)[]{$k_1$}
\Text(65,25)[]{$h$}
\Text(65,5)[]{$k_2$}
\DashArrowLine(40,35)(15,35){3}
\DashArrowLine(40,35)(65,55){3}
\DashArrowLine(40,35)(65,15){3}
\Text(85,35)[l]{$i{\anti g_{\phi h h}}\equiv$}
\Text(95,35)[l]{ $\phantom{{\anti g_{\phi h h}}\equiv} \hspace*{-.1in}
 {i\over \lphi}\biggl[\Bigl\{6b\xi(\gam(ad+bc)+cd)+ad^2\Bigr\}(k_1^2+k_2^2)$}
\Text(95,13)[l]{ $\phantom{ttttt}
+d\Bigl\{12ab\gam\xi+2bc+ad(6\xi-1)\Bigr\}k_3^2
-4d(ad+2bc)\mho^2-3\gam^{-1}c d^2\mho^2\biggr]$}
\end{picture}\\
\end{center}
}
\smallskip
{\color{blue}
\begin{center}
\begin{picture}(700,50)(10,0)
\Text(15,45)[]{$h^n_{\mu\nu}$}
\Text(15,25)[]{$k_3$}
\Text(65,65)[]{$h$}
\Text(65,45)[]{$k_1$}
\Text(65,25)[]{$\phi$}
\Text(65,5)[]{$k_2$}
\DashArrowLine(40,35)(15,35){3}
\DashArrowLine(40,35)(65,55){3}
\Gluon(40,35)(15,35){2}{3}
\DashArrowLine(40,35)(65,15){3}
\Text(85,35)[l]{$i{\anti g_{n\phi h}}k_{1\,\mu}k_{2\,\nu}\equiv$}
\Text(95,35)[l]{ $\phantom{{\anti g_{n\phi h}}k_{1\,\mu}k_{2\,\nu}:}
{i\over\lwh} 4\Bigl\{3\gam\xi\left[a(\gam b+d)+bc\right]+\half cd
\Bigr\}k_{1\,\mu}k_{2\,\nu}$}
\end{picture}\\
\end{center}
}
\smallskip
{\color{blue}
\begin{center}
\begin{picture}(700,50)(10,0)
\Text(15,45)[]{$h^n_{\mu\nu}$}
\Text(15,25)[]{$k_3$}
\Text(65,65)[]{$\phi$}
\Text(65,45)[]{$k_1$}
\Text(65,25)[]{$\phi$}
\Text(65,5)[]{$k_2$}
\DashArrowLine(40,35)(15,35){3}
\Gluon(40,35)(15,35){2}{3}
\DashArrowLine(40,35)(65,55){3}
\DashArrowLine(40,35)(65,15){3}
\Text(85,35)[l]{$i{\anti g_{n\phi \phi}}k_{1\,\mu}k_{2\,\nu}\equiv$}
\Text(95,35)[l]{ $\phantom{{\anti g_{n\phi \phi}}k_{2\,\nu}k_{3\,\mu}:}
{i\over\lwh} 4\Bigl\{3a\gam\xi\left[a\gam+2c\right]+\half c^2
\Bigr\}k_{1\,\mu}k_{2\,\nu}$}
\end{picture}\\
\end{center}
}
\smallskip
{\color{blue}
\begin{center}
\begin{picture}(700,50)(10,0)
\Text(15,45)[]{$h^n_{\mu\nu}$}
\Text(15,25)[]{$k_3$}
\Text(65,65)[]{$h$}
\Text(65,45)[]{$k_1$}
\Text(65,25)[]{$h$}
\Text(65,5)[]{$k_2$}
\DashArrowLine(40,35)(15,35){3}
\Gluon(40,35)(15,35){2}{3}
\DashArrowLine(40,35)(65,55){3}
\DashArrowLine(40,35)(65,15){3}
\Text(85,35)[l]{$i{\anti g_{nhh}}k_{1\,\mu}k_{2\,\nu}\equiv$}
\Text(95,35)[l]{ $\phantom{{\anti g_{nhh}}k_{1\,\nu}k_{2\,\mu}:}
{i\over\lwh} 4\Bigl\{3b\gam\xi\left[b\gam+2d\right]+\half d^2
\Bigr\}k_{1\,\mu}k_{2\,\nu}$}
\end{picture}\\
\end{center}
}
\caption{\it Feynman rules for the tri-linear vertices in the scalar sector. 
All momenta are outward flowing.
In the $h_{\mu\nu}^n$ vertices, 
we have made use of the symmetry of $h_{\mu\nu}^n$ under 
$\mu\leftrightarrow\nu$.}
\label{trilinearfig}
\end{figure}

\end{document}